\documentclass[3p]{elsarticle} 

\myfooter[R]{First version: May 22, 2020. This version: \today.}

\usepackage{amscd}      
\usepackage{amsbsy}     
\usepackage{amsmath}    
\usepackage{amsfonts}   
\usepackage{amsthm}     
\usepackage{enumerate}  
\usepackage{fancybox}   
\usepackage{geometry}   

\usepackage[linesnumbered,boxed,longend]{algorithm2e} 

\usepackage{tikz}       
\usetikzlibrary{arrows,calc,decorations,shapes}

\usepackage{cancel}     

\usepackage{caption}    
\usepackage{fancyhdr}   
\usepackage{float}      
\usepackage{graphics}   
\usepackage{graphicx}   
\usepackage{subcaption} 

\usepackage{bbm}        
\usepackage[OT1]{fontenc}

\usepackage{bookmark}   
\usepackage{hyperref}   
\usepackage{cleveref}   

\usepackage{hhline}     
\usepackage{longtable}  
\usepackage{tabularx}   

\numberwithin{equation}{section}	     

\DeclareMathOperator{\SP}{SP}

\newcommand{\xva}{\text{xVA}}

\DeclareMathOperator{\CVA}{CVA}

\DeclareMathOperator{\DVA}{DVA}

\DeclareMathOperator{\PD}{PD}

\DeclareMathOperator{\EPE}{EPE}

\DeclareMathOperator{\PnL}{P\&L}

\DeclareMathOperator{\PnLExplained}{\PnL_E}

\DeclareMathOperator{\PnLUnexplained}{\PnL_U}

\DeclareMathOperator{\PnLTrade}{\PnL_V}

\DeclareMathOperator{\PnLHedge}{\PnL_H}

\DeclareMathOperator{\PnLPortfolio}{\PnL_P}

\newcommand{\half}{\frac{1}{2}}
\newcommand{\bigOh}{\mathcal{ O}}       
\renewcommand{\d}{{\rm d}}
\newcommand{\e}{{\rm e}}                
\newcommand{\E}{\mathbb{ E}}            
\newcommand{\I}{\mathbbm{1}}            

\newcommand{\N}{\mathcal{N}}            
\renewcommand{\P}{\mathbb{ P}}          
\newcommand{\Q}{\mathbb{ Q}}            

\def\ds{{\d}s}
\def\dS{{\d}S}
\def\dt{{\d}t}

\newcommand{\pderiv}[2]{\frac{\partial#1}{\partial #2}}
\newcommand{\ppderiv}[2]{\frac{\partial^2 #1}{\partial {#2}^2}}
\newcommand{\pcrossderiv}[3]{\frac{\partial^2 #1}{\partial {#2} \partial {#3}}}

\newcommand{\indicator}[1]{\I_{\left\{#1\right\}}}
\newcommand{\expPower}[1]{\e^{#1}}
\newcommand{\expBrace}[1]{\exp{\left\{#1\right\}}}
\newcommand{\equalDistr}{\stackrel{\text{d}}{=}}

\newcommand{\rdef}{=:}
\newcommand{\ldef}{:=}

\newcommand{\var}{\mathbb{V}\text{ar}}      
\newcommand{\Var}{\var}                     

\theoremstyle{remark}

\newtheorem*{rem}{Remark}
\theoremstyle{definition}

\newcommand{\scaleSubfigure}{0.71}
\newcommand{\tradeColor}{blue}
\newcommand{\hedgeColor}{red}
\newcommand{\hedgeCvaColor}{cyan}
\newcommand{\interestColor}{teal}
\newcommand{\defaultColor}{brown}
\newcommand{\newTradeColor}{violet}
\newcommand{\bankColor}{blue!20}
\newcommand{\marketColor}{red!20}

\newcommand{\impliedVolFigureSize}{0.49\linewidth}
\newcommand{\mertonConvergenceFigureSize}{0.49\linewidth}
\newcommand{\cashFlowFigureSize}{\scaleSubfigure\linewidth}
\newcommand{\resultFigureSize}{0.49\linewidth}

\newcommand{\bank}{B}
\newcommand{\tradeCF}{c^{V}}
\newcommand{\tradeAllCF}{C^{V}}
\newcommand{\hedgeCF}{c^H}
\newcommand{\hedgeAllCF}{C^H}
\newcommand{\hedgeVal}{H}
\newcommand{\jump}{J}
\newcommand{\strike}{K}
\newcommand{\nrPathsMC}{L}
\newcommand{\shortRate}{r}
\newcommand{\recovRate}{R}
\newcommand{\stock}{S}
\newcommand{\tradeVal}{V}
\newcommand{\wealth}{w}
\newcommand{\brownian}{W}
\newcommand{\logStock}{X}

\newcommand{\marketData}{\gamma}
\newcommand{\tradePos}{\zeta}
\newcommand{\hedgePos}{\eta}
\newcommand{\intensity}{\lambda}
\newcommand{\drift}{\mu}
\newcommand{\mean}{\mu}
\newcommand{\hazardRate}{\xi}
\newcommand{\strategy}{\Pi}
\newcommand{\vol}{\sigma}
\newcommand{\default}{\tau}
\newcommand{\normPDF}{\phi}
\newcommand{\normCDF}{\Phi}

\newcommand{\withCCR}{\text{CCR}}
\newcommand{\withoutCCR}{\text{NoCCR}}
\newcommand{\trading}{\text{trading}}

\newcommand{\jumpIntensity}{\hazardRate_{\jump}}
\newcommand{\jumpMean}{\mean_{\jump}}
\newcommand{\jumpVol}{\sigma_{\jump}}
\newcommand{\jumpProcess}{X_{\jump}}
\newcommand{\poissonProcess}{X_{\mathcal{P}}}
\newcommand{\poissonRate}{\hazardRate_{\mathcal{P}}}
\newcommand{\strategyWithCCR}{\strategy^{\withCCR}}
\newcommand{\strategyWithoutCCR}{\strategy^{\withoutCCR}}
\newcommand{\wealthWithCCR}{\wealth^{\withCCR}}
\newcommand{\wealthWithoutCCR}{\wealth^{\withoutCCR}}
\newcommand{\strategyXVA}{\strategy^{\xva}}
\newcommand{\strategyTrading}{\strategy^{\trading}}
\newcommand{\wealthXVA}{\wealth^{\xva}}
\newcommand{\wealthTrading}{\wealth^{\trading}}
\newcommand{\impliedVol}{\vol_{\text{imp}}} 

\title{A Computational Approach to Hedging Credit Valuation Adjustment in a Jump-Diffusion Setting}

\begin{document}

\author[1,2]{Thomas van der Zwaard\corref{cor1}}
\ead{T.vanderZwaard@tudelft.nl}
\author[1,2]{Lech A.~Grzelak}
\ead{L.A.Grzelak@tudelft.nl}
\author[1,3]{Cornelis W.~Oosterlee}
\ead{C.W.Oosterlee@cwi.nl}
\cortext[cor1]{Corresponding author at Delft Institute of Applied Mathematics, TU Delft, Delft, the Netherlands.}
\address[1]{Delft Institute of Applied Mathematics, Delft University of Technology, Delft, the Netherlands}
\address[2]{Rabobank, Utrecht, the Netherlands}
\address[3]{CWI - National Research Institute for Mathematics and Computer Science, Amsterdam, the Netherlands}

\begin{abstract}
    \noindent This study contributes to understanding Valuation Adjustments ($\xva$) by focussing on the dynamic hedging of Credit Valuation Adjustment ($\CVA$), corresponding Profit \& Loss ($\PnL$) and the $\PnL$ explain.
    This is done in a Monte Carlo simulation setting, based on a theoretical hedging framework discussed in existing literature.
    We look at hedging $\CVA$ market risk for a portfolio with European options on a stock, first in a Black-Scholes setting, then in a Merton jump-diffusion setting.
    Furthermore, we analyze the trading business at a bank after including $\xva$s in pricing.
    We provide insights into the hedging of derivatives and their $\xva$s by analyzing and visualizing the cash-flows of a portfolio from a desk structure perspective.
    The case study shows that not charging $\CVA$ at trade inception results in an expected loss.
    Furthermore, hedging $\CVA$ market risk is crucial to end up with a stable trading strategy.
    In the Black-Scholes setting this can be done using the underlying stock, whereas in the Merton jump-diffusion setting we need to add extra options to the hedge portfolio to properly hedge the jump risk.
    In addition to the simulation, we derive analytical results that explain our observations from the numerical experiments.
    Understanding the hedging of $\CVA$ helps to deal with $\xva$s in a practical setting.
\end{abstract}

\maketitle

\section{Introduction}  \label{sec:introduction}

{\let\thefootnote\relax\footnotetext{The views expressed in this paper are the personal views of the authors and do not necessarily reflect the views or policies of their current or past employers.}}

Since the 2007-2008 global financial crisis, financial institutions have been required to apply and report Valuation Adjustments ($\xva$s) for over-the-counter (OTC) derivatives and hedge the associated risks.
These adjustments to the risk-neutral price of a derivative account for previously neglected risks that were revealed during the crisis.
Credit Valuation Adjustment ($\CVA$), corresponding to Counterparty Credit Risk (CCR), was the first of many $\xva$s.
$\CVA$ volatility is one of the major drivers behind the large losses observed during the crisis~\cite{BCBS201106}.
Linear contracts~\footnote{Linear contracts/instruments have contractual cash-flows that are a linear function of the underlying.} are no longer linear when CCR is included in the valuation.
Hence, not including CCR in pricing results in an incorrect hedging policy.
Next, $\xva$ pricing evolved with various other $\xva$s, see~\cite{Green201511,Gregory201509} for more information.
Calculating the increasing number of $\xva$s is computationally challenging, and has attracted significant academic and corporate interest.

The first literature on $\xva$ pricing appeared before and during the crisis~\cite{Cherubini200507,Gregory200902,PythkinZhu200707}.
To date, several books have addressed the topic~\cite{Green201511,Gregory201509,OosterleeGrzelak201911}.
There are three streams in literature on $\xva$ pricing, all describing the mathematical problem in different ways.
First of all, there approach where the $\xva$ is expressed as an expectation, which can numerically be approximated using a Monte Carlo approach.
A set of risk factors is simulated in a Monte Carlo engine, after which the future exposures that contribute to the $\xva$ metrics are evaluated along the simulated paths~\cite{BrigoCapponi200911,BrigoMasetti200511,BrigoPallaviciniPapatheodorou201107,PallaviciniPeriniBrigo201112,PallaviciniPeriniBrigo201212}.
The Monte Carlo approach allows for scalable computations.
Hence, this type of method is typically implemented by banks.
Second, there is the PDE approach that aims to solve the $\xva$ PDE directly~\cite{ArreguiSalvadorVazques201709,BurgardKjaer201109}.
Dimensionality is one of the downsides of this approach, though low-dimensional non-linear problems can be treated highly accurately.
Last, there is the BSDE approach, which may be non-trivial in a regression based Monte Carlo approach~\cite{BieleckiRutkowski201507,Crepey201501Part1,Crepey201501Part2,
HenryLabordere201207,HenryLabordereTanTouzi201402}.
Literature on $\xva$ has focussed on deriving the mathematical pricing equations, as well as addressing the computational challenges that arise when solving these pricing equations.
In our work, we closely examine $\xva$ pricing and market risk hedging.
We focus on the dynamic hedging of $\CVA$ market risk, where we study the cash-flows of a portfolio from a desk structure perspective.
This results in an improved understanding of the $\CVA$ hedging mechanics.
We show that the $\CVA$ market risk in the portfolio needs to be hedged to end up with a stable trading strategy.

The trading strategy is a combination of all trading positions with a counterparty and the corresponding hedging strategy~\cite{PallaviciniPeriniBrigo201112,PallaviciniPeriniBrigo201212}.
A wealth account~\cite{Duffie200111} is connected to this trading strategy, it tracks the total wealth of the trading strategy over time.
This is a cash account that accrues interest.
Bielecki and Rutkowski formalized this framework~\cite{BieleckiRutkowski201507}.
The generic trading strategy formulation provides a useful framework to analyze exchanges of assets and cash.
Furthermore, the generic formulation is useful for the hedging of $\xva$s.
We provide numerical examples and insights in a Monte Carlo simulation setting, starting in the Black-Scholes world.

The constant volatility is one of the known shortcomings of the Black-Scholes model.
The resulting flat implied volatility is in clear contrast with the market's implied volatility smile.
To properly manage this smile risk, several streams of modelling have emerged.
Among these approaches are the local volatility models (e.g., Dupire~\cite{Dupire1994}, Derman and Kani~\cite{DermanKani1998}) and the stochastic (local) volatility models (e.g., Heston~\cite{Heston199304}, Hagan \textit{et al.}~\cite{HaganKumarLesniewskiWoodward200209}, Lipton~\cite{Lipton200202}).
Although these models fit in the generic hedging framework, we choose to focus on jump-diffusion models, which exhibit heavy tails in the distribution of stock returns and fit well with the jump patterns observed for stocks.
Jump-diffusion models extend the Black-Scholes model with independently distributed jumps driven by a Poisson process.
Typical choices of jump size distributions are a double exponential distribution, Kou~\cite{Kou200208}, or normal distributed jumps, Merton~\cite{Merton197601}.
We choose to work with the latter.
Further research on the Merton jump-diffusion model has been on calibrating the model and hedging the jump risk that partially drives option prices under this model~\cite{HeKennedyColemanForsythLiVetzal200601,KennedyForsythVetzal200905}.

Therefore, after studying the dynamic hedging of $\CVA$ market risk in a Black-Scholes setting, we do the same in a Merton jump-diffusion setting.
In addition, we examine the impact of defaults on the portfolio.
The Profit and Loss of the trading strategy is examined to assess the hedge's performance.
We show that, for a portfolio of European options, not including $\CVA$ in the pricing results in an expected loss.
Hence, we interpret $\CVA$ as a fair compensation for the credit risk of the counterparty.
Charging $\CVA$ to the client at trade inception overcomes the expected loss.
$\CVA$ can be treated as a cash amount, however, this ignores the dependencies of the $\CVA$ on the market variables.
After including the $\CVA$ hedge in the strategy, we assess the impact of jumps in the underlying stock on the portfolio.
In particular, we find that in the context of $\CVA$, the jump risk can be mitigated to a large degree by adding extra hedging instruments to the strategy.
Rather than merely performing a simulation, we also derive analytic results to understand and explain our numerical observations.

This paper is organized as follows.
In Section~\ref{sec:pnl} we provide background information on Profit and Loss, which is used to examine the outcome of the hedging strategies.
In Section~\ref{sec:hedging} we introduce the trading strategy and the corresponding wealth account.
Next, the numerical simulation of the future market states and results of the various hedging strategies will be addressed in Sections~\ref{sec:marketSimulation} and~\ref{sec:results} respectively.
Finally, the work is concluded in Section~\ref{sec:conclusion}.

\section{Profit and Loss} \label{sec:pnl}

\textit{Profit and Loss} ($\PnL$) is a financial institution's income statement. This is an officially reported number that summarizes the change in \textit{Mark-to-Market} (MtM) value over a period of time, normally one business day.
During this period, the institution examines which part of the change in MtM can be attributed to market moves and other predefined effects, such as the passing of time.
This process is also called \textit{$\PnL$ attribution}, or \textit{$\PnL$ explain}, which explains the impact of the daily market movements on the value of the portfolio.
The goal is to verify that the modelled risk factors satisfactorily explain the change in portfolio value.
The $\PnL$ explain process runs after the closing of every business day, so it is a backward looking measure.
Typically, the variance between two consecutive days is also examined.

A financial institution wants to explain the $\PnL$ as well as possible.
Hence the residual $\PnL$, which from now on we address as \textit{$\PnL$ unexplained}, should be as small as possible.
Yet explaining all $\PnL$ is insufficient, as extreme numbers that are completely explained are also undesirable.
Therefore, institutions put thresholds/limits on both $\PnL$ and $\PnL$ unexplained at a portfolio level, where the thresholds depend on the portfolio composition.
In addition to these $\PnL$ limits, market risk limits on particular risk types are in place for both traders and portfolios.
The aim of these market risk limits is to make sure the trading activities remain within the institution's predefined risk appetite and that exposure to certain market drivers is bounded.
This naturally limits the $\PnL$ as well.
The portfolios are not always entirely flat in terms of risk, as in practice it is not feasible to rebalance the hedge daily, for example due to transaction costs.
Furthermore, having a flat portfolio is not always the goal as this may be a way to express a view on the market.
In practice, the residual risk of a portfolio is being monitored.

For $\xva$s a separate explain process exists, which is analogous to the one for the MtM value.
Limits on market risk drivers, $\PnL$ and $\PnL$ unexplained are present.

In literature the following approaches to $\PnL$ explain can be found~\cite{Green201511,AndersenPiterbarg2010Vol3}:
\begin{enumerate}
    \item Let $\tradeVal(t, \marketData(t))$ denote value $\tradeVal$ at time $t$ with a set \textit{market data} $\marketData(t)$. The first approach is a Taylor-based explain process where the partial derivatives of the instrument $\tradeVal$ with respect to $\marketData(t)$ are used to explain the difference in $\tradeVal(t_{k-1},\marketData(t_{k-1}))$ and $\tradeVal(t_k,\marketData(t_k))$, $t_{k-1} < t_k$.~\footnote{Green refers to this as ``risk-based explain''~\cite[Section 21.1.4]{Green201511}. Andersen and Piterbarg use the term ``second order $\PnL$ predict''~\cite[Section 22.2.1]{AndersenPiterbarg2010Vol3}.}
    In other words,
    \begin{align}
        \tradeVal(t_k,\marketData(t_k)) - \tradeVal(t_{k-1},\marketData(t_{k-1}))
            &= \left.\pderiv{\tradeVal}{t}\right|_{t_{k-1}}  \d t + \mathcal{A}(t_{k-1}) + \PnLUnexplained(t_k), \label{eq:explainRiskBased} \\
        \mathcal{A}(t_{k-1})
            &= \sum_{i=1}^n \left.\pderiv{\tradeVal}{\marketData_i}\right|_{t_{k-1}}  \d \marketData_i + \half \sum_{i=1}^n \sum_{j=1}^n \left.\pcrossderiv{\tradeVal}{\marketData_i}{\marketData_j}\right|_{t_{k-1}} \d \marketData_i \d \marketData_j. \nonumber
    \end{align}
    The left hand side of Equation~\eqref{eq:explainRiskBased} is $\PnL(t_k)$. $\PnLUnexplained(t_k)$ is the unexplained $\PnL$. The remaining terms can be interpreted as $\PnLExplained(t_k)$, i.e., the explained $\PnL$.

    \item In the second approach, the \textit{theta effect} $ \left.\pderiv{\tradeVal}{t}\right|_{t_{k-1}}  \d t $, is neglected and we are only interested in the change of value $\tradeVal$ in the interval $\left[t_{k-1},t_k\right]$ as a result of movements in the underlying market data $\marketData_i$.~\footnote{Green refers to this as ``orthogonal explain''~\cite[Section 21.1.4]{Green201511}. Andersen and Piterbarg use the term ``perturbed market data approach''~\cite[Section 22.1.5]{AndersenPiterbarg2010Vol3}.}
    In other words,
    \begin{align}
        \tradeVal(t_{k-1}, \marketData(t_k))  - \tradeVal(t_{k-1}, \marketData(t_{k-1}))
            &= \mathcal{A}(t_{k-1}) + \PnLUnexplained(t_k). \label{eq:explainOrthogonal}
    \end{align}
    The left hand side of Equation~\eqref{eq:explainOrthogonal} is $\PnL(t_k)$, and the $\mathcal{A}(t_{k-1})$ term is $\PnLExplained(t_k)$.
    The effect of ignoring the theta should not be too large, as $\PnL$ explain is a daily process, but the size of the effect will depend on the portfolio composition.
    This method considers instrument values at the same date, and the difference in value is caused by a perturbed market data set.

    \item Another approach is a high-level attribution using effects to explain the $\PnL$ in a consecutive manner.~\footnote{Green refers to this as ``step-wise explain''~\cite[Section 21.1.4]{Green201511}. Andersen and Piterbarg use the term ``bump and do not reset'' type of $\PnL$ explain, a.k.a. the ``waterfall explain''~\cite[Section 22.2.2.1]{AndersenPiterbarg2010Vol3}.}
    Before starting the explain process, a list of effects is composed, which are applied one-by-one.
    After applying an effect, the portfolio is revalued and the difference is attributed to the particular effect.
    These changes will be permanent, so when applying a new effect, the old effects stay in place.
    This implies that cross-effects are introduced, which cannot be allocated properly, as well as an order dependence on the effects.
    Yet this approach should yield similar results as the previous ones.
    Examples of the effects used in the $\PnL$ attribution can be: time decay (including holiday effects, carry effects as well as shifts in market data), market data changes (IR / FX / inflation / bonds / credit), turns, fixings, and trade closing.
\end{enumerate}

When a financial institution observes jumps in $\PnL$, these are investigated.
Real-world jumps are typically caused by trade insertions or cancellations (captured via the theta term), rather than by underlying market movements.
Though theta contributions will be present in practice, they are not relevant for our discussion, which focusses on the effect of market movements on the $\PnL$ of a portfolio.
Hence we use the orthogonal explain approach as per Equation~\eqref{eq:explainOrthogonal}, which makes use of perturbed market data.
Here we can simply take the next day's market data as the perturbed market data.

\section{Hedging framework} \label{sec:hedging}

We start with a risk-neutral pricing framework, where we consider the various contributions to the price of a derivative.
Using the standard replicating portfolio argument~\cite{OosterleeGrzelak201911}, the price of a derivative is equivalent to the price of a replicating portfolio containing other securities.

We define a trading strategy as a combination of positions in a set of available trading instruments, accompanied by the hedging positions in hedging instruments, which mitigate the market risk associated with the positions in the trading instruments.
We assume that the short-selling of all assets is possible.
In practice, hedges do not necessarily mitigate all the market risk, but certain limits are placed per market risk factor on a portfolio level.
A trader will make sure that the exposure of a portfolio to certain risk factors will remain below these predefined thresholds by taking the necessary positions in market instruments.
Here we will assume that all the market risk is always reduced by the dynamic hedging strategy, i.e., we choose the hedging positions that together replicate the risk profile of the trading positions as good as possible.

The \textit{trading strategy} $\strategy(t)$ is generically defined for $N$ \textit{trading instruments} $\tradeVal_i(t)$ in which \textit{trading positions} $\tradePos_i(t)$ are taken, which are assumed to be exogenously provided.
Analogously, we consider $M$ \textit{hedging instruments} $\hedgeVal_i(t)$ in which \textit{hedging positions} $\hedgePos_i(t)$ are taken.
The trading strategy $\strategy(t)$ is then summarized as follows:
\begin{align}
    \strategy(t)
      &=  \sum_{i = 1}^{N}  \tradePos_i(t) \tradeVal_i(t) + \sum_{i = 1}^{M}  \hedgePos_i(t) \hedgeVal_i(t) . \label{eq:strategy}
\end{align}

We start in a Black-Scholes setting, where European options are considered for the sake of clarity and ease of computation.
However, the generic formulation allows for much more flexibility, for example, for the case of a large portfolio of interest rate swaps.
The underlying interest rate risks are then eliminated by taking hedging positions in the par instruments used to build the underlying yield curve(s) that are in turn required to value the portfolio of interest rate swaps.

Define the \textit{cash-flows} for the trading and hedging instruments respectively by $\tradeCF_{i,j}(t)$ and $\hedgeCF_{i,j}(t)$, denoting the time $t$ value of the $j$-th cash-flow paid at time $T_j$, corresponding to respectively instruments $\tradeVal_i$ and $\hedgeVal_i$, which in total generate respectively $n_i$ and $m_i$ cash-flows. We then define the time $t$ value of the cumulative cash-flows  corresponding to instrument $\tradeVal_i$ and $\hedgeVal_i$ respectively as
\begin{align}
    \tradeAllCF_i(t)
        &= \sum_{j=1}^{n_i} \tradePos_i(T_j) \tradeCF_{i,j}(t)\indicator{T_j \leq t}, \ \
    \hedgeAllCF_i(t)
        = \sum_{j=1}^{m_i} \hedgePos_i(T_j) \hedgeCF_{i,j}(t) \indicator{T_j \leq t},  \label{eq:cumulativeCF}
\end{align}
which is to be interpreted as the quantity representing all cash-flows paid up and till time $t$, taking into account the time value of money, $\bank(t)$, being the \textit{bank account}.
This is the solution of $\d \bank(t) = \shortRate(t) \bank(t) \dt$, where $\bank(t_0) = 1$ and with \textit{risk-free rate} $\shortRate(t)$.

In parallel with the trading strategy $\strategy(t)$, we define a \textit{wealth process} $\wealth(t)$ that represents the total wealth realized over time, obtained by summing up all the profits and losses over time.
There are two sources for changes in wealth: one is the rebalancing of positions in instruments, the other results from cash-flows associated with the various instruments.
We assume that $\tradeVal_i(t)$ and $\hedgeVal_i(t)$ denote the value after exchange of all cash-flows at time $t$.
For example, if a cash-flow takes place at date $t$, the values $\tradeVal_i(t)$ and $\hedgeVal_i(t)$ do not contain the value of this cash-flow.
Rebalancing the trading positions, and consecutively the hedging positions, takes place after the exchange of cash-flows.
The wealth is a cash amount that accrues interest over time at the risk-free rate, which means that we assume the institution can borrow and lend at the risk-free rate.

At time $t_0$, the wealth will be composed of the cost to enter the positions in all instruments, including any potential cash-flows taking place at $t_0$:
\begin{align}
    \wealth(t_0)
        &= - \strategy(t_0)
        + \sum_{i=1}^N \tradeAllCF_i(t_0)
        + \sum_{i=1}^M \hedgeAllCF_i(t_0). \label{eq:wealthInit}
\end{align}

The rebalancing of the trading and hedging positions is denoted by $\d \tradePos_i(t)$ and $\d \hedgePos_i(t)$ respectively.
We write the following (recursive) expression for the wealth at $t_0 \leq t_{k-1} < t_k$:
\begin{align}
    \wealth(t_k)
        &= \wealth(t_{k-1})\frac{\bank(t_k)}{\bank(t_{k-1})}
        - \sum_{i=1}^N \bank(t_k) \int_{t_{k-1}}^{t_k} \frac{\tradeVal_i(u)}{\bank(u) }\d\tradePos_i(u)
        + \sum_{i=1}^N \bank(t_k) \int_{t_{k-1}}^{t_k} \frac{\d\tradeAllCF_i(u)}{\bank(u) } \nonumber \\
        &\quad - \sum_{i=1}^M \bank(t_k) \int_{t_{k-1}}^{t_k} \frac{\hedgeVal_i(u)}{\bank(u) }\d\hedgePos_i(u)
        + \sum_{i=1}^M \bank(t_k) \int_{t_{k-1}}^{t_k} \frac{\d\hedgeAllCF_i(u)}{\bank(u) }. \label{eq:wealthContRecursive}
\end{align}
The first term in Equation~\eqref{eq:wealthContRecursive} is the wealth at the previous point in time $t_{k-1}$ that has accrued interest.
This is followed by the re-balancing and cash-flows of respectively the trading and hedging instruments.

The continuous time formulation from Equation~\eqref{eq:wealthContRecursive} is discretized in time $t_0 < t_1 < \ldots < t_{k-1} < t_k \ldots < t_K$:~\footnote{For a generic function $f(\cdot)$ we define $\d f(t_k) = f(t_k) - f(t_{k-1})$}:
\begin{align}
    \wealth(t_k)
        &\approx \wealth(t_{k-1})\frac{\bank(t_k)}{\bank(t_{k-1})}
        - \sum_{i=1}^N \tradeVal_i(t_k) \d\tradePos_i(t_k)
        + \sum_{i=1}^N \d \tradeAllCF_i(t_k) \nonumber \\
        &\quad - \sum_{i=1}^M \hedgeVal_i(t_k)\d\hedgePos_i(t_k)
        + \sum_{i=1}^M \d\hedgeAllCF_i(t_k). \label{eq:wealthDiscrRecursive}
\end{align}
In this context, the initial wealth from Equation~\eqref{eq:wealthInit} remains valid.
Furthermore, $t_K$ is the final time at which either all trading instruments have matured, or at which all positions are closed.
We keep the positions in trading and hedging instruments constant over a time interval $(t_{k-1}, t_k]$.
The terms $\d \tradeAllCF_i(t_k)$ and $\d \hedgeAllCF_i(t_k)$ respectively represent the trading and hedging instruments' cash-flows present in time interval $(t_{k-1},t_k]$:
\begin{align}
     \d \tradeAllCF_i(t_k)
        &=  \sum_{j=1}^{n_i} \tradePos_i(T_j) \tradeCF_{i,j}(t_k) \indicator{t_{k-1} < T_j \leq t_k}, \ \
     \d \hedgeAllCF_i(t_k)
        =\sum_{j=1}^{n_i} \hedgePos_i(T_j) \hedgeCF_{i,j}(t_k) \indicator{t_{k-1} < T_j \leq t_k}. \label{eq:cumulativeCFDifferential}
\end{align}

Strategy~\eqref{eq:strategy} is by design \textit{self-financing}, i.e., there is no cash in- and/or outflow during the lifetime of the strategy.
In other words, $\displaystyle  \strategy(t) + \wealth(t) = 0$ holds on average.
At $t_0$ we assume that any funding required to set up the trading strategy is obtained from the internal Treasury department.

\subsection{Output metrics} \label{sec:outputMetrics}

Now that the trading strategy $\strategy(t)$ and wealth $\wealth(t)$ are defined, we define several output metrics to evaluate the performance of the trading strategy.
In a perfect world, the self-financing constraint $\displaystyle  \strategy(t) + \wealth(t) = 0$ holds on average for all $t_k\in[t_0, t_{K}]$, i.e., \\
$\displaystyle \E_{t_0}\left[\strategy(t_k) + \wealth(t_k)\right] = 0$.
In particular, we are interested in this condition at final time $t_{K}$, where all open positions are closed such that $\strategy(t_K)=0$.
So, we evaluate the \textit{expected terminal wealth}: $\displaystyle \E_{t_K}\left[ \wealth(t_0)\right]$.
Assuming that one can continuously rebalance the hedging positions, the terminal wealth should be zero on average.

In addition to the terminal wealth, we also examine $\PnL$.
The relevant market information at time $t$ is represented by $\marketData(t)$ and we rewrite $\tradeVal_i(t)$ as follows: $\tradeVal_i(t) = \tradeVal_i(t, \marketData(t))$.
This allows us to price product $\tradeVal_i$ at time $t$ with a set of market data $\marketData(t)$.
In a similar fashion, we write $\hedgeVal_i(t) = \hedgeVal_i(t, \marketData(t))$.
Define the following $\PnL$ quantities corresponding to trading strategy~\eqref{eq:strategy}:
\begin{align}
    \PnLTrade(t_k)
        &= \sum_{i = 1}^{N}  \tradePos_i(t_{k-1})\left[ \tradeVal_i(t_{k-1},\marketData(t_k)) - \tradeVal_i(t_{k-1},\marketData(t_{k-1}))\right],\label{eq:pnlTrade} \\
    \PnLHedge(t_k)
        &= \sum_{i = 1}^{M}  \hedgePos_i(t_{k-1})\left[ \hedgeVal_i(t_{k-1},\marketData(t_k)) - \hedgeVal_i(t_{k-1},\marketData(t_{k-1}))\right], \label{eq:pnlHedge} \\
    \PnLPortfolio(t_k)
        &= \PnLTrade(t_k) + \PnLHedge(t_k). \label{eq:pnlPortfolio}
\end{align}
$\PnLTrade$ in Equation~\eqref{eq:pnlTrade} is the $\PnL$ generated by the trading positions.
When the hedging quantities $\hedgePos(t)$ are based on sensitivities of the trading positions, $\PnLHedge$ in Equation~\eqref{eq:pnlHedge} can be seen as a first-order $\PnL$ explain.
By choosing the same instruments for hedging as for the daily $\PnL$ explaining process, we are in fact looking at how well the hedging strategy is able to explain changes in value of the trading strategy as a result of changes in the market data.

The residual risk, after combining trading and hedging positions, is represented by $\PnLPortfolio$ in Equation~\eqref{eq:pnlPortfolio}.
As the hedge is constructed to eliminate all desired sources of randomness, the daily change portfolio value is automatically equal to the $\PnLPortfolio$, i.e., we can rewrite $\PnLPortfolio$ from Equation~\eqref{eq:pnlPortfolio} by means of Equations~\eqref{eq:pnlTrade} and~\eqref{eq:pnlHedge}:
\begin{align}
    \PnLPortfolio(t_k)
        &= \strategy(t_{k-1}, \marketData(t_k)) - \strategy(t_{k-1}, \marketData(t_{k-1})) . \label{eq:pnlPortfolio2}
\end{align}
We then use the orthogonal $\PnL$ attribution process from Equation~\eqref{eq:explainOrthogonal} to assess which portion of the residual risk can be explained.
Note the absence of cash-flows in this discussion.
This is because we want to treat cash-flows in a consistent manner, such that only the market movements are attempted to be explained.

\section{Market simulation} \label{sec:marketSimulation}

The hedging framework from Section~\ref{sec:hedging} will be used in a numerical simulation setting to assess Valuation Adjustments and the hedging thereof.
The starting point of the numerical analysis is choosing a portfolio of trading instruments and the corresponding hedging instruments.
The economic value of a trading instrument is the combination of the risk-free value and Valuation Adjustments, where we choose to look at $\CVA$ only.
We consider only European options for the trading instruments.
$\DVA$ turns out to be trivial for this portfolio composition, so $\DVA$ is ignored in the analysis.
$\CVA$ computations require the following two main components: market exposure and default probability.
The former is introduced in Section~\ref{sec:europeanOptionExposure}, the latter is discussed here directly.

We model the jump-to-default by a Poisson process $\poissonProcess(t)$ with constant hazard rate $\intensity(t) = \poissonRate$, i.e. a homogenous Poisson process.~\footnote{
To gain some intuition on how this Poisson processes works, look at its expected value
$\displaystyle \E \left[ \poissonProcess(t)\right] = \poissonRate t$,
which is the expected number of events in a time interval with length $t$.
So say we have an interval $[0,T]$ of length $T$ where we expect one event every $2T$, then we must set $\poissonRate = \frac{1}{2T}$.}
The deterministic hazard rate (e.g., constant or piece-wise constant) implies that credit events~\footnote{
A credit event is considered to be the first event of a Poisson counting process which occurs at some random time $\default$ with probability
$\displaystyle  \P\left( \poissonProcess(\default + \dt) - \poissonProcess(\default) \left| \poissonProcess(\default) = 0 \right) \right. $,
i.e., the probability of default in interval $[\default, \default+ \dt)$ conditional on survival until time $\default$.}
are independent of the interest rates and deterministic recovery rates.
In addition, assume that the \textit{recovery rate} $R$ is deterministic.
\textit{Survival probability} $\SP(t,T)$ is the probability that the counterparty will survive until time $T$, conditional on survival till time $t$:
\begin{align*}
    \SP(t,T)
        &= \E_t \left[ \expBrace{-\int_t^T \intensity(s) \ds} \right]
        = \expBrace{-\int_t^T \intensity(s) \ds}
        = \expPower{-\poissonRate (T-t)}.
\end{align*}
A credit curve is the credit-analogue of the yield curve, where we do not extract discount factors but survival probabilities from the curve.
In a market implied setting, survival probabilities $\SP(t_0,T)$ are extracted from the market using quotes of highly standardized CDSs.
For $\CVA$ calculations we require knowledge about the \textit{probability of default} $\PD(t,T)$ of a counterparty in a certain time interval $[t,T]$.
The probability of default is related to the survival probability: $\PD(t,T) = 1 - \SP(t,T)$.

We choose to hedge all the market risk introduced by the risk-free value and $\CVA$ corresponding to the trading instruments.
We do not consider the credit risk component introduced by the $\CVA$.
This means that jump risk upon default is not hedged dynamically, meaning that credit risk warehousing takes place.
CDS positions can be used to hedge this residual risk~\cite{Green201511,Gregory201509}, which fits within the current framework.
In Section~\ref{sec:simulationFramework} we look in more detail at the setup of the trading strategy.

\subsection{Model-free European option exposures} \label{sec:europeanOptionExposure}

Given a call option $\tradeVal$ that runs from $t_0$ till maturity $t_K$, we create a grid of \textit{monitoring dates} $t_0 < t_1 < \ldots < t_k \ldots < t_K$ at which we compute \textit{exposures}. The following convenient result can easily be derived for the \textit{discounted expected positive exposure ($\EPE$)} of a call/put option:
\begin{align}
    \EPE(t_0,t_k) &= \tradeVal(t_0).  \label{eq:epeCallPut}
\end{align}

Using the result from Equation~\eqref{eq:epeCallPut} in the formula for $\CVA$, assuming no \textit{Wrong Way Risk (WWR)}, yields:
\begin{align}
    \CVA(t_0)
        &= (1-\recovRate) \sum_{k=1}^{K} \EPE(t_0,t_k) \PD(t_{k-1},t_k) \nonumber \\
        &= (1-\recovRate) \sum_{k=1}^{K} \tradeVal(t_0) \PD(t_{k-1},t_k) \nonumber \\
        &= (1-\recovRate) \tradeVal(t_0) \PD(t_0,t_K). \label{eq:cvaCallPut}
\end{align}
The results in Equations~\eqref{eq:epeCallPut} and~\eqref{eq:cvaCallPut} are model-free up to the level how $\tradeVal(t)$ is computed.~\footnote{$\CVA$ is in essence a compound option on the value of a portfolio.
A compound option refers to an option on an option. Say that we consider a portfolio of a single option, then the $\max$ function in the expected exposure in the $\CVA$ formula makes the $\CVA$ a compound option.}

\subsection{Simulation framework} \label{sec:simulationFramework}
For our trading instruments we choose a constant long unit position (i.e., buy) in a European call option $\tradeVal_1(t)$, i.e., $N=1$ and $\tradePos_1(t)=1 \ \forall t$.~\footnote{This reverse Black-Scholes hedge is an academic setup, as buying a call option and hedging the delta risk is not what one usually encounters.
However, for illustrative purposes this particular setting is chosen.}
The option is assumed to be an OTC deal where the underlying asset $\stock(t)$ is a third-party asset, i.e., $\stock(t)$ is not the asset of the option seller.
This implies the absence of WWR, meaning we assume that the creditworthiness of the option seller and buyer and the underlying asset move independently.
In addition, we assume a constant credit curve over time, i.e., no stochasticity for the credit is used.
Furthermore, we assume the option has a cash-settled payoff.
The market risk of the option is hedged by buying/selling the underlying stock from/to an exchange, i.e., $M=1$ and $\hedgeVal_1(t) = \stock(t)$:
\begin{align}
    \strategy(t)
        &=  \tradeVal_1(t) +  \hedgePos_1(t) \stock(t), \label{eq:tradeStrategyExample}
\end{align}
where $\hedgePos_1(t)$ depends on a model chosen by the financial institution.
Hedging positions $\hedgePos_1(t)$ are rebalanced on a daily basis.
We assume no dividends are paid, though they can easily be added to the framework in the form of cash-flows.
The trading instrument will generate one cash-flow, namely the payoff at maturity in case the option is in the money and the counterparty has not defaulted.

The trading instruments $\tradeVal_i(t)$ can represent the risk-neutral value, which corresponds to the case without Counterparty Credit Risk (CCR).
On the other hand, it can also represent the economic value, which is the sum of the risk-neutral value and $\xva$s that are taken into account, and corresponds to the case where CCR is taken into account, i.e.,
\begin{align}
\left\{
    \begin{array}{rll}
      \tradeVal_1(t) &\ldef \tradeVal(t), &\text{(the case without CCR)}\\
      \tradeVal_1(t) &\ldef \tradeVal(t) - \CVA(t) &\text{(the case with CCR)} \\
                     &\ = \tradeVal(t) \left( 1 - (1-\recovRate) \PD(t,t_K) \right), &
    \end{array}
\right. \label{eq:product}
\end{align}
where for the case with CCR we used Equation~\eqref{eq:cvaCallPut}.
Recall that $\tradeVal(t)$ represents the risk-free value of the option, whereas $\tradeVal_1(t)$ represents the risky value of the option.
This composition of two terms also needs to be taken into account when determining hedging position $\hedgePos_1(t)$: do we hedge only $\tradeVal(t)$ or also $\CVA(t)$?

First, the market ($\stock(t)$, $\tradeVal_i(t)$ and $\hedgeVal_i(t)$) is simulated using a model (e.g., Black-Scholes or Merton jump-diffusion).
While valuing the portfolio, these numbers are assumed to be exogenous, meaning there is no longer a model dependence.
The model used to compute the hedging quantities and perform the $\PnL$ explain needs to be calibrated to the market.
When this model is the same as the model used to simulate the market, the calibration is trivial.

We use two strategies, $\strategyWithoutCCR$ and $\strategyWithCCR$, to assess the effect of CCR.
The former corresponds to the portfolio without simulation of defaults, while the latter includes CCR by simulating default times.
In all cases, the hedges are rebalanced daily and assumed to be free of CCR.
The simulated \textit{default times} $\default=t_d$ are drawn from the same distribution that drives the credit curve.
So, the simulated default times are the first jumps of $\poissonProcess(t)$.
In the experiments, we assume a \textit{risk-free closeout}, where we give back $\tradeVal_1$ to the defaulted counterparty, and in return receive $\recovRate \cdot \tradeVal(\default)$ in case this value is positive.
We assume stock $\stock(t)$ to be independent of any default events of the counterparty.
At default, we re-enter the same deal, at zero additional costs, with another counterparty which is assumed to be credit risk free, for example a clearing house.
This approach is in line with considering the $\CVA$ as the cost of hedging counterparty credit risk, regardless of a counterparty default.
If the $\CVA$ market risk is hedged, this hedge position is closed at default.

\subsection{Black-Scholes dynamics} \label{sec:blackScholesStrategy}

So far we have not assumed any model for $\stock(t)$ and $\tradeVal_1(t)$, only a choice of credit curve and simulation of defaults was made.
The next step is to assume a model for $\stock(t)$ and $\tradeVal_1(t)$.
Our first choice is the Black-Scholes model that allows for analytical option prices and derivatives, in a deterministic interest rate setting.
Recall the Black-Scholes SDE under real-world measure $\P$:
\begin{align}
    \d\stock(t)
        &= \drift \stock(t) \dt + \vol \stock(t) \d\brownian^{\P}(t). \label{eq:bs}
\end{align}
For the simulation of the market scenarios, the SDE~\eqref{eq:bs} is discretized using an Euler scheme.
The risk introduced by the underlying stock is eliminated when choosing the Black-Scholes delta hedging quantity:
\begin{align}
    \hedgePos_1(t)
        &= - \pderiv{\tradeVal_1(t)}{\stock}. \label{eq:hedgePositionBS}
\end{align}
In the risk-free case where $\tradeVal_1(t) = \tradeVal(t)$, hedging quantity~\eqref{eq:hedgePositionBS} holds directly.~\footnote{We do not go into the way the delta hedge is set up, for example via the repo market, as we do not consider funding effects in this paper.}
On the other hand, in the risky case $\tradeVal_1(t) = \tradeVal(t) - \CVA(t)$, where $\CVA$ is hedged, we can write the following using result~\eqref{eq:cvaCallPut}:
\begin{align}
    \hedgePos_1(t)
        &= - \pderiv{\tradeVal(t)}{\stock} \left( 1 - (1-\recovRate) \PD(t,t_K) \right). \label{eq:hedgePositionCVA}
\end{align}

In our experiments, we simulate the synthetic market and do all pricing under the $\Q$ dynamics.
Indeed, the stock dynamics in Equation~\eqref{eq:bs} are under the $\P$ measure.
Under this measure, delta neutrality is imposed and the hedging quantities are derived.
The pricing of the derivatives and calculating risks as in Equations~\eqref{eq:hedgePositionBS} and~\eqref{eq:hedgePositionCVA} is then done under $\Q$ (after imposing a no-arbitrage argument), which is in line with risk-neutral pricing theory.
Then, a synthetic market is created by means of a real-world evolution under the $\P$ measure with drift $\drift$.
In particular, in our numerical experiments we choose the two measures to collapse to the same through the choice of $\drift$ and using the risk-neutral Brownian motion.
For the simulated defaults we assume $\Q$ defaults, which can for example be obtained through a calibration of the chosen credit model to CDSs.
Alternatively, one could plug in historic $\P$ data and still do the valuation and risk calculation under $\Q$.

The simulation is illustrated by a series of schematic drawings that indicate the flow of cash and instruments, see~\ref{sec:deskStructureCashFlows}.
There, we first consider the case without CCR to get a basic understanding of the mechanics of the trading strategy from the perspective of a bank.
Then, CCR is introduced, and the bank's trading desk and $\xva$ desk are represented as a single entity, referred to as the trading desk.
Finally, we remove this assumption by examining the internal exchange of cash-flows and products between the desks.

\subsection{Merton jump-diffusion dynamics} \label{sec:mertonStrategy}

\textit{Jump-diffusion models} aim to overcome the Black-Scholes assumption of constant implied volatility, by introducing independently distributed jumps in the dynamics.
They can generically be defined as~\cite{OosterleeGrzelak201911}:
\begin{align}
    \d \logStock(t)
        &= \drift \dt + \vol \d\brownian^{\P}(t) + \jump(t) \d \jumpProcess^{\P}(t), \ \
    \stock(t)
        = \expPower{\logStock(t)}. \label{eq:mertonDynamics}
\end{align}
The \textit{jumps} $\jump$ arrive according to \textit{Poisson process} $\jumpProcess^{\P}(t)$ that is assumed to be independent of the Brownian process $\brownian^{\P}(t)$.
Typical choices of jump size $\jump$ distributions are a double exponential distribution as introduced by Kou~\cite{Kou200208} or normally distributed jumps as introduced by Merton~\cite{Merton197601}. We choose to work with the latter. Jump magnitudes in this model follow distribution $\jump\sim \N\left( \jumpMean, \jumpVol^2\right)$.

For the simulation of a synthetic market we use an Euler discretization of the $\P$ dynamics from Equation~\eqref{eq:mertonDynamics}.
On the other hand, for pricing options and calculating risks we need $\Q$ dynamics.
To obtain these, in Equation~\eqref{eq:mertonDynamics} we should choose the drift as follows~\footnote{In this case we can use the known result that for $\displaystyle A\sim \N\left(\mean,\vol^2\right)$ we know $\displaystyle \E\left[e^A\right] = \expPower{\mean + \half \vol^2}$.}:
\begin{align*}
    \drift
        &= \shortRate - \jumpIntensity \E\left[ e^{\jump} -1 \right] - \half \vol^2
        = \shortRate - \jumpIntensity \left(\expPower{\jumpMean+ \half \jumpVol^2} - 1 \right)  - \half \vol^2.
\end{align*}
Furthermore, we use a risk-neutral Brownian motion and jump process in the $\Q$ dynamics, which are again assumed to be independent.
For European option prices under the Merton jump-diffusion model an analytic expression exists, see~\ref{sec:mertonOptionPrice}.

After simulating the market $\stock(t) = \expPower{\logStock(t)}$ with the Merton model, we compute the Merton option price using Equation~\eqref{eq:mertonPrice}.
Here we do not make an assumption on $\tradeVal_1(t)$ being a risk-free or risky option value (without $\CVA$ versus with $\CVA$).
Distinguishing between the two cases can be done in a similar fashion as discussed in Section~\ref{sec:blackScholesStrategy}.
For setting up the hedging strategy, we consider the following three approaches.

The first approach is applicable to the situation where the institution's pricing model is misaligned with the market.
In our setup, this is represented by using a Black-Scholes delta hedge as in Section~\ref{sec:blackScholesStrategy}, even though the market does not follow these dynamics.
This is done by extracting the Black-Scholes implied volatility from the Merton option prices observed in the market, and setting up a Black-Scholes delta hedge using the underlying stock. We confirm the results by Naik and Lee~\cite{NaikLee199010} that this hedging strategy is not suitable for the case of an underlying asset driven by both diffusion and jump risk.

The second approach represents the case in which one's pricing model is perfectly aligned with the market.
In this context this means that the Merton delta, as in Equation~\eqref{eq:dVdS}, can be used to compute the hedging quantity.
Equations~\eqref{eq:hedgePositionBS} and~\eqref{eq:hedgePositionCVA} still hold in this case, given that the Black-Scholes deltas are replaced by Merton deltas.

Up to this point, no attempt has been made to hedge the jump risk introduced by the Merton model.
The option pricing formula~\eqref{eq:mertonPrice} contains an infinite sum of scaled Black-Scholes option prices, which is a direct result of the jump size following a continuous distribution.
Thus, in an attempt to hedge the jump risk introduced by the model, one would theoretically need infinitely many options, which is practically infeasible.
Hedging the jump risk has been addressed by adding a number of options to the hedging portfolio~\cite{HeKennedyColemanForsythLiVetzal200601,KennedyForsythVetzal200905}.
This significantly reduces the variance of the portfolio.
In particular, a local minimal variance hedging strategy was examined, combined with a delta position in the underlying stock.
In this paper, we use the analytical jump parameter sensitivities from~\ref{sec:mertonOptionPrice} to determine the hedging positions that aim to eliminate the underlying jump risk.

The third approach includes the hedging of jump risk by adding single option to the set of hedging instruments, i.e., we have $\hedgeVal_1(t)=\stock(t)$ and $\hedgeVal_2(t)$, which is a European option different from option $\tradeVal_1(t)$ we aim to hedge.
Option $\hedgeVal_2(t)$ is a risk-free option, such that no additional $\xva$s are introduced.
To summarize, we have:
\begin{align}
    \strategy(t)
        &= \tradeVal_1(t) + \hedgePos_1(t) \stock(t) + \hedgePos_2(t) \hedgeVal_2(t). \nonumber
\end{align}
We choose stock position $\hedgePos_1(t)$ such that the portfolio is delta-neutral, i.e., such that $\pderiv{\strategy(t)}{S}=0$:
\begin{align}
    \hedgePos_1(t) &=
        - \pderiv{\tradeVal_1(t)}{\stock} - \hedgePos_2(t) \pderiv{\hedgeVal_2(t)}{\stock}. \nonumber
\end{align}
This leaves the question of how to choose $\hedgePos_2(t)$.
We use analytical jump parameter sensitivities to determine the remaining hedging position.
As the strategy contains one hedging option, we must choose one of the jump parameters that we consider most important.
All jump parameters have a level effect, $\jumpMean$ also affects the skew, and $\jumpVol$ also affects the curvature, see~\ref{sec:mertonJumpImpact}.
The level effect from $\jumpIntensity$ is more significant than that for $\jumpMean$ and $\jumpVol$.
Hence, we choose $\jumpIntensity$ to set up the hedge.
So, for $\hedgePos_2(t)$ we have:
\begin{align}
    \hedgePos_2(t) &=
        - \pderiv{\tradeVal_1(t)}{\jumpIntensity}\left[\pderiv{\hedgeVal_2(t)}{\jumpIntensity}\right]^{-1}. \label{eq:mertonSingleOptionHedge}
\end{align}
The partial derivatives w.r.t. $\jumpIntensity$ are computed analytically using the result in Equation~\eqref{eq:dVdXiJ3}.
The hedging strategy as presented here is equivalent to first taking a position in the stock $\hedgePos_1(t)$ to hedge the trading instrument (so Equation~\ref{eq:hedgePositionBS} but with the Merton delta), then taking position~\eqref{eq:mertonSingleOptionHedge} to hedge the jump risk, and then updating $\hedgePos_1(t)$ to account for the additional delta risk generated by this position in the hedging option.

\section{Numerical results} \label{sec:results}
We implement the market simulation, as introduced in Section~\ref{sec:marketSimulation}, in a Monte Carlo setting.
The algorithm used to obtain the numerical results is summarized below (Algorithm~\ref{algo:cvaHedge}).
All results are obtained with the following parameters: $t_0 = 0$, $T = 1$, $\stock(t_0) = 100$, $\shortRate = 0.1$, $\vol = 0.2$, $\strike=95$, $\poissonRate = 0.2$, and $\recovRate = 0.5$.
For the Merton jump-diffusion parameters we choose $\jumpVol = 0.1$, $\jumpMean = -0.125$, and $\jumpIntensity = 0.1$, which can be interpreted as a jump of average size $-11.3\%$ that is expected every 10 years.
We use $\nrPathsMC=10^5$ Monte Carlo paths and $200$ time steps per year to create the set of monitoring dates.
The results are displayed using a number of $100$ shares for the option, such that the controlled notional by the option is $10^4$.
As a result, the vertical axes of the plots can be interpreted as errors in bps.
Furthermore, in the results the bank is represented as a single entity, meaning that the trading desk and $\xva$ desk do not have separate trading strategies and wealth accounts.
This split in the results can easily be made using the schematic drawings in~\ref{sec:deskStructureCashFlows}, but here we do not do this for sake of brevity.

\begin{algorithm}[h]
    \footnotesize
    \SetAlgoLined 
    \DontPrintSemicolon 


    \KwIn{Trading strategies $\strategyWithoutCCR$ and $\strategyWithCCR$, risk factors $\marketData$, number of simulation paths $\nrPathsMC$ and dates $K$}
    \KwOut{Numerical results of the $\CVA$ hedging exercise}
    \BlankLine
    Initialize two portfolios $\strategyWithoutCCR$ and  $\strategyWithCCR$ \;
    Initialize simulation grid of risk factors $\marketData$ \;

    \For{$l \gets 1$ \KwTo $\nrPathsMC$}{
        \For{$k \gets 1$ \KwTo $K$}{
            Simulate all risk factors $\marketData(t_k)$ for path $l$ \;
            Re-value $\strategyWithoutCCR(t_k)$ and  $\strategyWithCCR(t_k)$ for path $l$ \;
            \For{$i \gets 1$ \KwTo $N$}{
                Re-value $\tradeVal_i(t_{k-1}, \marketData(t_k))$ for path $l$ to be used later in the $\PnL$ calculations \;
            }
            \For{$j \gets 1$ \KwTo $M$}{
                Re-value $\hedgeVal_j(t_{k-1}, \marketData(t_k))$ for path $l$ to be used later in the $\PnL$ calculations \;
            }
        }
        Simulate default time $\default_l$ \;
        Perform closeout at default time $\default_l$ for $\strategyWithCCR$ \;
        \For{$k \gets 1$ \KwTo $K$}{
            Compute the relevant $\PnL$ quantities at time $t_k$, prepare $\PnLExplained$ for path $l$ \;
            Compute wealth $\wealthWithoutCCR$ and $\wealthWithCCR$ for path $l$ \;
        }
    }
    Compute required output metrics and visualize results \;
    \caption{$\CVA$ hedging algorithm}
    \label{algo:cvaHedge}
\end{algorithm}

\subsection{Hedging CVA in a Black-Scholes setting} \label{sec:bsHedge}

The first numerical results correspond to the Black-Scholes hedging setting as discussed in Section~\ref{sec:blackScholesStrategy}.
The Black-Scholes model is used for market simulation, computing hedging quantities, and $\PnL$ explain.
First, we find that $\CVA$ must be charged to the client at trade inception to prevent an expected loss of the strategy.
However, this $\CVA$ charge should not be treated as merely a cash amount, which ignores the impact of changes in the underlying market on the $\CVA$ through time.
The $\PnLPortfolio$ volatility resulting from the simulation seems to explode as the maturity date is approached.
This behaviour can be understood through a set of analytical results that allow to interpret this phenomenon.
We conclude that hedging the market risk of the $\CVA$ significantly reduces the $\PnLPortfolio$ volatility, and that therefore $\CVA$ market risk must be hedged.

\subsubsection{CVA as a cash amount} \label{sec:cvaAsCash}
\begin{figure}[!h]
  \centering
  \begin{subfigure}[b]{0.49\linewidth}
    \includegraphics[width=\linewidth]{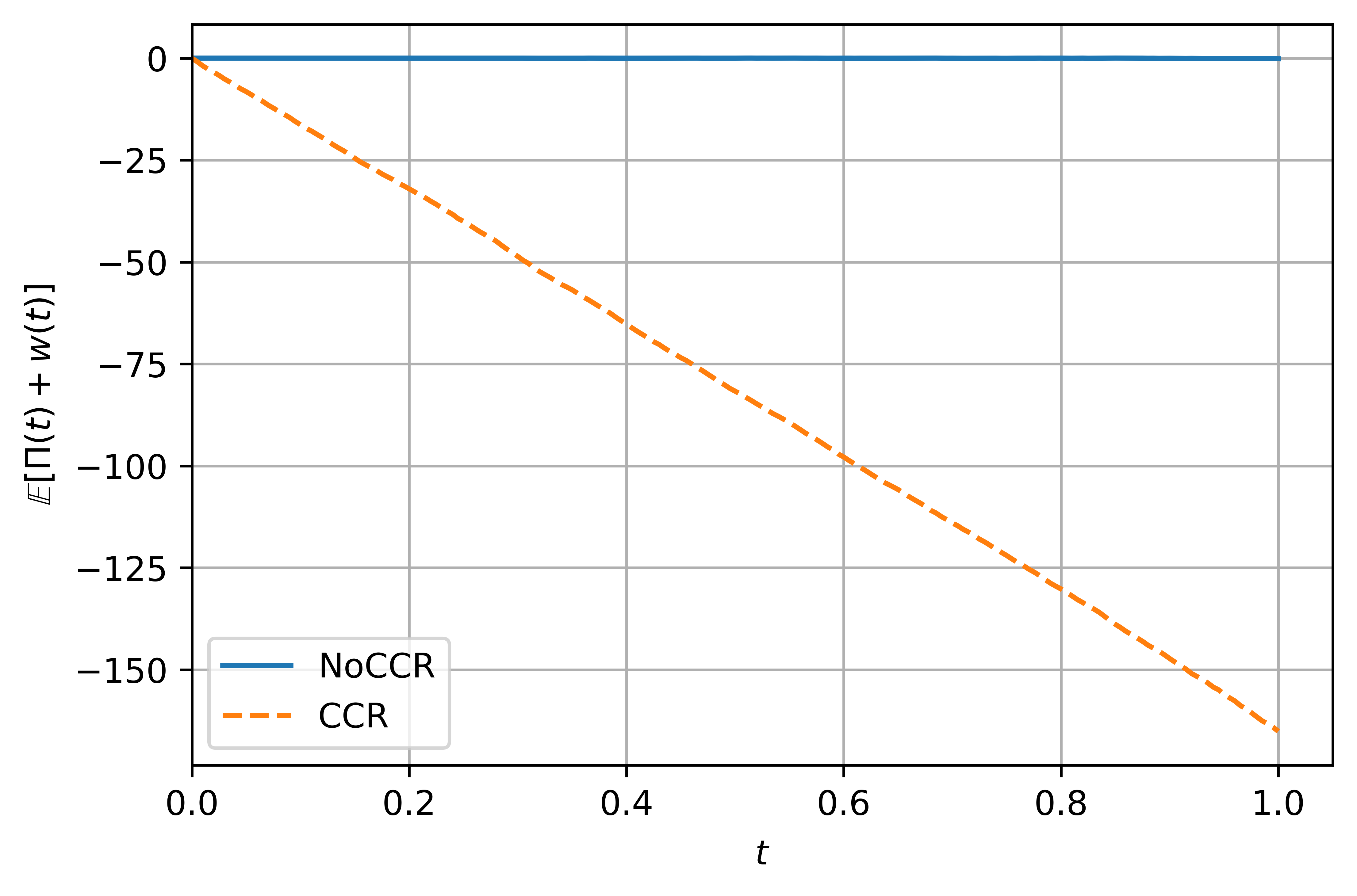}
    \caption{Average $\strategy(t)$ and $\wealth(t)$.}
    \label{fig:bsNoHedgeCvaExcludedTotalAverage}
  \end{subfigure}
  \begin{subfigure}[b]{0.49\linewidth}
    \includegraphics[width=\linewidth]{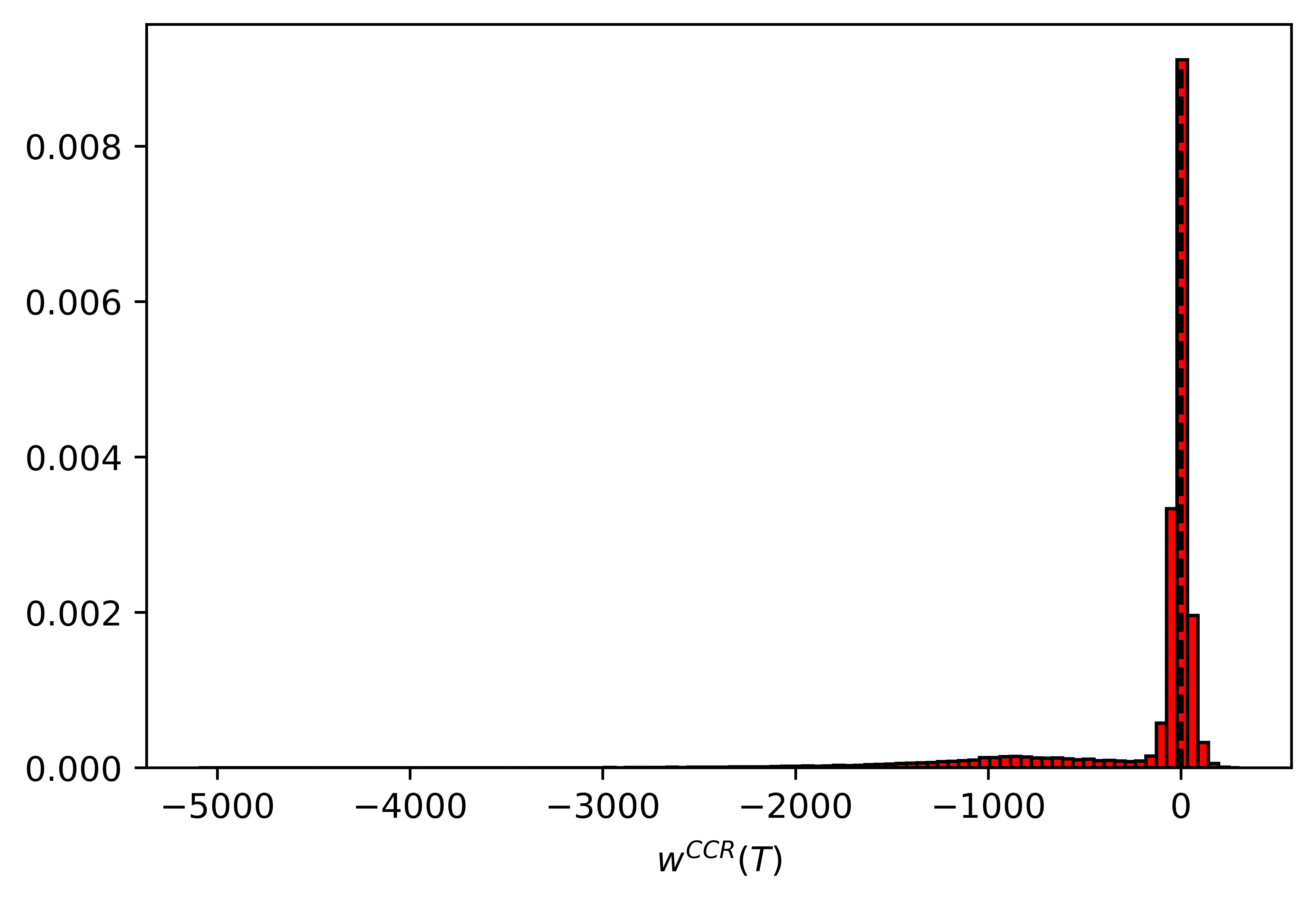}
    \caption{Distribution of $\wealthWithCCR$ at maturity.}
    \label{fig:bsNoHedgeCvaExcludedRiskyWealthDistribution}
  \end{subfigure}
  \caption{$\CVA$ not included in the portfolio.}
  \label{fig:bsNoHedgeCvaExcluded}
\end{figure}

In Figure~\ref{fig:bsNoHedgeCvaExcludedTotalAverage} we confirm an expected loss at maturity if the counterparty could default, but no $\CVA$ is charged.
This is done by examining the impact of simulated defaults on the portfolio, where an expected loss is represented by $\E_{t_0}[\strategyWithCCR(t_K) + \wealthWithCCR(t_K)]<0$.
For $\strategyWithoutCCR$ we have the desired result, namely $\E_{t_0}[\strategyWithoutCCR(t_K) + \wealthWithoutCCR(t_K)] \approx 0$, with some residual noise coming from the Monte Carlo simulation.
The terminal wealth distribution in Figure~\ref{fig:bsNoHedgeCvaExcludedRiskyWealthDistribution} can be interpreted as a bimodal distribution.
The large peak corresponds to the paths without default, whereas the low and wide peak on the left corresponds to the losses encountered as a result of defaults.

\begin{figure}[!h]
  \centering
  \begin{subfigure}[b]{0.49\linewidth}
    \includegraphics[width=\linewidth]{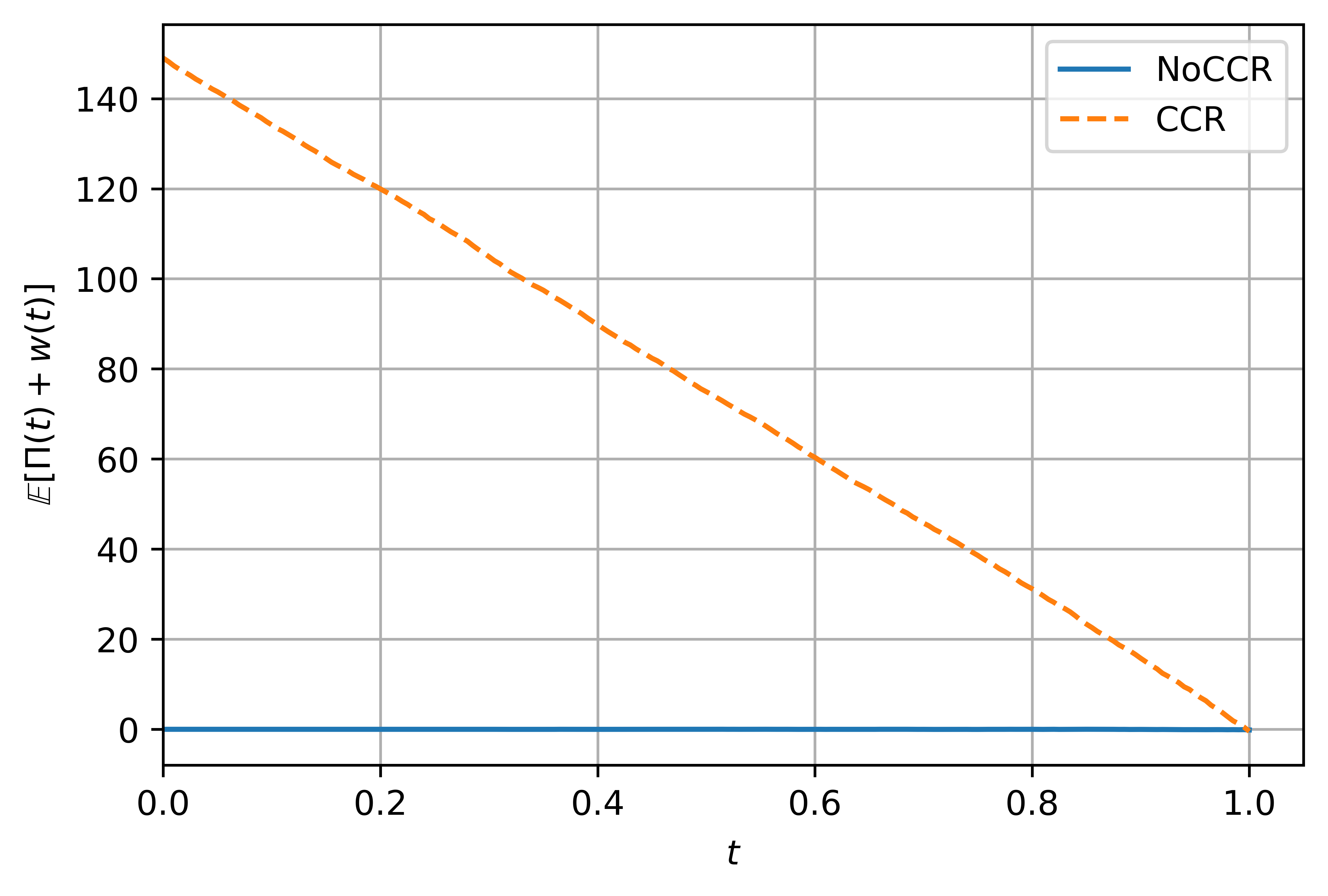}
    \caption{Average $\strategy(t)$ and $\wealth(t)$.}
    \label{fig:bsNoHedgeCvaIncludedTotalAverage}
  \end{subfigure}
  \begin{subfigure}[b]{0.49\linewidth}
    \includegraphics[width=\linewidth]{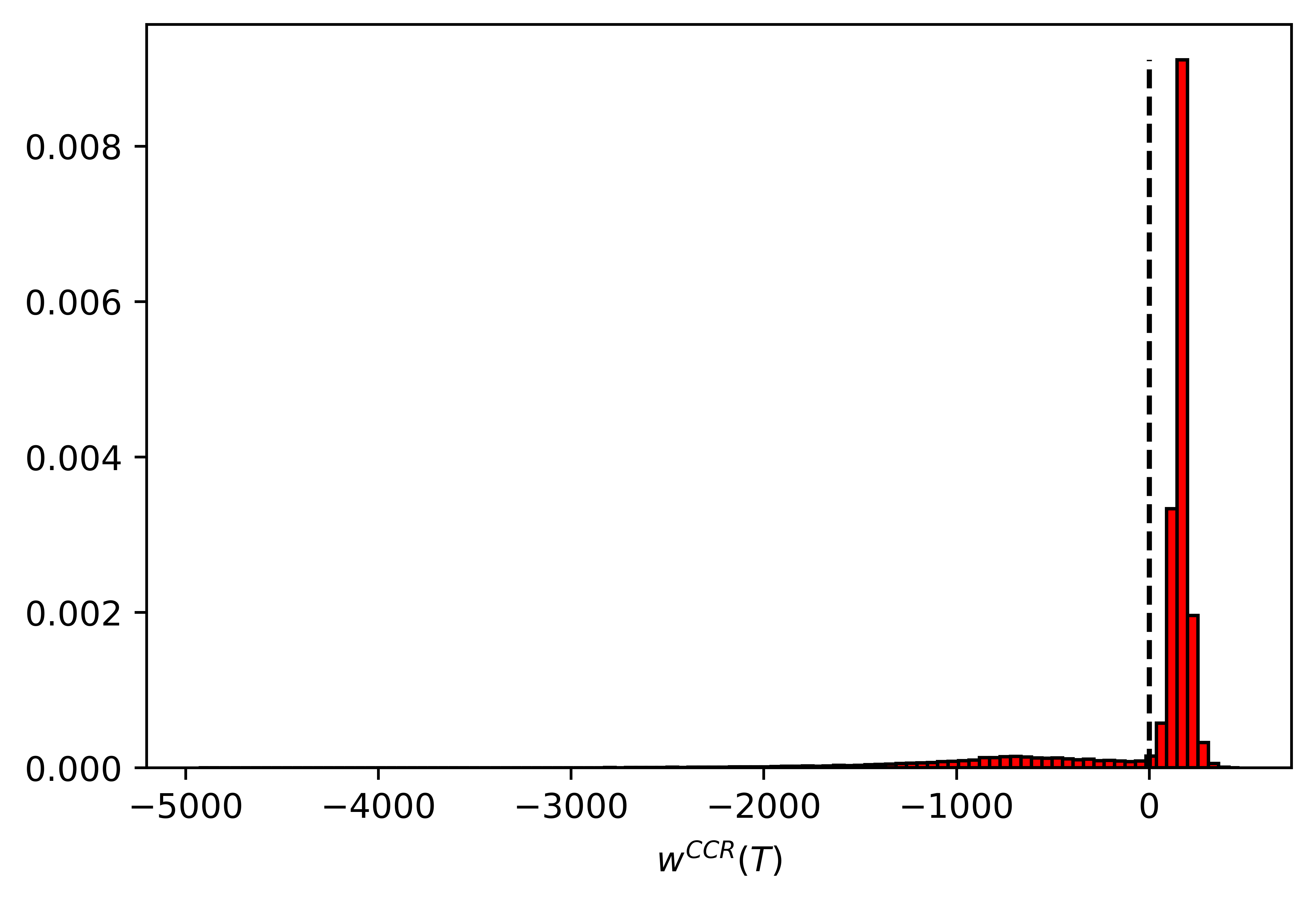}
    \caption{Distribution of $\wealthWithCCR$ at maturity.}
    \label{fig:bsNoHedgeCvaIncludedRiskyWealthDistribution}
  \end{subfigure}
  \caption{$\CVA$ included in the portfolio as a cash amount.}
  \label{fig:bsNoHedgeCvaIncluded}
\end{figure}

Intuitively, $\displaystyle \E_{t_0}[\strategyWithCCR(t_K) + \wealthWithCCR(t_K)] \approx \CVA(t_0)\frac{\bank(t_K)}{\bank(t_0)}$ should hold.
Hence, we add the $\CVA$ charge at inception as a cash amount to the wealth account, meaning that we perform a linear shift in initial wealth.
This should result in $\E_{t_0}[\strategyWithCCR(t_K) + \wealthWithCCR(t_K)] \approx 0$, which is indeed the case, see Figure~\ref{fig:bsNoHedgeCvaIncludedTotalAverage}.
So, the $\CVA$ charge proves to be a fair compensation of the credit riskiness of the counterparty.
Furthermore, we see in Figure~\ref{fig:bsNoHedgeCvaIncludedRiskyWealthDistribution} that the distribution of $\wealthWithCCR(t_K)$ gets shifted to the right due to the $\CVA$ charge added at inception.
The shift is precisely the required amount such that the mean of the distribution is around zero.

We see that charging the $\CVA$ at inception to the client and putting it on the wealth account overcomes the issue of an expected loss.
However, treating the $\CVA$ charge as a cash number with no dependencies on underlying market variables is not useful in practice and naive.
In practice, the $\CVA$ is charged to the client at trade inception.
As time passes, changes in the market result in a change in $\CVA$.

\subsubsection{CVA driven by market risk} \label{sec:cvaWithMarketRisk}

Next, we consider the $\CVA$ to be driven by market risk, but we decide not to hedge this risk.
This means we use the hedging quantity from Equation~\eqref{eq:hedgePositionBS} with the risk-free Black-Scholes delta, even though we have $\tradeVal_1(t) = \tradeVal(t) - \CVA(t)$.
From now on, we refer to the case of Black-Scholes paths and delta as the pure Black-Scholes case.

\begin{figure}[!h]
  \centering
  \begin{subfigure}[b]{\resultFigureSize}
    \includegraphics[width=\linewidth]{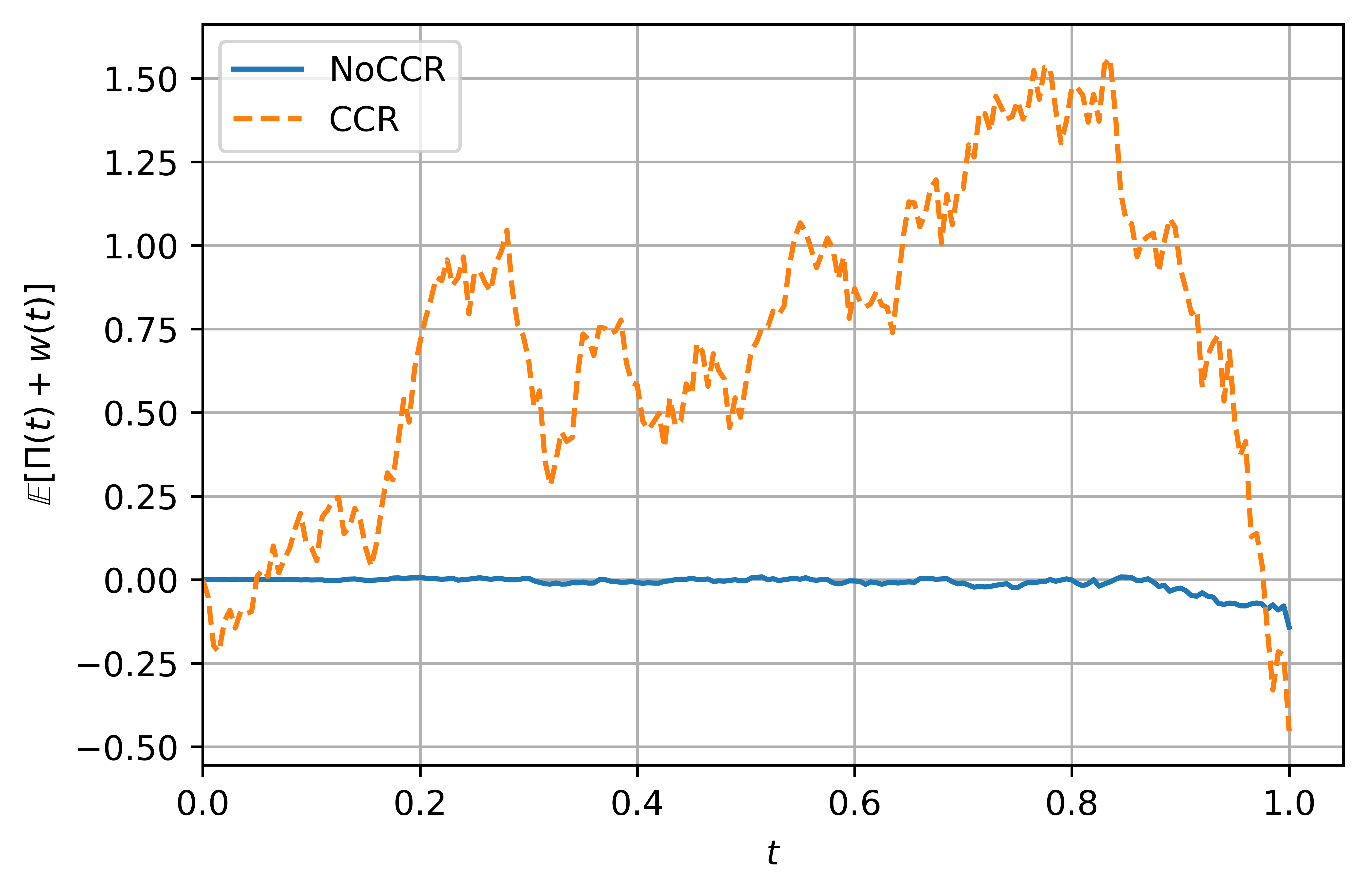}
    \caption{Average $\strategy(t)$ and $\wealth(t)$.}
    \label{fig:bsNoHedgeNewDealTotalAverage}
  \end{subfigure}
  \begin{subfigure}[b]{\resultFigureSize}
    \includegraphics[width=\linewidth]{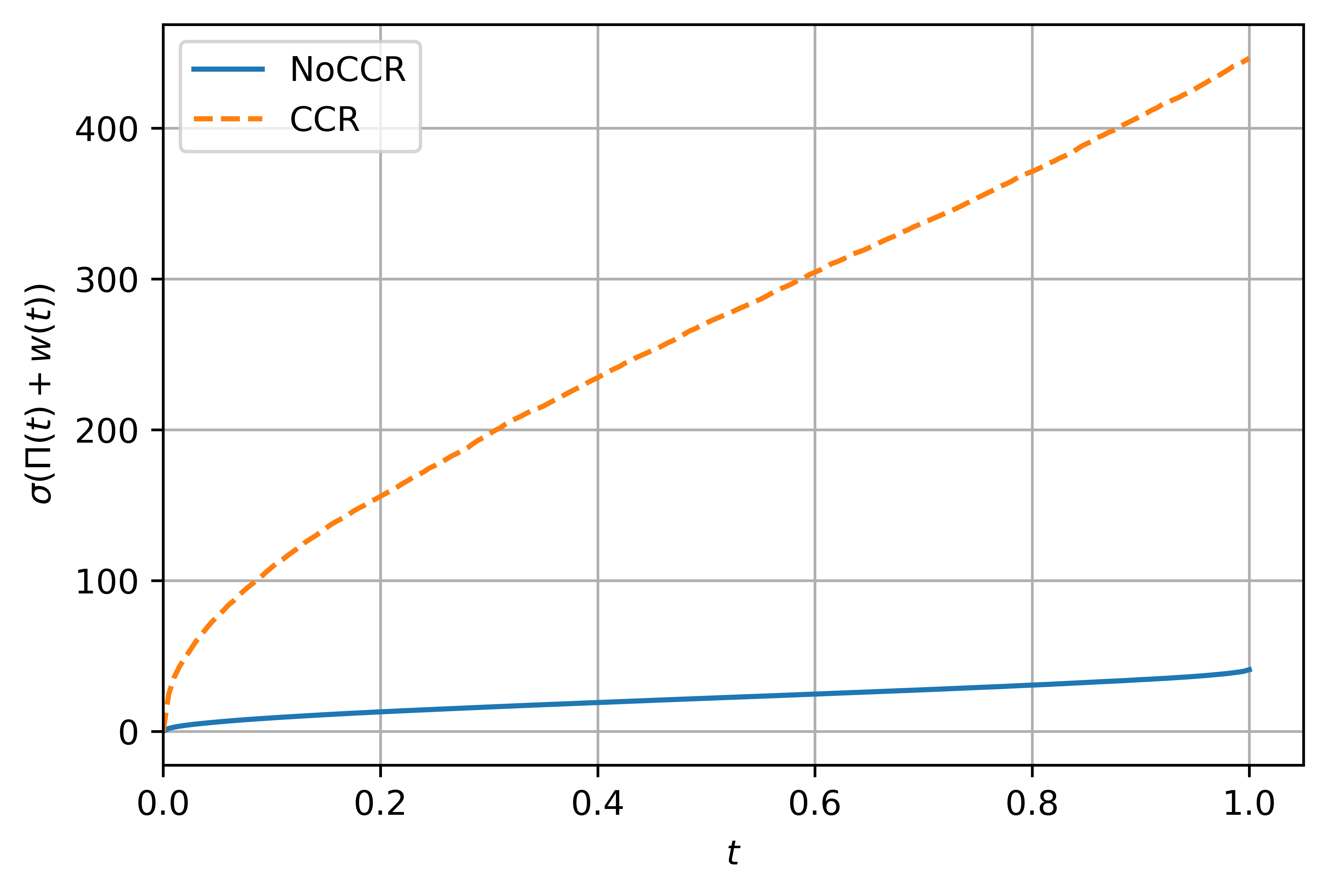}
    \caption{Volatility of $\strategy(t)$ and $\wealth(t)$.}
    \label{fig:bsNoHedgeNewDealTotalStd}
  \end{subfigure}
  \caption{Comparison of $\strategyWithoutCCR$ and $\strategyWithCCR$ using a Black-Scholes market and valuation model. $\CVA$ is not hedged.}
  \label{fig:bsNoHedgeNewDealPart1}
\end{figure}
\begin{figure}[!h]
  \begin{subfigure}[b]{\resultFigureSize}
    \includegraphics[width=\linewidth]{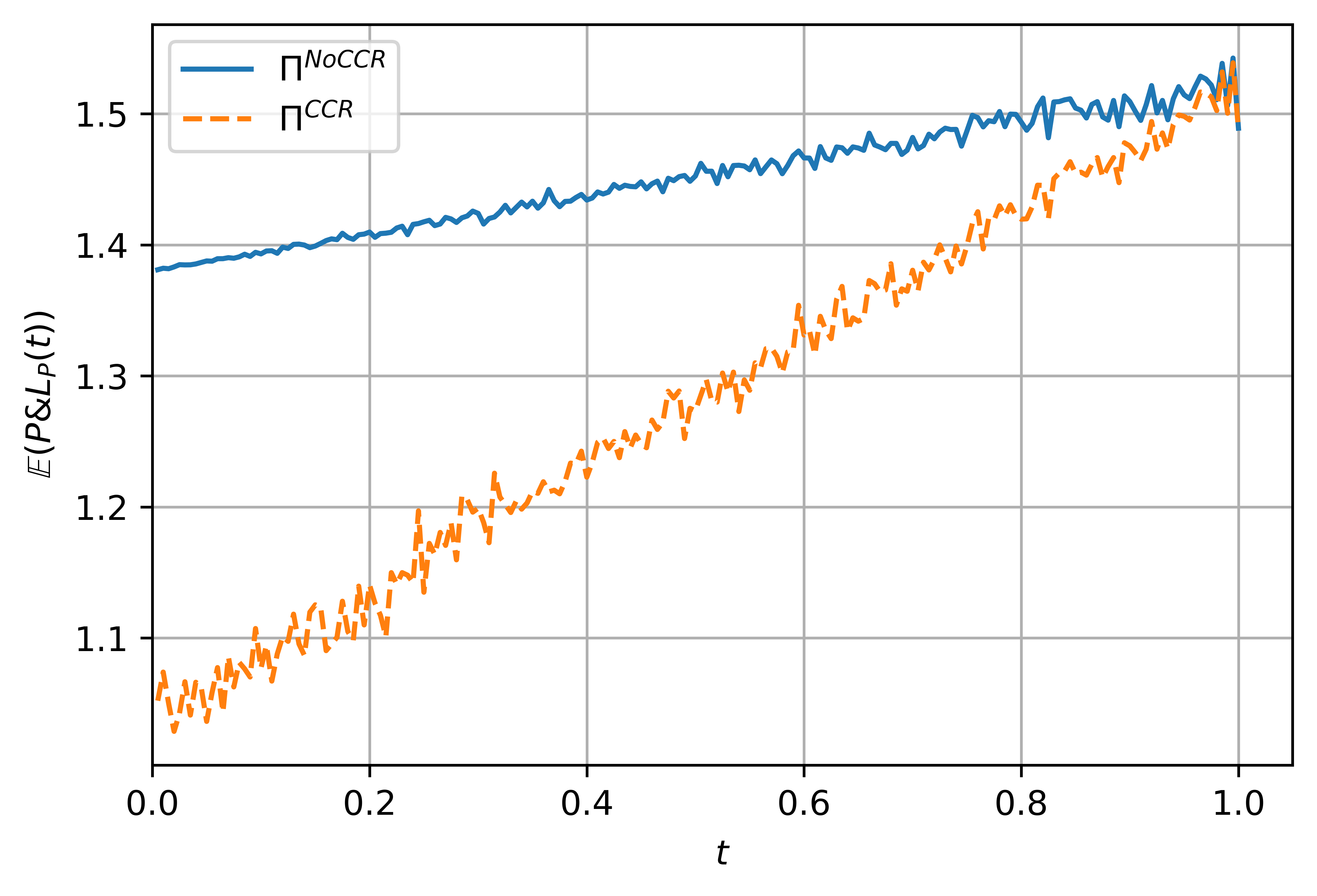}
    \caption{Average $\PnLPortfolio(t)$.}
    \label{fig:bsNoHedgeNewDealPnLPortfolioMean}
  \end{subfigure}
  \begin{subfigure}[b]{\resultFigureSize}
    \includegraphics[width=\linewidth]{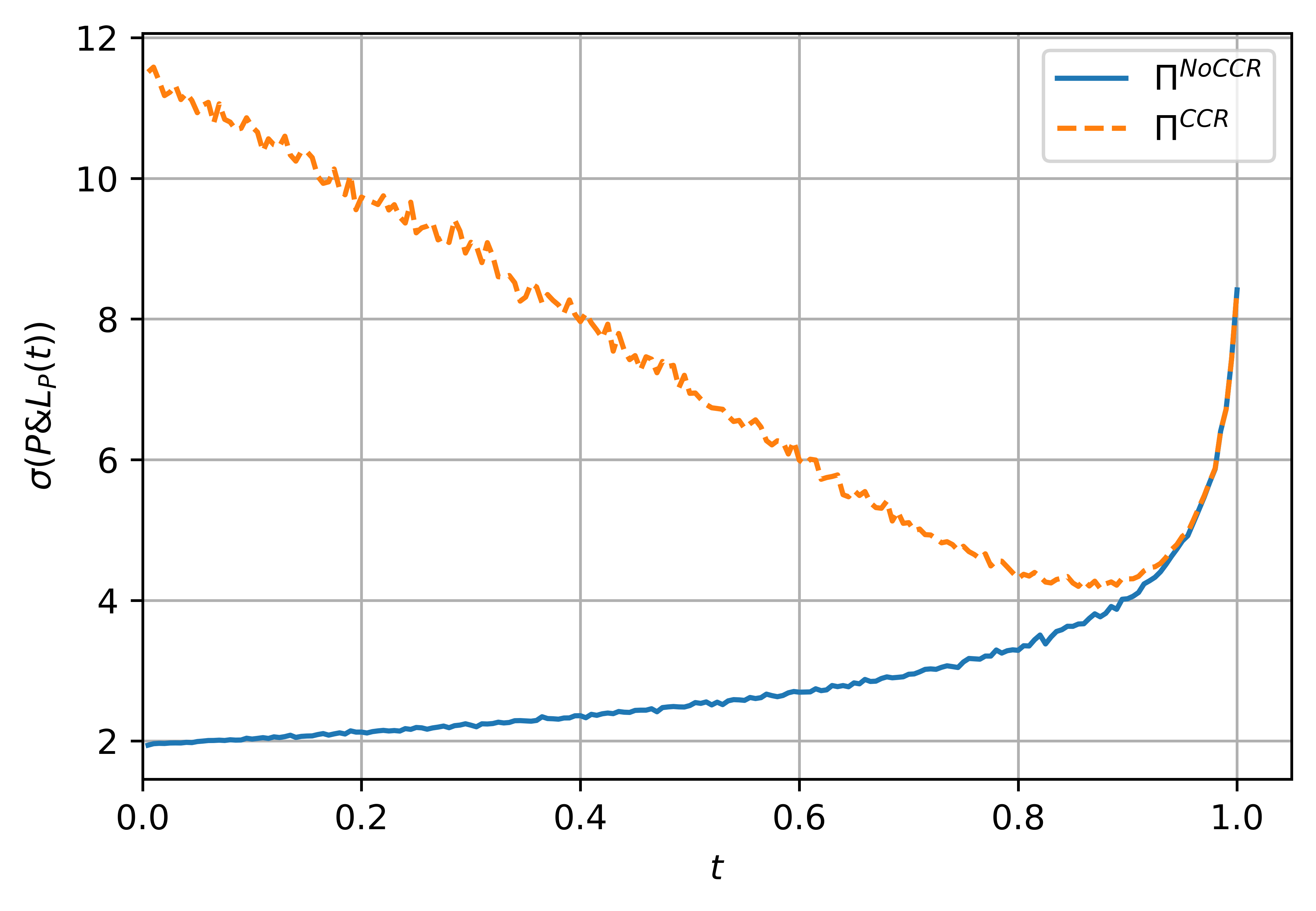}
    \caption{Volatility of $\PnLPortfolio(t)$.}
    \label{fig:bsNoHedgeNewDealPnLPortfolioVolatility}
  \end{subfigure}
  \begin{subfigure}[b]{\resultFigureSize}
    \includegraphics[width=\linewidth]{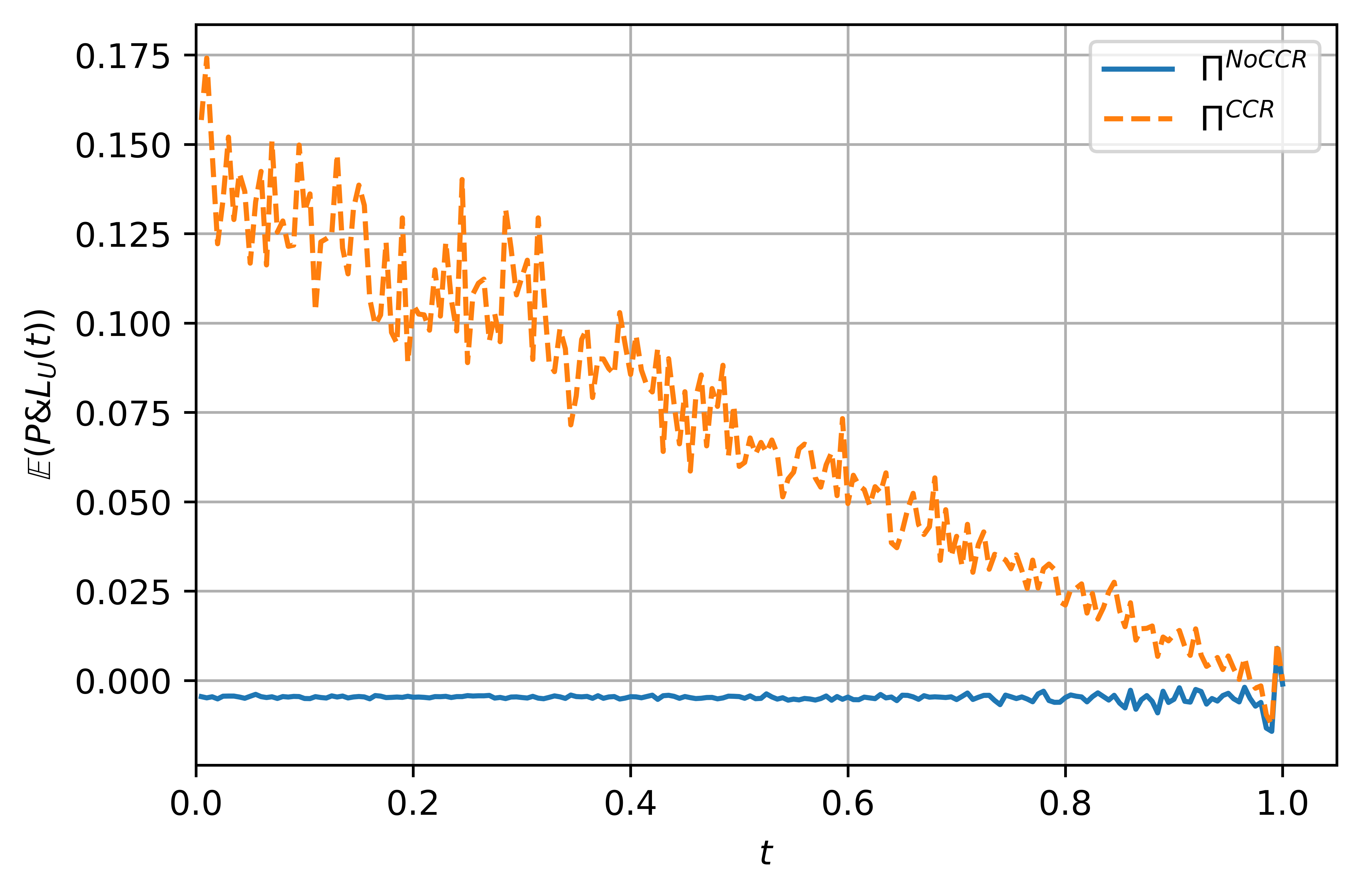}
    \caption{Average $\PnLUnexplained(t)$.}
    \label{fig:bsNoHedgeNewDealPnLUnexplainedMean}
  \end{subfigure}
  \begin{subfigure}[b]{\resultFigureSize}
    \includegraphics[width=\linewidth]{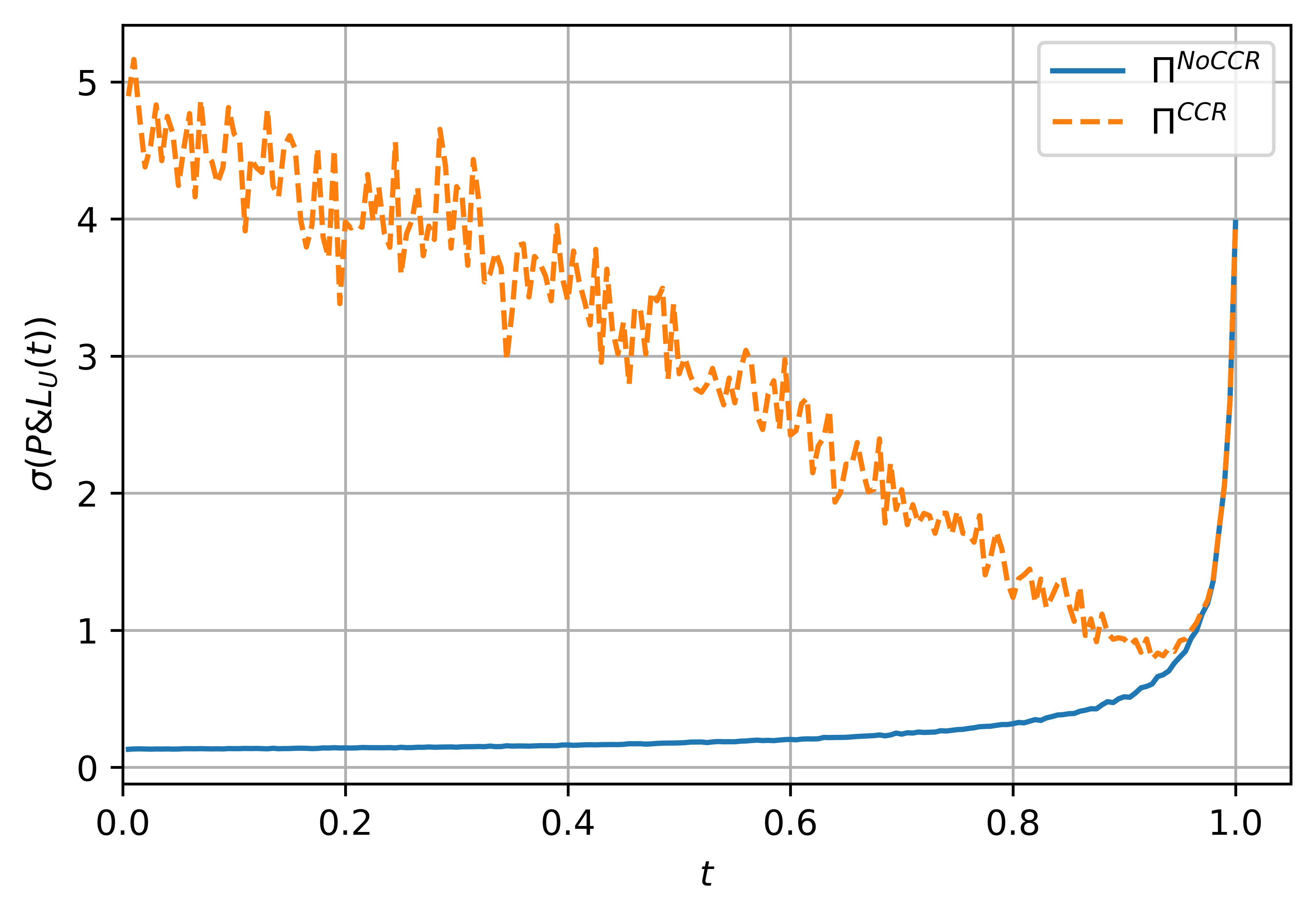}
    \caption{Volatility of $\PnLUnexplained(t)$.}
    \label{fig:bsNoHedgeNewDealPnLUnexplainedVolatility}
  \end{subfigure}
  \caption{Comparison of $\strategyWithoutCCR$ and $\strategyWithCCR$ using a Black-Scholes market and valuation model. $\CVA$ is not hedged.}
  \label{fig:bsNoHedgeNewDealPart2}
\end{figure}

Figure~\ref{fig:bsNoHedgeNewDealTotalAverage} represents the case in which $\CVA$ was added to the portfolio but not hedged.
We see that the expected terminal wealth condition $\displaystyle \E_{t_0}\left[\strategy(t_K) + \wealth(t_K)\right] = 0$ is satisfied.
In Figure~\ref{fig:bsNoHedgeNewDealTotalStd}, we see that the variability of $\strategy(t) + \wealth(t)$ for $\strategyWithCCR$ is much higher compared to that of $\strategyWithoutCCR$.
This is expected as a result of the additional source of randomness introduced by the simulated defaults in $\strategyWithCCR$.
$\strategy(t) + \wealth(t)$ must always be examined first when assessing the performance of the portfolio, but it does not guarantee optimal performance.

Therefore, consider Figures~\ref{fig:bsNoHedgeNewDealPnLPortfolioMean} and~\ref{fig:bsNoHedgeNewDealPnLPortfolioVolatility}, where we plot the mean and volatility of $\PnLPortfolio(t)$.
One can clearly see that even though the mean $\PnLPortfolio$ for $\strategyWithCCR$ is slightly lower compared to $\strategyWithoutCCR$, the volatility is significantly higher over the majority of the lifetime of the option.
Thus, this hedging strategy is incomplete.
The two lines in Figure~\ref{fig:bsNoHedgeNewDealPnLPortfolioVolatility} overlap close to maturity, because the binarity of the payoff takes over.
Furthermore, most of the simulated defaults have occurred by this time.
Because after a default the same deal is entered with a credit risk free counterparty, the two portfolios eventually exhibit similar behaviour.
From Figures~\ref{fig:bsNoHedgeNewDealPnLUnexplainedMean} and~\ref{fig:bsNoHedgeNewDealPnLUnexplainedVolatility} we see that a significant portion of the $\PnLPortfolio(t)$ can be explained using the underlying stock.~\footnote{In case a delta hedge was employed to hedge first order risks, this term was ignored in the $\PnL$ explain process to avoid taking the delta effect into account twice.}
We see that the potential defaults through the lifetime of the option contribute to a significant initial difference in $\PnLUnexplained$.
This difference diminishes over time, as fewer defaults are expected to occur before the maturity of the option.
As the strong increase in variance is not promising at a first glance, in Section~\ref{sec:hedingError} a thorough analysis of this behaviour can be found.
All in all, the results indicate that $\CVA$ needs to be hedged to eliminate the majority of the underlying risk in the portfolio.

\begin{figure}[!h]
  \centering
  \begin{subfigure}[b]{\resultFigureSize}
    \includegraphics[width=\linewidth]{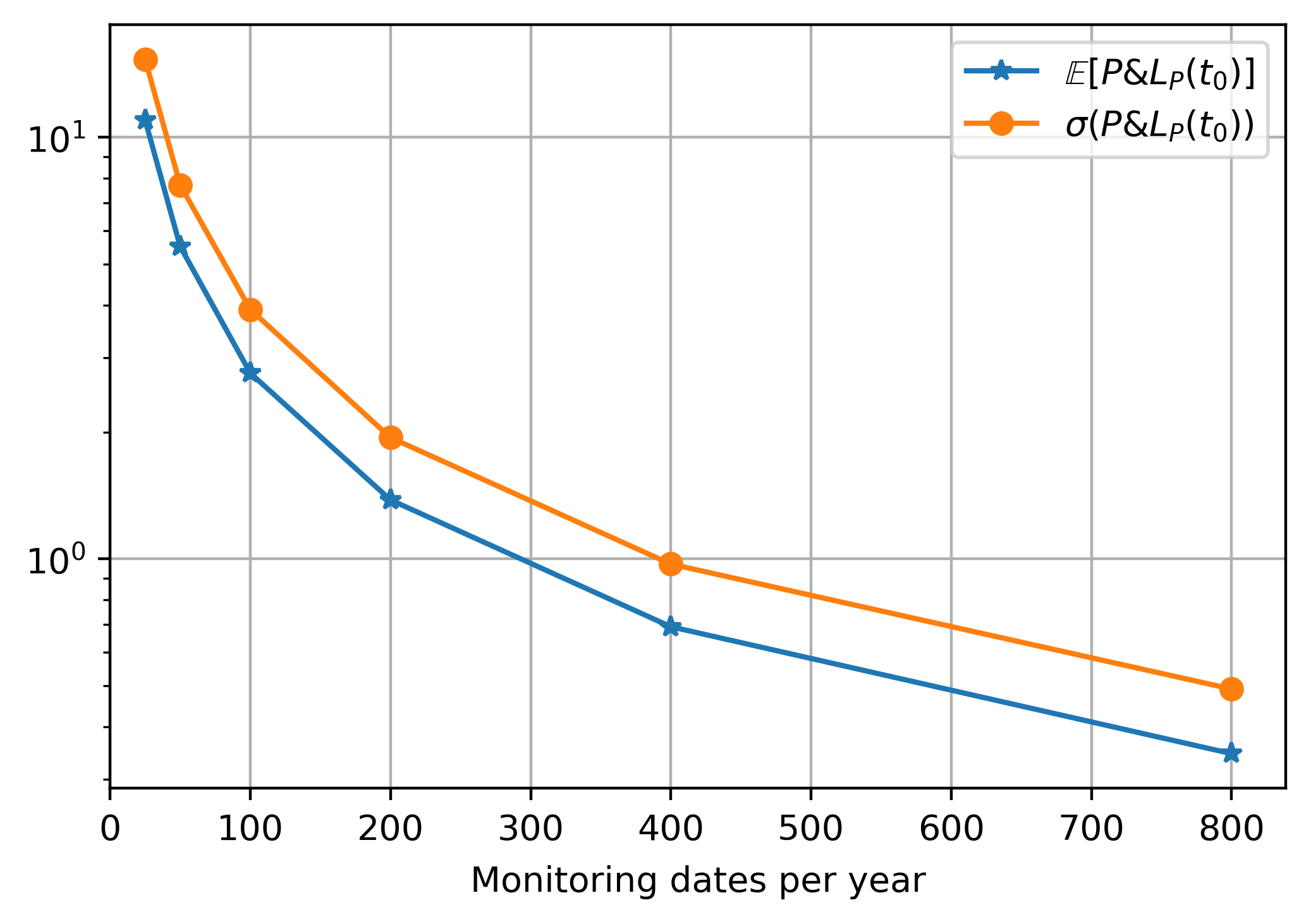}
    \caption{$\strategyWithoutCCR$ results of $\PnLPortfolio(t_0)$ average and volatility for different number of monitoring dates, $L=10^5$.}
    \label{fig:pnlPortfolioConvergence}
  \end{subfigure}
  \begin{subfigure}[b]{\resultFigureSize}
    \includegraphics[width=\linewidth]{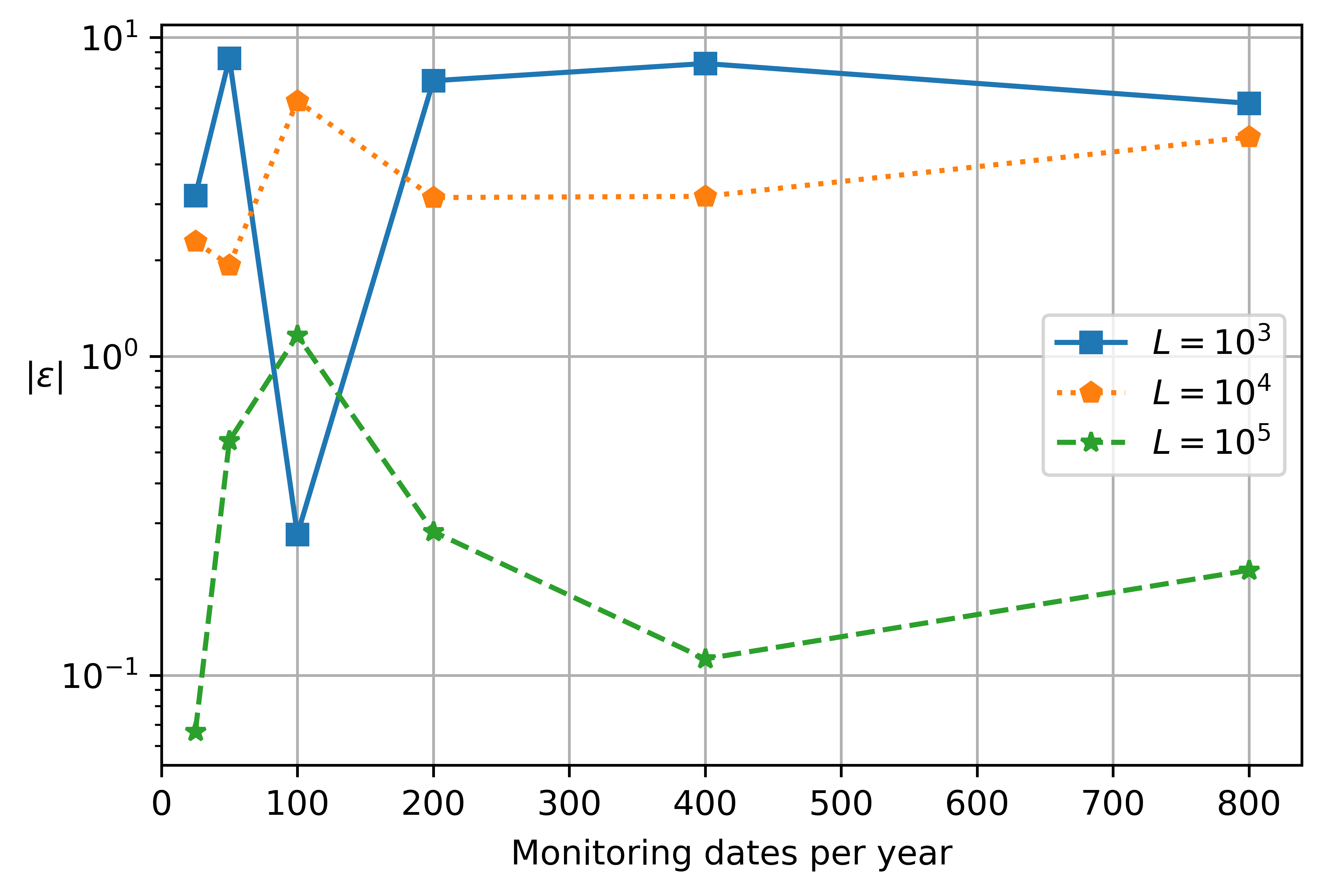}
    \caption{$|\varepsilon(t_0)|$ for different number of monitoring dates, per different number of Monte Carlo paths $L$.}
    \label{fig:epsByPaths}
  \end{subfigure}
  \caption{}
  \label{fig:bsAdditionalTests}
\end{figure}

Looking at Figure~\ref{fig:bsNoHedgeNewDealPnLPortfolioMean}, one might question why the average $\PnLPortfolio(t_0)$ for $\strategyWithoutCCR$ is not equal to zero, as this would mean that a Black-Scholes delta hedge is unable to hedge all the risk of the option.
Figure~\ref{fig:pnlPortfolioConvergence} shows that this number converges to zero if $\dt \to 0$, so our observations are merely the result of the discretization.
The volatility of $\PnLPortfolio(t_0)$ for $\strategyWithoutCCR$ also converges to zero if the number of time-steps in the discretization is increased.

We also confirm that the $\CVA$ charged initially covers the otherwise experienced expected loss at default.
We do this by showing that the following error measure is approximately zero:
\begin{align}
    \varepsilon(t_0)
        &= \E_{t_0} \left[\indicator{\default \leq t_K} \left[\frac{\bank(t_0)}{\bank(\default)} \left(\recovRate \cdot \tradeVal(\default) - V(\default) \right)+ \CVA(t_0) \right]  + \indicator{\default > t_K} \CVA(t_0) \right]\nonumber \\
        &\approx \frac{1}{\nrPathsMC} \sum_{l=1}^{\nrPathsMC} \indicator{\default_l \leq t_K} \left[ \frac{\bank(t_0)}{\bank(\default_l)} \left(\recovRate \cdot\tradeVal(\default_l) - \tradeVal(\default_l) \right)+ \CVA(t_0) \right]  + \indicator{\default_l > t_K} \CVA(t_0),  \label{eq:errorMeasure}
\end{align}
where $\nrPathsMC$ denotes the number of Monte Carlo paths used in the simulation.
We analyze the behaviour of this error measure by changing the number of Monte Carlo paths and number of time-steps used in the simulation.
From Figure~\ref{fig:epsByPaths} we see that for a given number of monitoring dates per year, increasing the number of paths results in $|\varepsilon(t_0)|\rightarrow 0$.
Here we clearly see that using $\nrPathsMC=10^3$ provided unstable results, but we do observe converging behaviour when increasing $\nrPathsMC$.
As defaults occur infrequently, the number of paths must be large to properly approximate the numerical expectation from Equation~\eqref{eq:errorMeasure}.
Furthermore, the choice of $200$ monitoring dates per year appears to be a very fine tradeoff between speed and accuracy.

\subsubsection{Variance analysis} \label{sec:hedingError}

In Figure~\ref{fig:bsNoHedgeNewDealPnLPortfolioVolatility} we see an exploding behaviour in the volatility of $\PnLPortfolio$ as $t\to T$, for both $\strategyWithoutCCR$ and $\strategyWithCCR$, which we want to understand.
Therefore, we explain the distribution features we observe by looking at the $\PnLPortfolio$ mean and variance for $\strategyWithoutCCR$.
Thus, we examine $\PnLPortfolio$ for the portfolio in Equation~\eqref{eq:tradeStrategyExample}, with the Black-Scholes delta hedging quantity as in Equation~\eqref{eq:hedgePositionBS}.
$\PnLPortfolio$ from Equation~\eqref{eq:pnlPortfolio2} with $\marketData(t) = \stock(t)$ can be rewritten as follows:

\begin{align}
    \PnLPortfolio(t_k)
        &= \strategy(t_{k-1}, S(t_k)) - \strategy(t_{k-1}, S(t_{k-1}))  \nonumber \\
        &= \left[ \tradeVal_1(t_{k-1}, S(t_k)) - \tradeVal_1(t_{k-1}, S(t_{k-1})) \right] +  \hedgePos_1(t_{k-1}) \left[ \stock(t_k) - \stock(t_{k-1}) \right]. \label{eq:pnlPortfolio1}
\end{align}
Dividing both sides of Equation~\eqref{eq:pnlPortfolio1} by $\dS = \stock(t_k) - \stock(t_{k-1})$ , and using the definition of a forward finite difference approximation yields:
\begin{align}
    \frac{\PnLPortfolio(t_k)}{\stock(t_k) - \stock(t_{k-1})}
        &= \frac{ \tradeVal_1(t_{k-1}, S(t_k)) - \tradeVal_1(t_{k-1}, S(t_{k-1})) }{\stock(t_k) - \stock(t_{k-1})}  - \pderiv{\tradeVal_1(t_{k-1})}{\stock} \nonumber \\
        &= \frac{\dS}{2} \ppderiv{\tradeVal_1(t_{k-1},S(t_{k-1}))}{S} + \bigOh\left((\dS)^2\right), \label{eq:pnlPortfolioFD1}\\
    \Rightarrow \PnLPortfolio(t_k)
        &= \frac{(\d\stock)^2}{2} \ppderiv{\tradeVal_1(t_{k-1},S(t_{k-1}))}{\stock} + \bigOh\left((\d\stock)^3\right). \label{eq:pnlPortfolioFD2}
\end{align}

Equation~\eqref{eq:pnlPortfolioFD2} shows that two types of errors drive $\PnLPortfolio$.
First, there is the truncation error of the forward finite differences which are used in Equation~\eqref{eq:pnlPortfolioFD1}.
Furthermore, there is a discretization error that disappears if $\dS \to 0$, which will happen if $\dt = t_k-t_{k-1} \to 0$.

In Equation~\eqref{eq:pnlPortfolioFD2}, we recognize the second partial derivative as the option gamma, which under the Black-Scholes model is given by the following analytic expression:
\begin{align}
    \ppderiv{\tradeVal_1(t,\stock)}{\stock}
        &= \strike \expPower{-\shortRate(T-t)} \frac{\normPDF(d_2(t,\stock))}{\stock^2\vol\sqrt{T-t}}, \label{eq:gamma1} \\
    d_2(t,\stock)
        &= \frac{\ln \frac{S}{K} + \left(\shortRate - \half \vol^2\right) [T - t]}{\vol \sqrt{T-t}}. \nonumber
\end{align}
Define $X \sim \N(\mu_X, \sigma^2_X)$ such that $\stock(t_k) \equalDistr \stock(t_{k-1}) \expPower{X}$, see~\ref{sec:hedgeErrorDerivations} for further details.
Using this and the Black-Scholes gamma~\eqref{eq:gamma1}, we rewrite Equation~\eqref{eq:pnlPortfolioFD2} as follows:
\begin{align}
    \PnLPortfolio(t_k)
        &= \frac{(\d\stock)^2}{2} \strike \expPower{-\shortRate\tau} \frac{\normPDF(d_2(t_{k-1}, S(t_{k-1})))}{S^2(t_{k-1})\vol\sqrt{\tau}} + \bigOh\left((\d\stock)^3\right) \nonumber \\
        &= \frac{\stock^2(t_{k-1})\left[\expPower{X} - 1\right]^2}{2}  \strike \expPower{-\shortRate\tau} \frac{\normPDF(d_2(t_{k-1}, \stock(t_{k-1})))}{\stock^2(t_{k-1})\vol\sqrt{\tau}} + \bigOh\left((\d\stock)^3\right) \nonumber \\
        &= \frac{\strike \expPower{-\shortRate\tau} }{2\vol\sqrt{\tau}}  \left[\expPower{X} - 1\right]^2 \normPDF(d_2)  + \bigOh\left((\d\stock)^3\right), \label{eq:pnlPortfolioFD5}
\end{align}
where $\tau \ldef T - t_{k-1}$ and for ease of notation we write $d_2 = d_2(t_{k-1}, \stock(t_{k-1}))$.

Define $d_2 \sim \N(\mu_{d_2}, \sigma^2_{d_2})$, see~\ref{sec:hedgeErrorDerivations}.
Then, using definitions~\eqref{eq:pnlPortfolioFDVar2f} and~\eqref{eq:pnlPortfolioFDVar2g} for respectively $f(\mu_X, \sigma_X)$ and $g(\mu_X, \sigma_X)$, we obtain this expression of the variance of $\PnLPortfolio$:
\begin{align}
    \Var_{t_0}\left(\PnLPortfolio(t_k)\right)
        \approx \frac{\strike^2 \expPower{-2\shortRate\tau}}{8\pi\vol^2\tau} & \left[
        f(\mu_X, \sigma_X) \cdot \frac{\expPower{-\frac{\mu^2_{d_2}}{1 + 2\sigma^2_{d_2}}}}{\sqrt{1+2\sigma^2_{d_2}}} -  g(\mu_X, \sigma_X) \cdot \frac{\expPower{-\frac{\mu^2_{d_2}}{1+\sigma^2_{d_2}}}}{1+\sigma^2_{d_2}} \right]. \label{eq:pnlPortfolioFDVar2}
\end{align}
See~\ref{sec:hedgeErrorDerivations} for a full derivation of this result.

\begin{figure}[!h]
  \centering
  \includegraphics[scale=0.6]{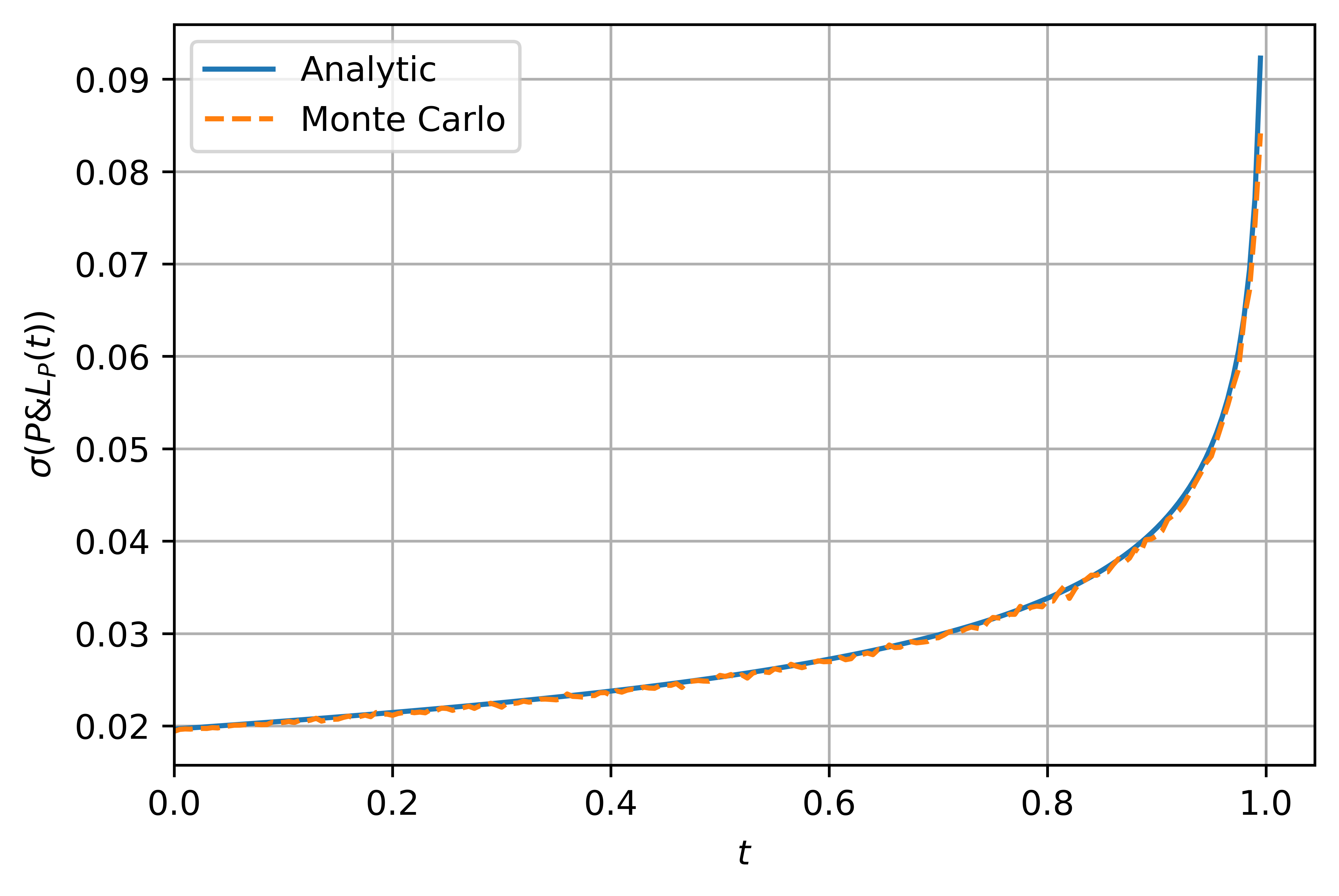}
  \caption{Numerical confirmation that the analytical variance for $\strategyWithoutCCR$ from Equation~\eqref{eq:pnlPortfolioFDVar2} is in line with the Monte Carlo simulation.}
  \label{fig:pnlPortfolioAnalyticStd}
\end{figure}

From Figure~\ref{fig:pnlPortfolioAnalyticStd} we see that the analytical result from Equation~\eqref{eq:pnlPortfolioFDVar2} and volatility from the Monte Carlo simulation are in line.
The strong increase of $\PnLPortfolio$ volatility close to maturity can be understood from the perspective of $\PnLPortfolio$  being driven by the option gamma, see Equation~\eqref{eq:pnlPortfolioFD2}.
It is known for the option delta to be unstable near maturity, causing an increased gamma.
This explains the increase in variance of $\PnLPortfolio(t_k)$ as $t_k \to T$.
Especially when the option is close to the ATM point, the gamma is large, causing a large gamma volatility.

Looking at Equation~\eqref{eq:pnlPortfolioFDVar2}, the strong increase in volatility towards maturity can be explained by looking at the variance as a scaled difference between the factors $\left(\sqrt{1+2\sigma^2_{d_2}}\right)^{-1}$ and $ \left(1+\sigma^2_{d_2}\right)^{-1}$.
The first term is non-linear, while the second is linear.
Approaching maturity, the non-linearity of $ \left(\sqrt{1+2\sigma^2_{d_2}}\right)^{-1}$ increases, causing the increased variance.

\subsubsection{Hedging CVA} \label{sec:bsHedgeCVA}
In Section~\ref{sec:cvaWithMarketRisk} we concluded that $\CVA$ market risk needs to be hedged.
Hence, we do so using the underlying stock, summarized by the hedging quantity Equation~\eqref{eq:hedgePositionCVA}.
\begin{figure}[!h]
  \centering
  \begin{subfigure}[b]{\resultFigureSize}
    \includegraphics[width=\linewidth]{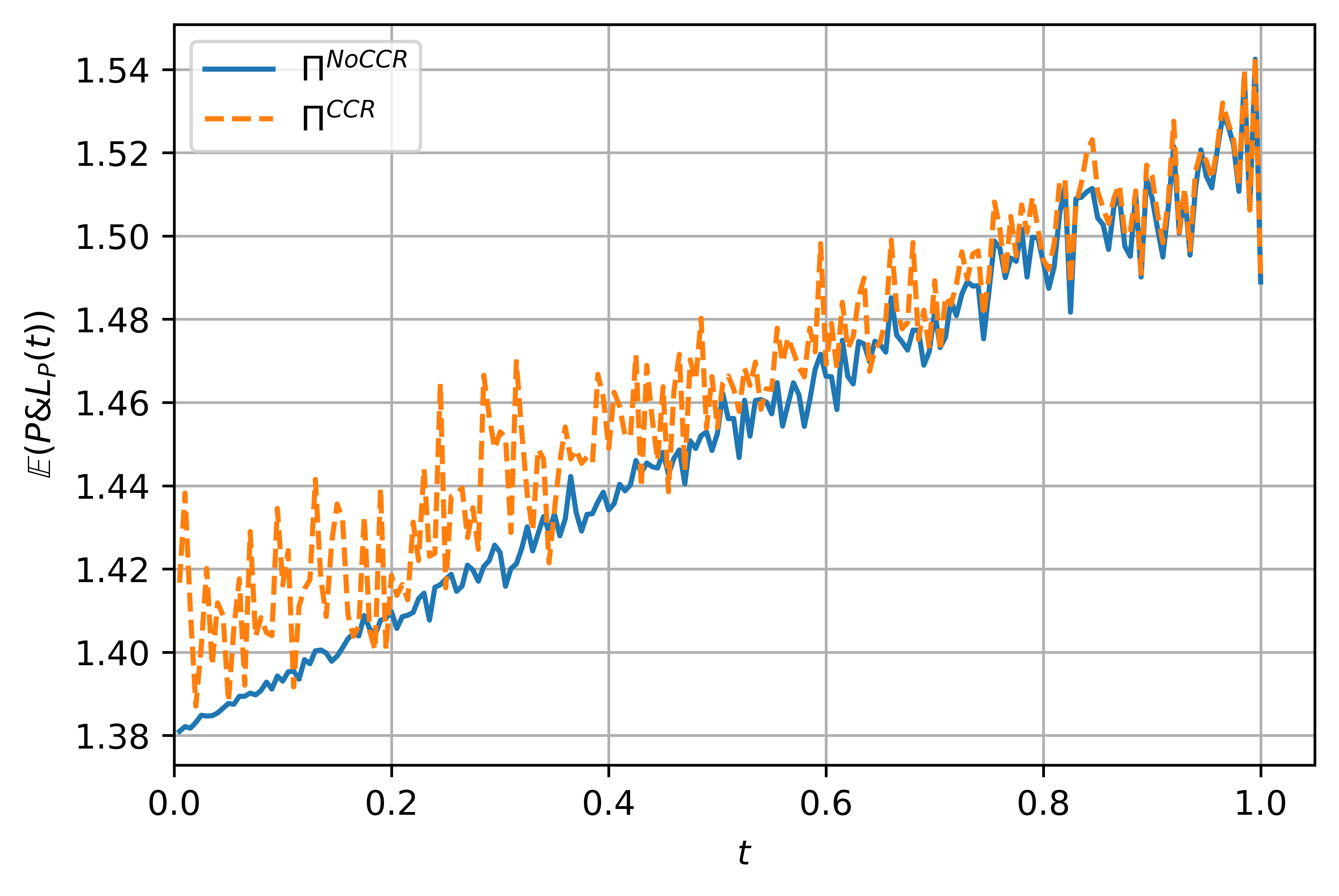}
    \caption{Average $\PnLPortfolio(t)$.}
    \label{fig:bsHedgeNewDealPnLPortfolioMean}
  \end{subfigure}
  \begin{subfigure}[b]{\resultFigureSize}
    \includegraphics[width=\linewidth]{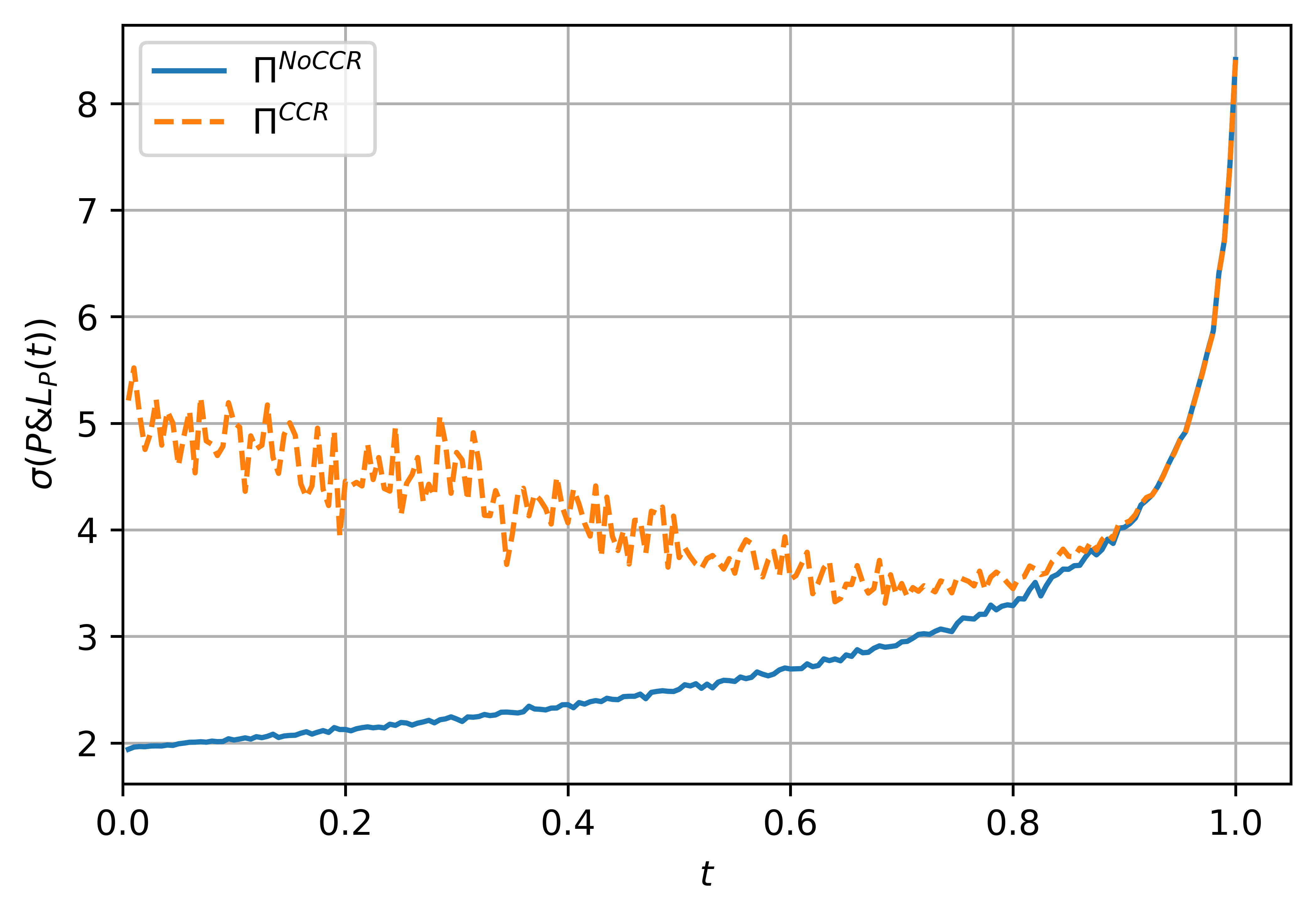}
    \caption{Volatility of $\PnLPortfolio(t)$.}
    \label{fig:bsHedgeNewDealPnLPortfolioVolatility}
  \end{subfigure}
  \caption{Comparison of $\strategyWithoutCCR$ and $\strategyWithCCR$ using a Black-Scholes market and valuation model. $\CVA$ is hedged using the underlying stock.}
  \label{fig:bsHedgeNewDeal}
\end{figure}
In terms of the zero average return, hedging the $\CVA$ yields the same results as not hedging.
Comparing Figures~\ref{fig:bsNoHedgeNewDealPnLPortfolioMean} and~\ref{fig:bsHedgeNewDealPnLPortfolioMean} shows that the average $\PnLPortfolio$ for $\strategyWithCCR$ is initially lower for the case where we do not hedge the $\CVA$ than when we do.
However, this difference is not significant, especially not when taking the size of the volatility into consideration.
In other words, the volatility appears to be a dominating factor in these results.
Comparing Figures~\ref{fig:bsNoHedgeNewDealPnLPortfolioVolatility} and~\ref{fig:bsHedgeNewDealPnLPortfolioVolatility} shows a benefit of hedging the $\CVA$, as this yields a significantly lower volatility in $\PnLPortfolio$.
Hence, we confirm that $\CVA$ market risk must be hedged.

Regarding the $\PnLUnexplained$, the results from Figures~\ref{fig:bsNoHedgeNewDealPnLUnexplainedMean} and~\ref{fig:bsNoHedgeNewDealPnLUnexplainedVolatility} also hold for the situation with $\CVA$ market risk hedge.
This makes sense, as the $\CVA$ market risk is either hedged and then does not need to be explained, or it is not hedged but then could be explained.
So, hedging $\CVA$ market risk does not affect the result of how much $\PnL$ can be explained.
The residual risk that remains after the $\PnL$ explain is illustrated by the differences between $\strategyWithoutCCR$ and $\strategyWithCCR$ in Figures~\ref{fig:bsNoHedgeNewDealPnLUnexplainedMean} and~\ref{fig:bsNoHedgeNewDealPnLUnexplainedVolatility}.
As suggested at the start of Section~\ref{sec:marketSimulation}, CDSs can be added to the set of hedging instruments to hedge this jump risk at default, rather than only warehousing the credit risk.

\subsection{Hedging CVA in a Merton jump-diffusion setting} \label{sec:jumpsHedge}
The numerical results in this section correspond to the Merton jump-diffusion hedging setting from Section~\ref{sec:mertonStrategy}.
The market is simulated using the Merton jump-diffusion model.
The hedging quantities and $\PnLExplained$ are computed with various models, depending on the experiment.

\subsubsection{Black-Scholes delta hedge} \label{sec:bsDeltaHedgeMertonPaths}
Here, a Black-Scholes delta hedge takes place.
Introducing jumps in the stock dynamics results in an extra source of randomness.
In particular, $\strategy(t) + \wealth(t)$ for $\strategyWithoutCCR$ shows roughly twice as much volatility through the option's lifetime.
For $\strategyWithCCR$, the simulated defaults appear to be dominating as there are no significant differences compared to the pure Black-Scholes case.

\begin{figure}[!h]
  \centering
  \begin{subfigure}[b]{\resultFigureSize}
    \includegraphics[width=\linewidth]{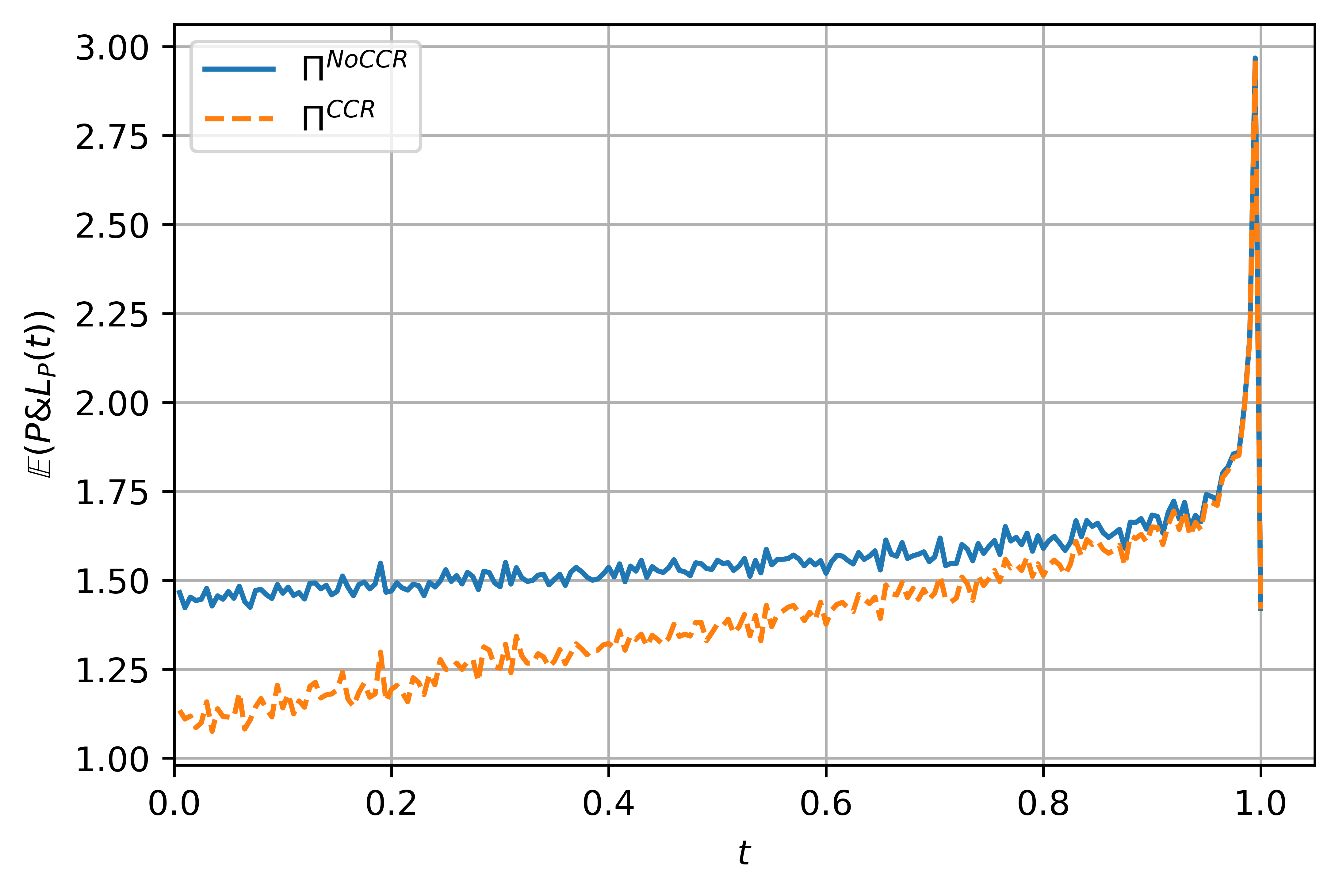}
    \caption{Average $\PnLPortfolio(t)$.}
    \label{fig:mertonPathsBSDeltaNoHedgeNewDealPnLPortfolioMean}
  \end{subfigure}
  \begin{subfigure}[b]{\resultFigureSize}
    \includegraphics[width=\linewidth]{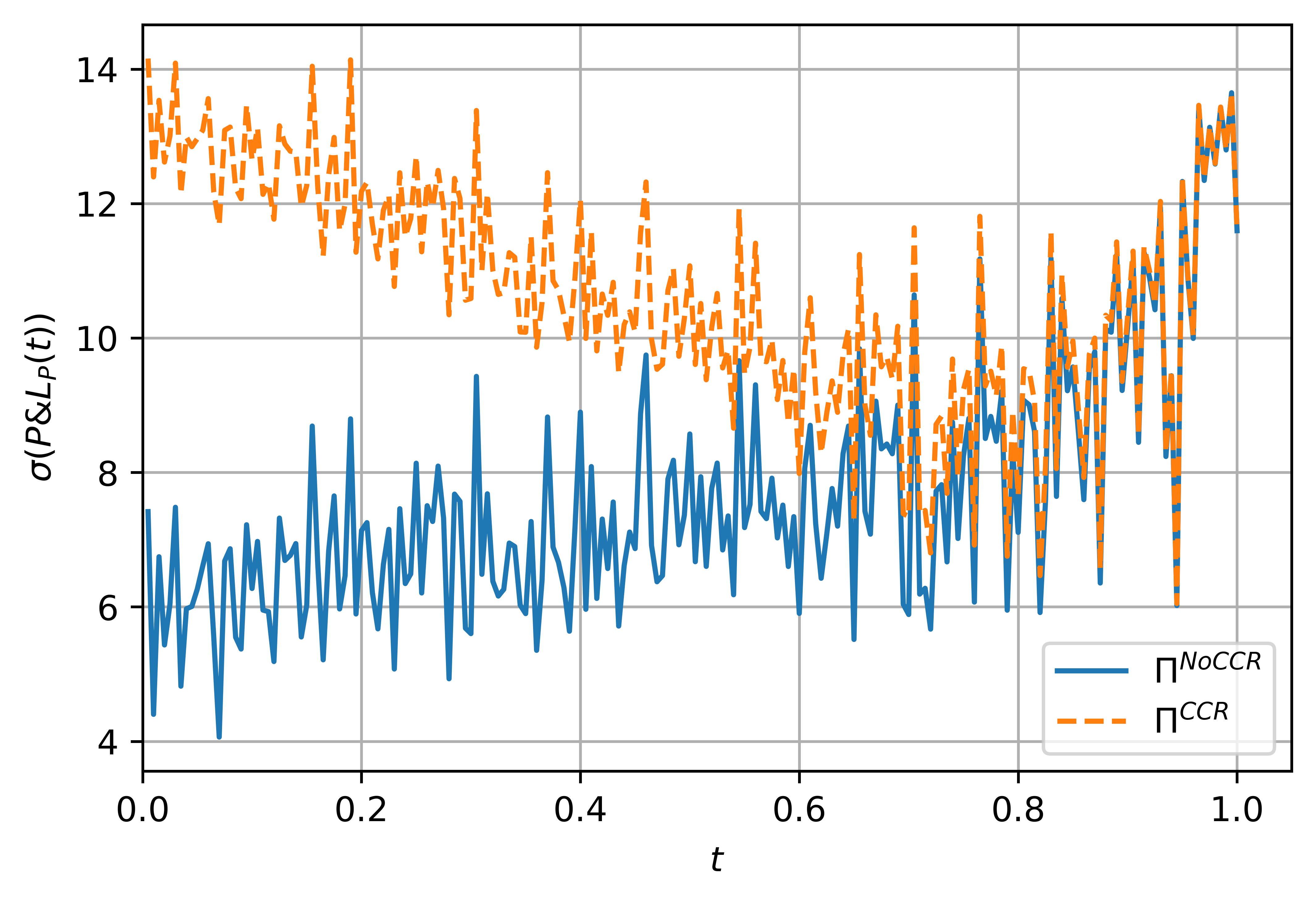}
    \caption{Volatility of $\PnLPortfolio(t)$.}
    \label{fig:mertonPathsBSDeltaNoHedgeNewDealPnLPortfolioVolatility}
  \end{subfigure}
  \begin{subfigure}[b]{\resultFigureSize}
    \includegraphics[width=\linewidth]{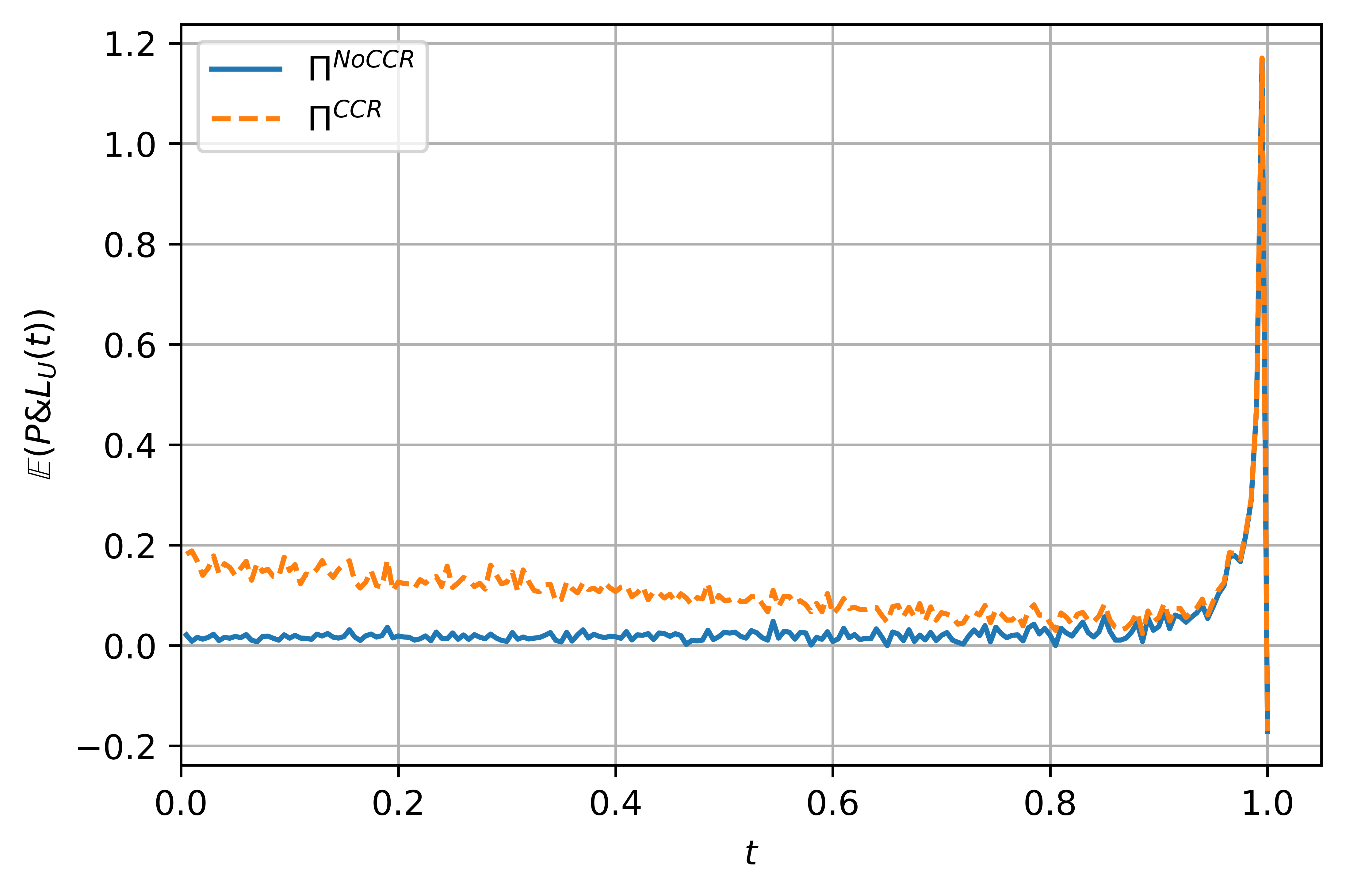}
    \caption{Average $\PnLUnexplained(t)$.}
    \label{fig:mertonPathsBSDeltaNoHedgeNewDealPnLUnexplainedMean}
  \end{subfigure}
  \begin{subfigure}[b]{\resultFigureSize}
    \includegraphics[width=\linewidth]{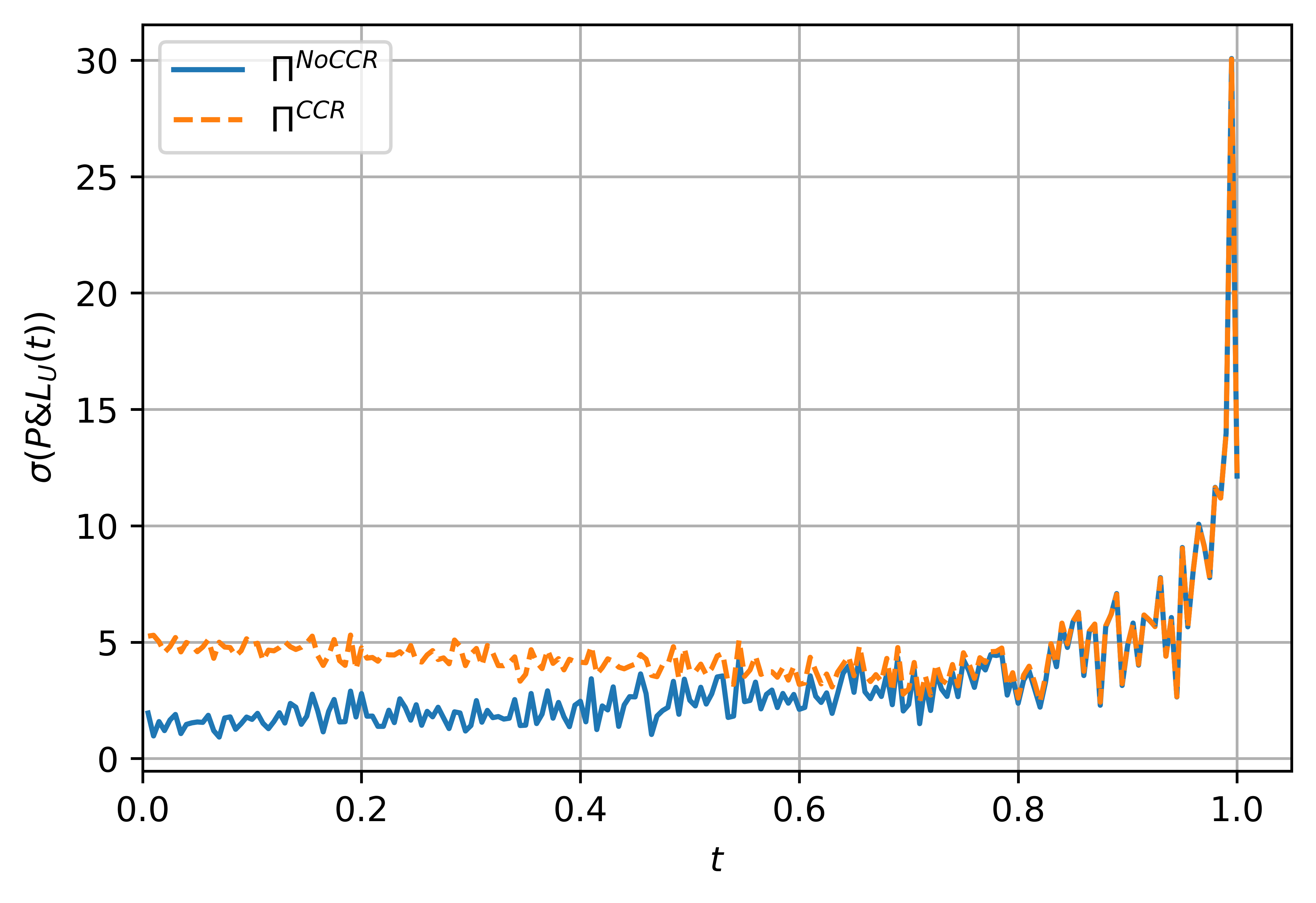}
    \caption{Volatility of $\PnLUnexplained(t)$.}
    \label{fig:mertonPathsBSDeltaNoHedgeNewDealPnLUnexplainedVolatility}
  \end{subfigure}
  \caption{Comparison of $\strategyWithoutCCR$ and $\strategyWithCCR$ using a Merton market and a Black-Scholes valuation model. $\CVA$ is not hedged.}
  \label{fig:mertonPathsBSDeltaNoHedgeNewDeal}
\end{figure}

There are two observations when comparing the $\PnLPortfolio$ mean in Figure~\ref{fig:mertonPathsBSDeltaNoHedgeNewDealPnLPortfolioMean} with the pure Black-Scholes results from Figure~\ref{fig:bsNoHedgeNewDealPnLPortfolioMean}. First, the level of the average is higher during the lifetime of the option.
Second, a peak is observed close to maturity.
This is the result of the Black-Scholes delta's inability to cope with jumps just before maturity.
In particular, a jump in the underlying stock for the paths where the option is close to the ATM point before maturity can result in a change of the option being in or out of the money.
The volatility of $\PnLPortfolio$ from Figure~\ref{fig:mertonPathsBSDeltaNoHedgeNewDealPnLPortfolioVolatility} is larger as well as significantly more variable over time when comparing with the pure Black-Scholes case (see Figure~\ref{fig:bsNoHedgeNewDealPnLPortfolioVolatility}), though the shape stays roughly the same.
These results are in line with our expectations, as the Black-Scholes delta does not take into account the additional source of randomness from the stock jumps, so this risk is not hedged.

The average $\PnLUnexplained$ in Figure~\ref{fig:mertonPathsBSDeltaNoHedgeNewDealPnLUnexplainedMean} is slightly lower than the average $\PnLPortfolio$.
Yet the peak before maturity observed for $\PnLPortfolio$ remains, hence the hedging strategy in combination with the chosen model yields undesired results.
Recall that this peak is not present in the pure Black-Scholes case (see Figure~\ref{fig:bsNoHedgeNewDealPnLUnexplainedMean}).
For the volatility of $\PnLUnexplained$ in Figure~\ref{fig:mertonPathsBSDeltaNoHedgeNewDealPnLUnexplainedVolatility}, the same reduction versus the $\PnLPortfolio$ volatility is observed as in the pure Black-Scholes case (see Figure~\ref{fig:bsNoHedgeNewDealPnLUnexplainedVolatility}).
However, close to maturity the $\PnLUnexplained$ volatility is even higher than the $\PnLPortfolio$ volatility.
This effect is the result of significant peaks in the gamma explain volatility just before maturity.
The option gamma is the rate of change in the option delta w.r.t. changes in the underlying.
For the ATM cases, the delta is extremely sensitive to changes in the underlying asset.
So, paths around the ATM level just before maturity cause this increase in $\PnLUnexplained$ volatility.
Furthermore, we observe significant movement in the volatility for the Merton case just before the maturity of the option.
Together, this indicates that the jump effects are missing in the explain process.

The $\CVA$ hedge in this context has no effect on either $\strategy(t) + \wealth(t)$ or $\PnLUnexplained$, which is in line with our observations for the pure Black-Scholes case.
For the $\PnLPortfolio$ the same effect of the $\CVA$ hedge is observed as in the pure Black-Scholes case: the average $\PnLPortfolio$ of $\strategyWithCCR$ overlaps significantly with that of $\strategyWithoutCCR$ after the introduction of the $\CVA$ hedge.
Furthermore, the initial volatility in the $\strategyWithCCR$ is much lower in the case of a $\CVA$ hedge, and the volatilities overlap much earlier in the case of a $\CVA$ hedge.

Introducing jumps in the Merton dynamics indeed results in additional randomness when comparing with the pure Black-Scholes case, already before considering the CCR.
For the case with CCR, the introduction of jumps in the stock seems to dominate the effect of the CCR in this case study.
The $\CVA$ hedge has the same desired effects for both Black-Scholes and Merton paths, of course ignoring the additional randomness introduced by the Merton jumps.
We can conclude that the Black-Scholes delta hedge in the case of a Merton market has its shortcomings.
Therefore, as a next step, we compare these results with a Merton delta hedge.

\subsubsection{Merton delta hedge} \label{sec:mertonDeltaHedgeMertonPaths}
We now use the Merton delta hedge in a Merton market, and do not hedge the $\CVA$.
The only difference compared to the Black-Scholes delta hedge is in the average $\PnLPortfolio$ and $\PnLUnexplained$.
For the $\PnLPortfolio$, the peak close to maturity we observed in the case of a Black-Scholes delta hedge, see Figure~\ref{fig:mertonPathsBSDeltaNoHedgeNewDealPnLPortfolioMean}, has disappeared.
For the $\PnLUnexplained$, the big upward peak just before maturity, see Figure~\ref{fig:mertonPathsBSDeltaNoHedgeNewDealPnLUnexplainedMean}, has disappeared too.
A minor downward peak remains due to instability in the gamma explain.
The Merton Greeks are not capable of mitigating these instabilities.

\begin{figure}[!h]
  \centering
  \begin{subfigure}[b]{\resultFigureSize}
    \includegraphics[width=\linewidth]{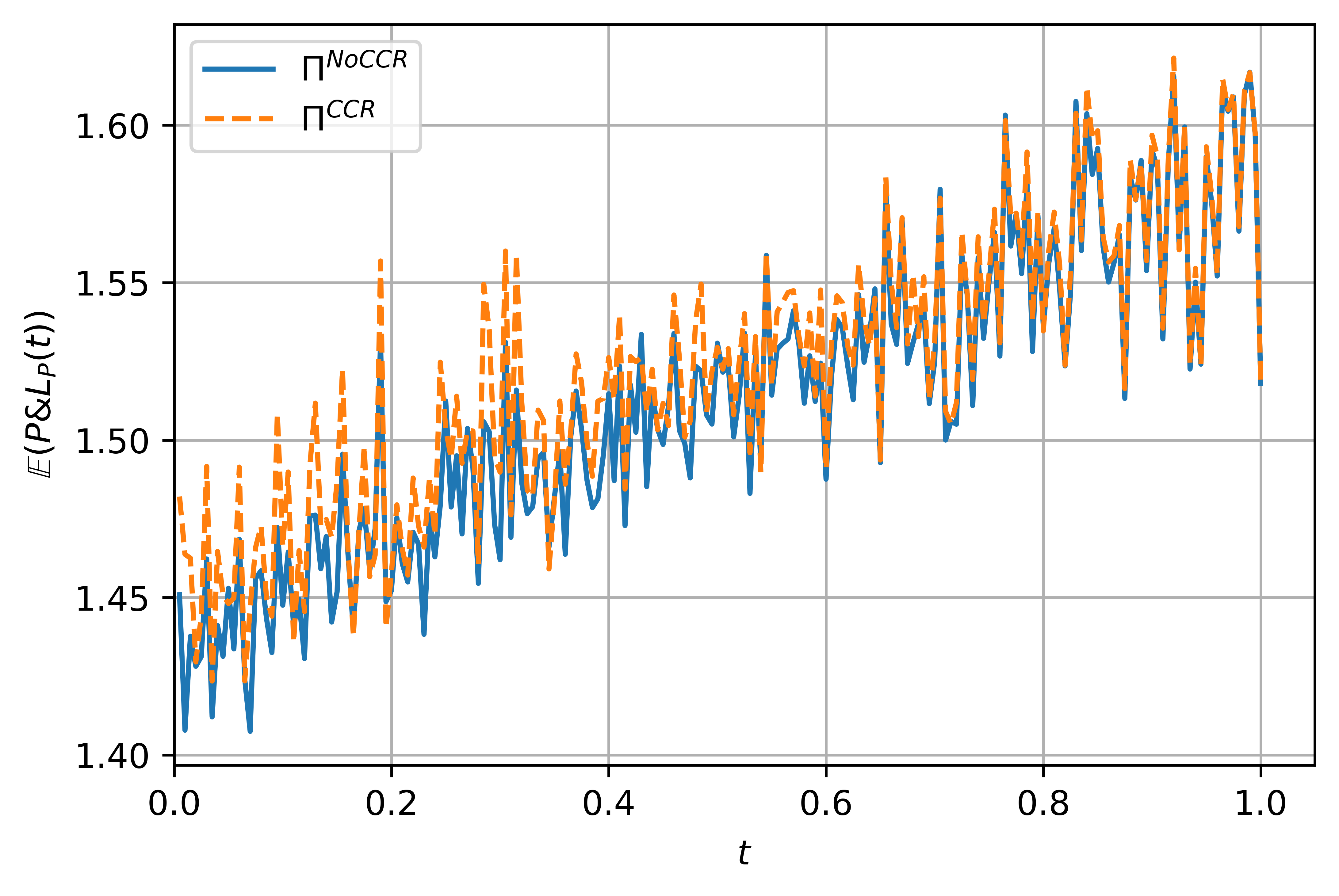}
    \caption{Average $\PnLPortfolio(t)$.}
    \label{fig:mertonDeltaHedgeNewDealPnLPortfolioMean}
  \end{subfigure}
  \begin{subfigure}[b]{\resultFigureSize}
    \includegraphics[width=\linewidth]{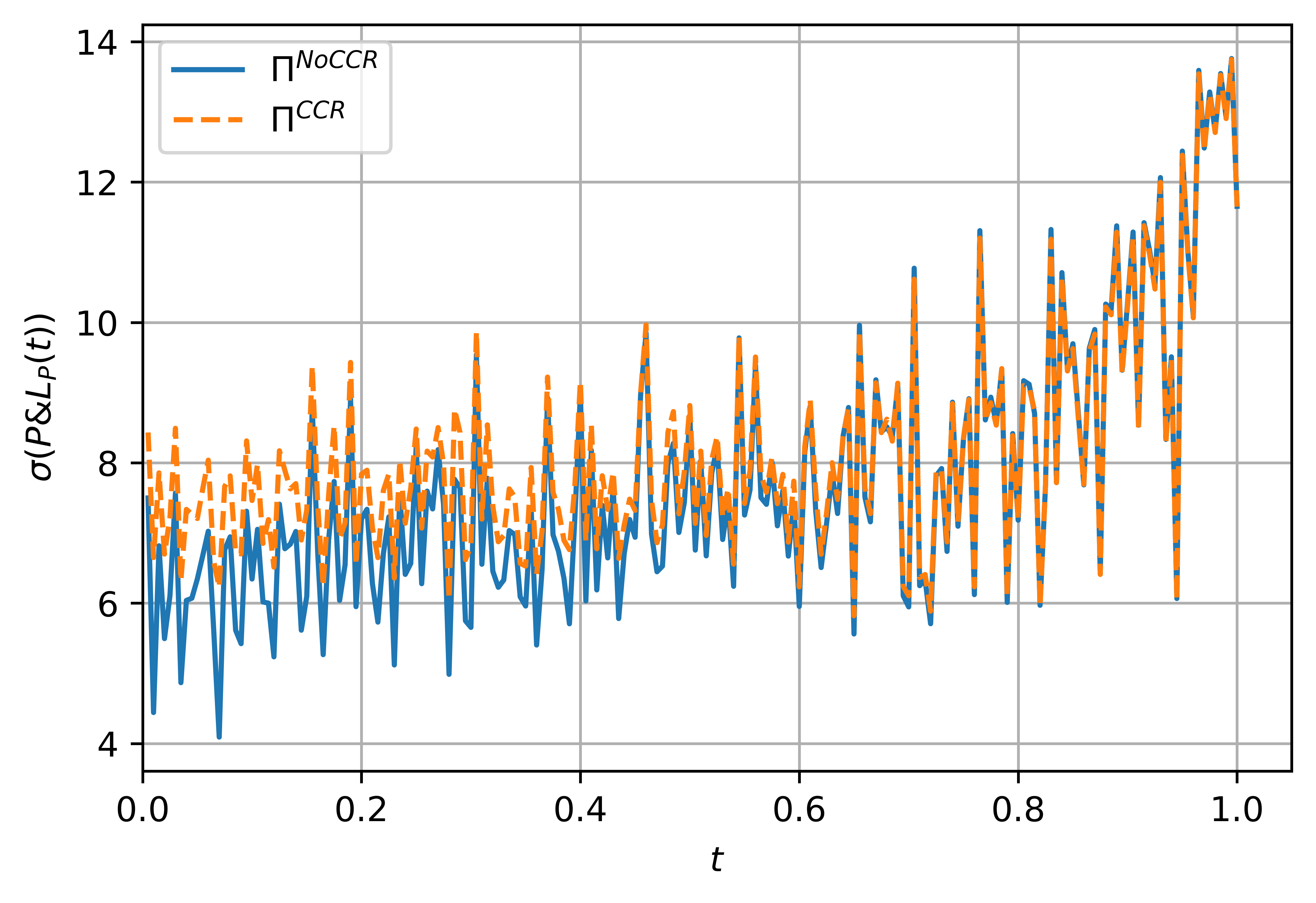}
    \caption{Volatility of $\PnLPortfolio(t)$.}
    \label{fig:mertonDeltaHedgeNewDealPnLPortfolioVolatility}
  \end{subfigure}
  \caption{Comparison of $\strategyWithoutCCR$ and $\strategyWithCCR$ using a Merton market and valuation model. $\CVA$ is hedged using the underlying stock.}
  \label{fig:mertonDeltaHedgeNewDeal}
\end{figure}

The $\CVA$ hedge effect is clearly visible in Figures~\ref{fig:mertonDeltaHedgeNewDealPnLPortfolioMean} and~\ref{fig:mertonDeltaHedgeNewDealPnLPortfolioVolatility}.
After filtering out the Merton delta effects, the $\CVA$ hedge yields the same conclusion as for the Black-Scholes delta hedge.

Overall, the impact of moving from the Black-Scholes delta hedge to the Merton delta hedge was significant, especially in the average $\PnLPortfolio$.
Yet some issues and undesired results remained, for example due to the instability of the gamma explain.
Clearly we need another way to hedge away more of the jump risk associated with the Merton model.

\subsubsection{Hedging the jump risk using an option} \label{sec:optionHedgeSingle}
Therefore, we hedge the residual jump risk from the Merton model using an extra instrument in the hedging portfolio: one extra option.
Typically, OTM options are chosen in the hedging portfolio as they are cheap.
One of the difficulties resulting from Equation~\eqref{eq:mertonSingleOptionHedge} is that $\pderiv{\hedgeVal_2(t)}{\jumpIntensity}$ tends to zero close to maturity, due to the option being OTM.
$\pderiv{\tradeVal_1(t)}{\jumpIntensity}$ may behave the same, however, it is likely that $\pderiv{\hedgeVal_2(t)}{\jumpIntensity}$ will decay faster due to the option being deeper OTM.
As a result, $\hedgePos_2(t)$ in Equation~\eqref{eq:mertonSingleOptionHedge} will increase drastically close to maturity of the option.
For the hedging option we choose an OTM put with strike $\strike = 90$, maturity $T = 1$ and controlling the same number of shares as the original option we hedge.~\footnote{Alternatively, a hedging strategy that rolls over a short term option position to hedge a long term option can be used, see for example~\cite{HeKennedyColemanForsythLiVetzal200601}.}
In case we hedge the $\CVA$, this is done using both the underlying stock and same option we use to hedge the product itself.

\begin{figure}[!h]
  \centering
  \begin{subfigure}[b]{\resultFigureSize}
    \includegraphics[width=\linewidth]{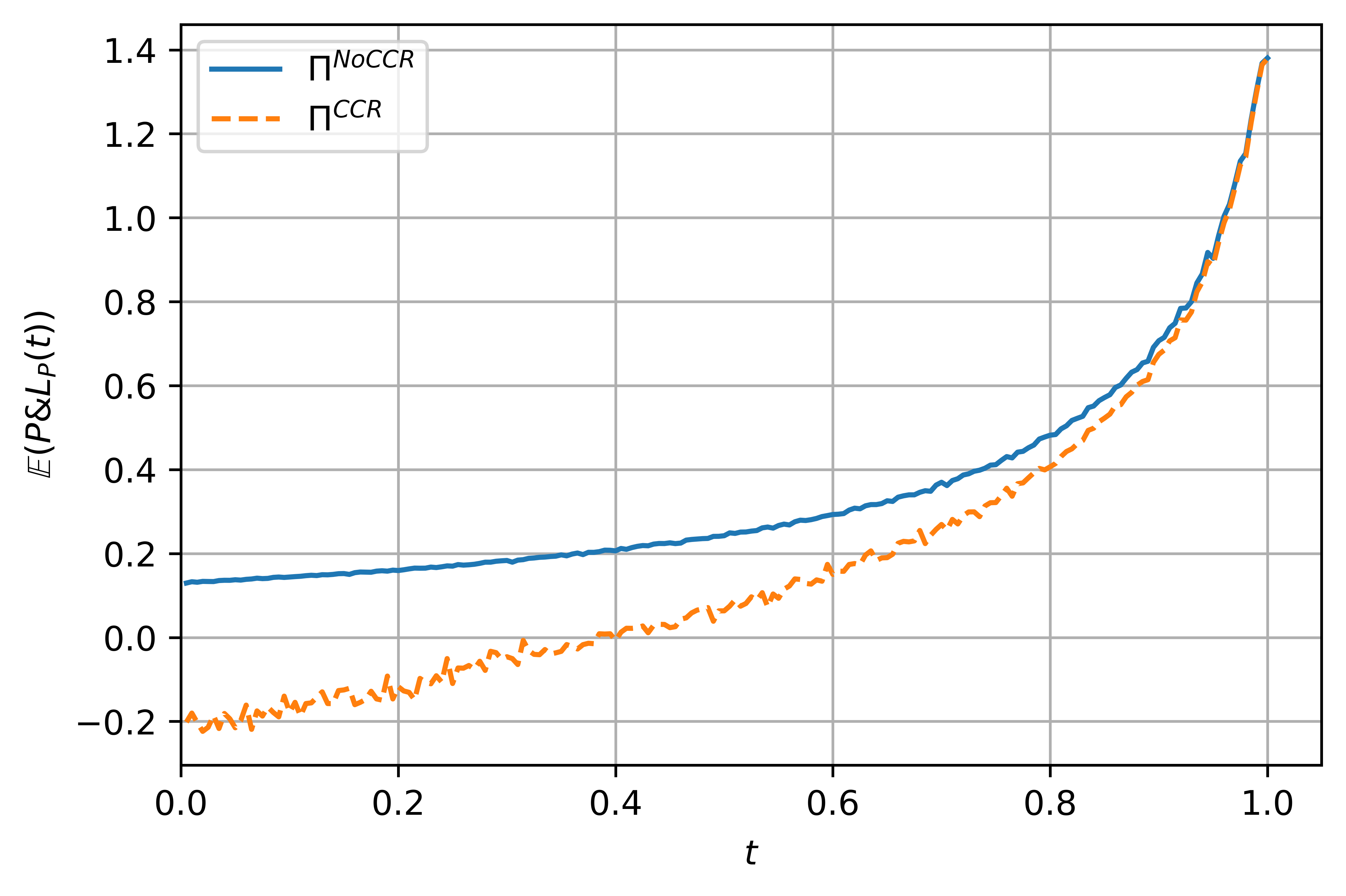}
    \caption{Average $\PnLPortfolio(t)$.}
    \label{fig:mertonSingleOptNoHedgeNewDealPnLPortfolioMean}
  \end{subfigure}
  \begin{subfigure}[b]{\resultFigureSize}
    \includegraphics[width=\linewidth]{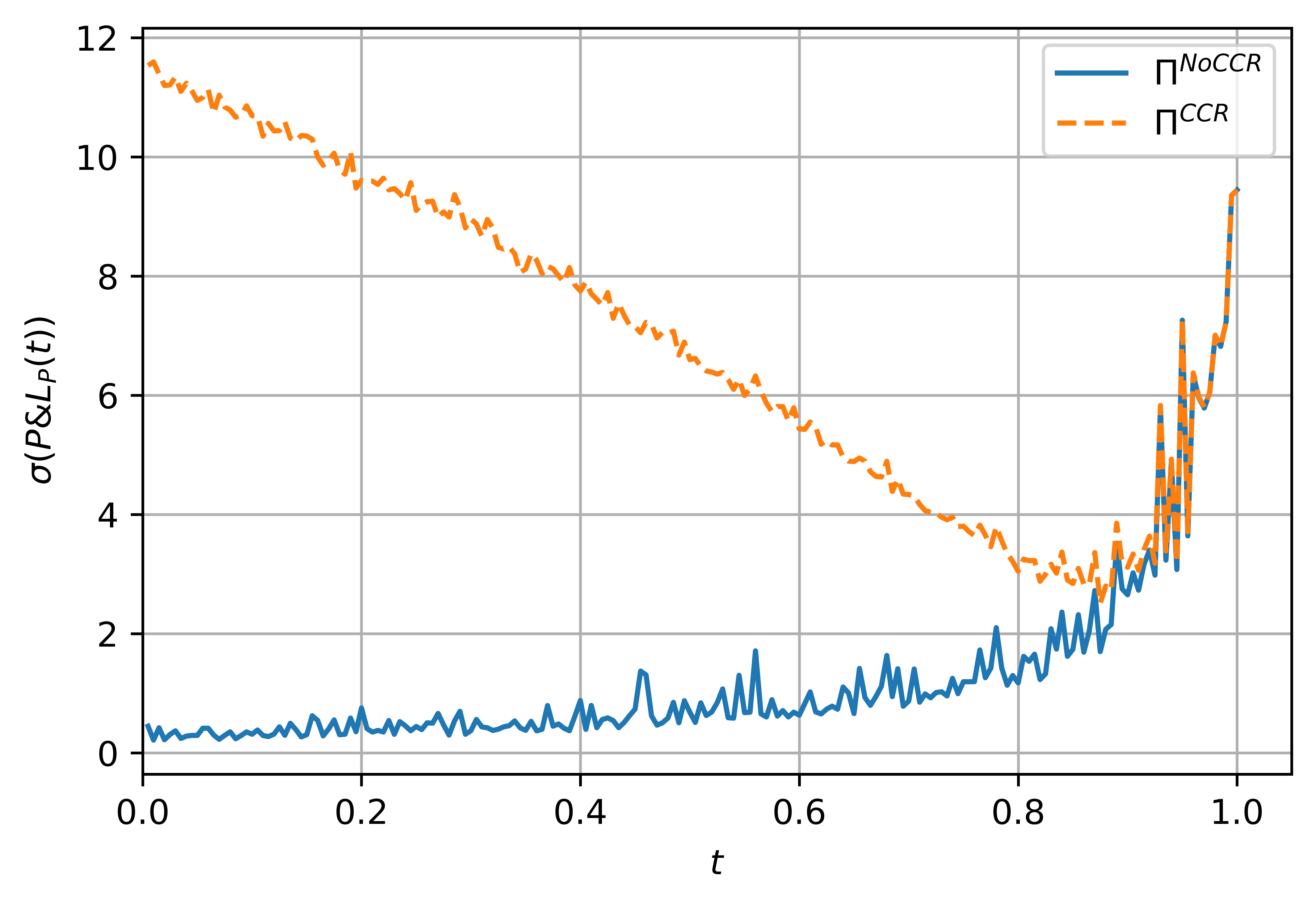}
    \caption{Volatility of $\PnLPortfolio(t)$.}
    \label{fig:mertonSingleOptNoHedgeNewDealPnLPortfolioVolatility}
  \end{subfigure}
  \begin{subfigure}[b]{\resultFigureSize}
    \includegraphics[width=\linewidth]{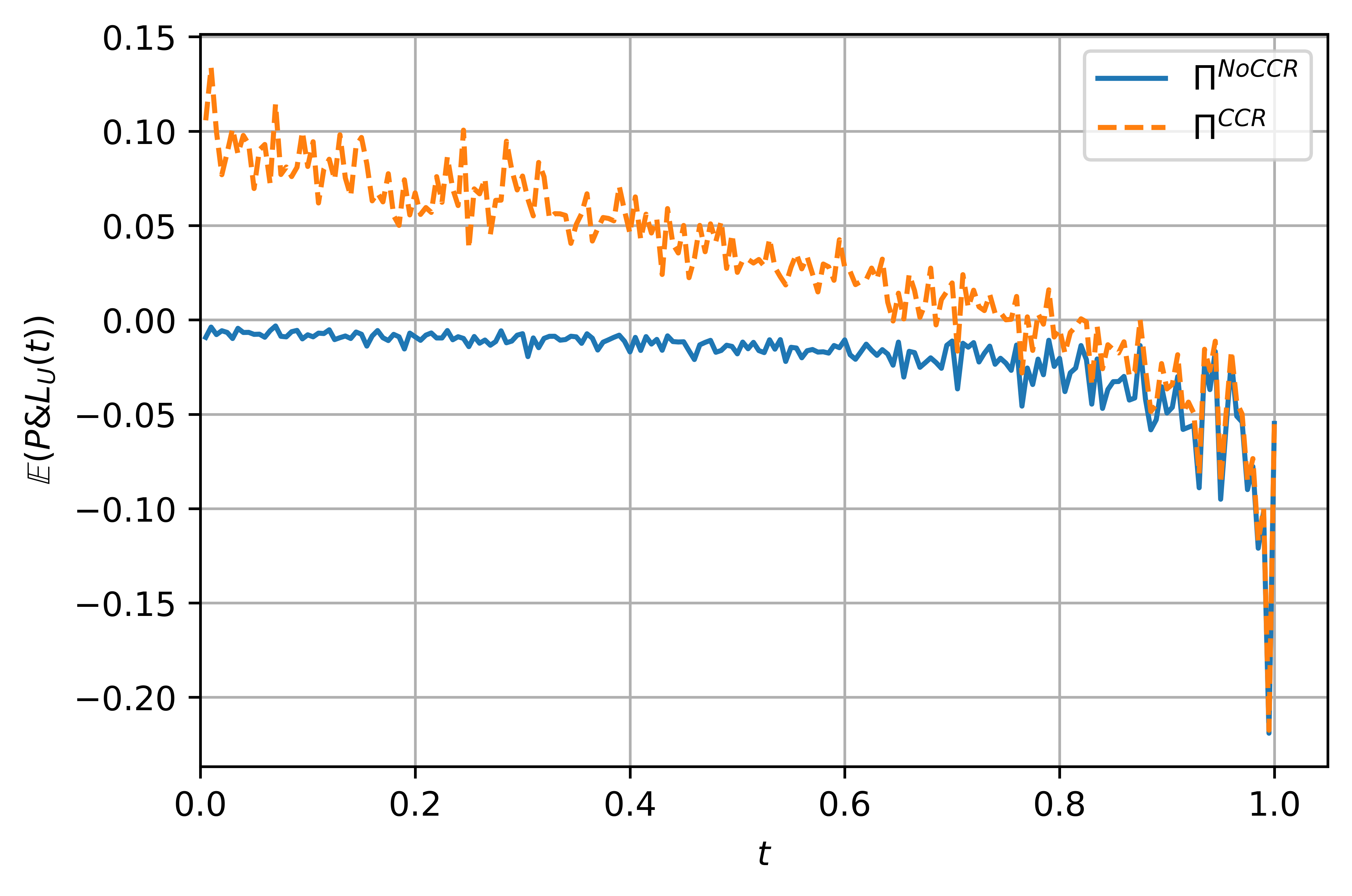}
    \caption{Average $\PnLUnexplained(t)$.}
    \label{fig:mertonSingleOptNoHedgeNewDealPnLUnexplainedMean}
  \end{subfigure}
  \begin{subfigure}[b]{\resultFigureSize}
    \includegraphics[width=\linewidth]{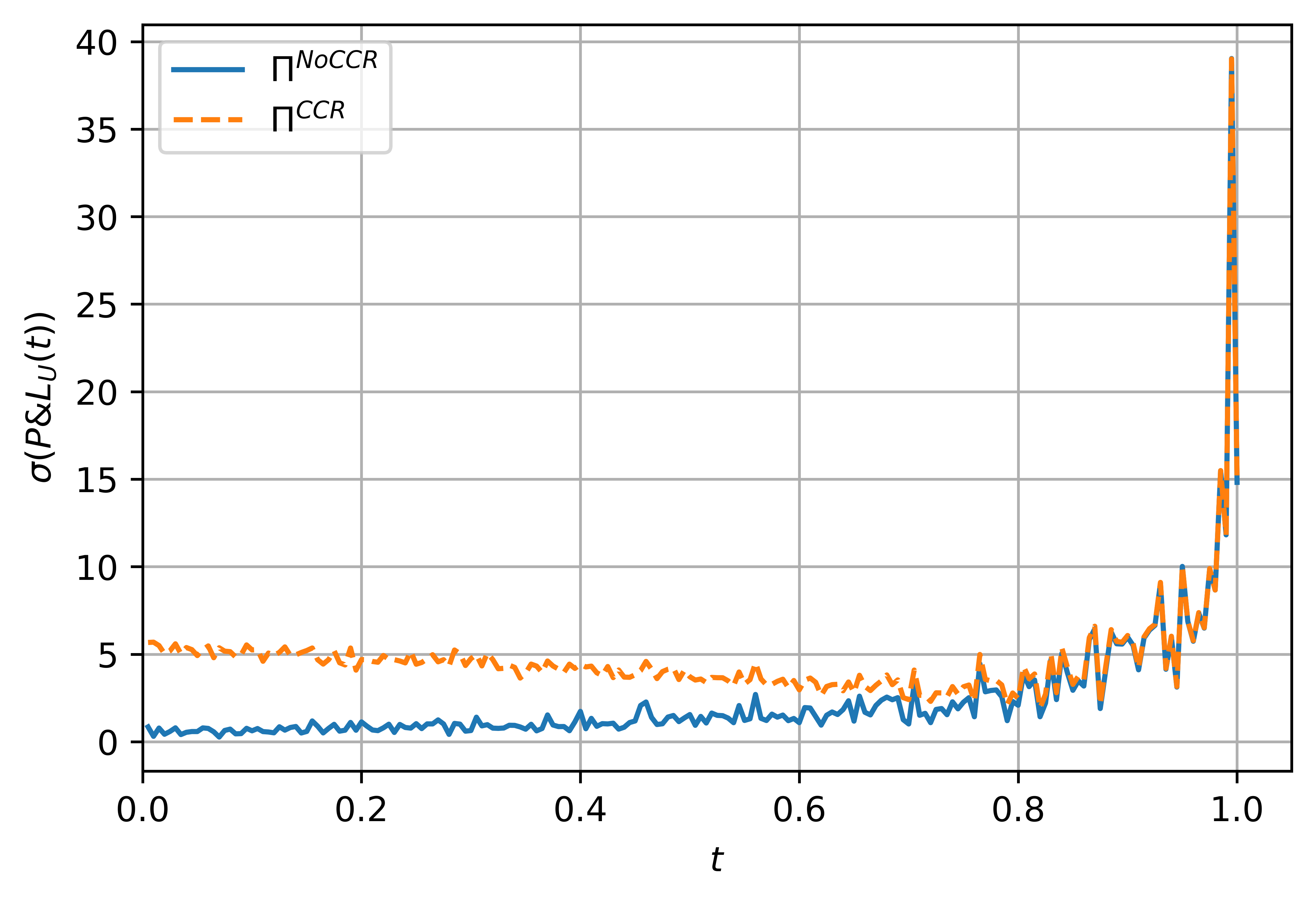}
    \caption{Volatility of $\PnLUnexplained(t)$.}
    \label{fig:mertonSingleOptNoHedgeNewDealPnLUnexplainedVolatility}
  \end{subfigure}
  \caption{Comparison of $\strategyWithoutCCR$ and $\strategyWithCCR$ using a Merton market and valuation model. An additional option is added to the hedging portfolio. $\CVA$ is not hedged.}
  \label{fig:mertonSingleOptNoHedgeNewDeal}
\end{figure}

The effect of the single option hedge is clearly visible in the volatility of $\strategy(t) + \wealth(t)$ for $\strategyWithoutCCR$, as we observe results similar to the pure Black-Scholes case.
So the single option hedge yields the desired effects.
For $\strategyWithCCR$, the CCR still dominates the jumps from the stock.

The mean $\PnLPortfolio$ in Figure~\ref{fig:mertonSingleOptNoHedgeNewDealPnLPortfolioMean} exhibits non-linear behaviour due to the additional optionally in the portfolio.
The $\PnLPortfolio$ volatility in Figure~\ref{fig:mertonSingleOptNoHedgeNewDealPnLPortfolioVolatility} is much lower compared to before, see Figure~\ref{fig:mertonPathsBSDeltaNoHedgeNewDealPnLPortfolioVolatility}.
Again we observe the beneficial effects of the hedging option, as the results look more like the pure Black-Scholes case.

The average $\PnLUnexplained$ in Figure~\ref{fig:mertonSingleOptNoHedgeNewDealPnLUnexplainedMean}, compared to the Merton delta hedge, is comparable up to the same non-linear effect we also observe for the $\PnLPortfolio$.
Overall, the volatility in Figure~\ref{fig:mertonSingleOptNoHedgeNewDealPnLUnexplainedVolatility} is comparable to the previous results in Figure~\ref{fig:mertonPathsBSDeltaNoHedgeNewDealPnLUnexplainedVolatility}, except for a significantly higher peak just before maturity.
This is the result of an exploding gamma explain volatility close to maturity.

\begin{figure}[!h]
  \centering
  \begin{subfigure}[b]{\resultFigureSize}
    \includegraphics[width=\linewidth]{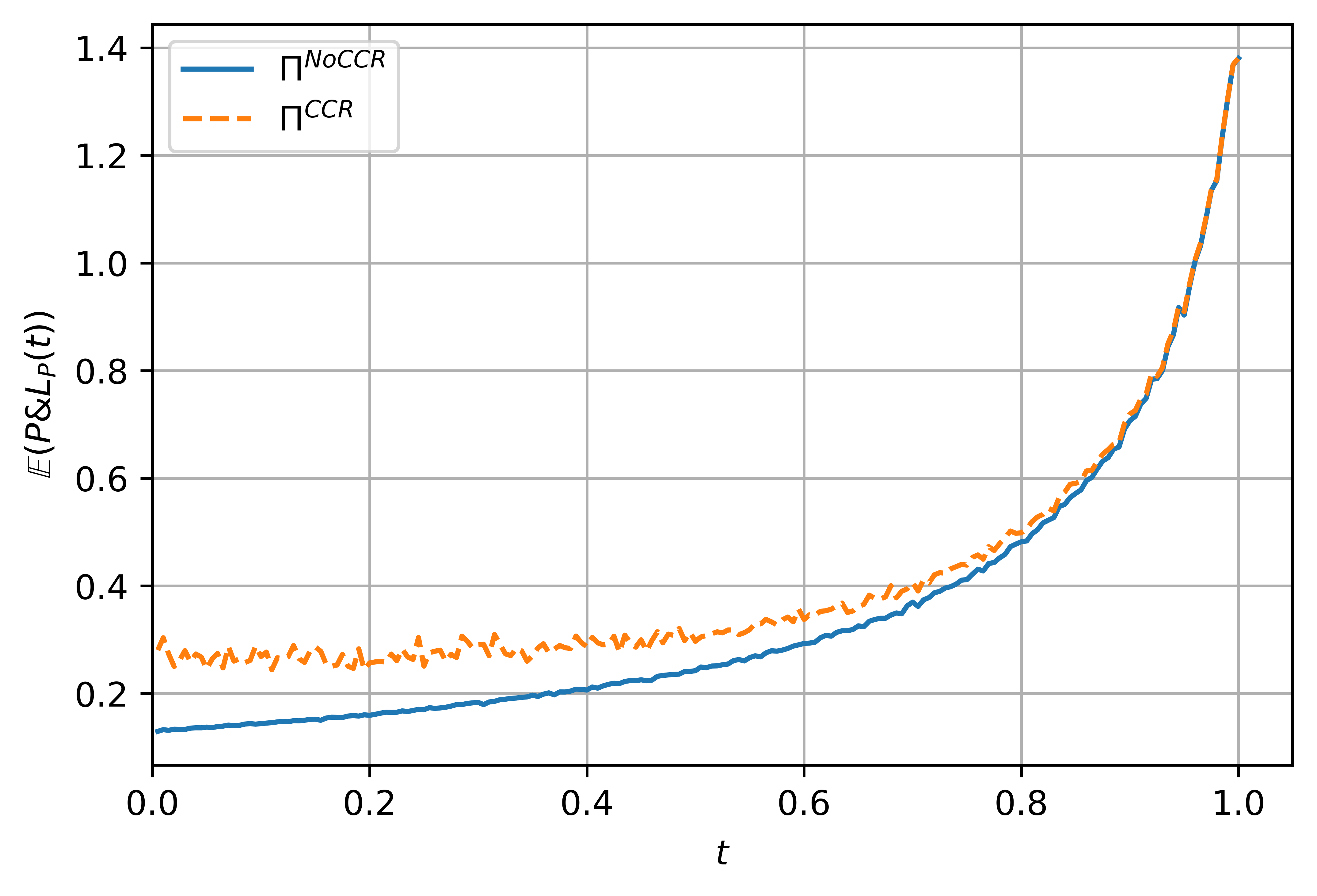}
    \caption{Average $\PnLPortfolio(t)$.}
    \label{fig:mertonSingleOptHedgeNewDealPnLPortfolioMean}
  \end{subfigure}
  \begin{subfigure}[b]{\resultFigureSize}
    \includegraphics[width=\linewidth]{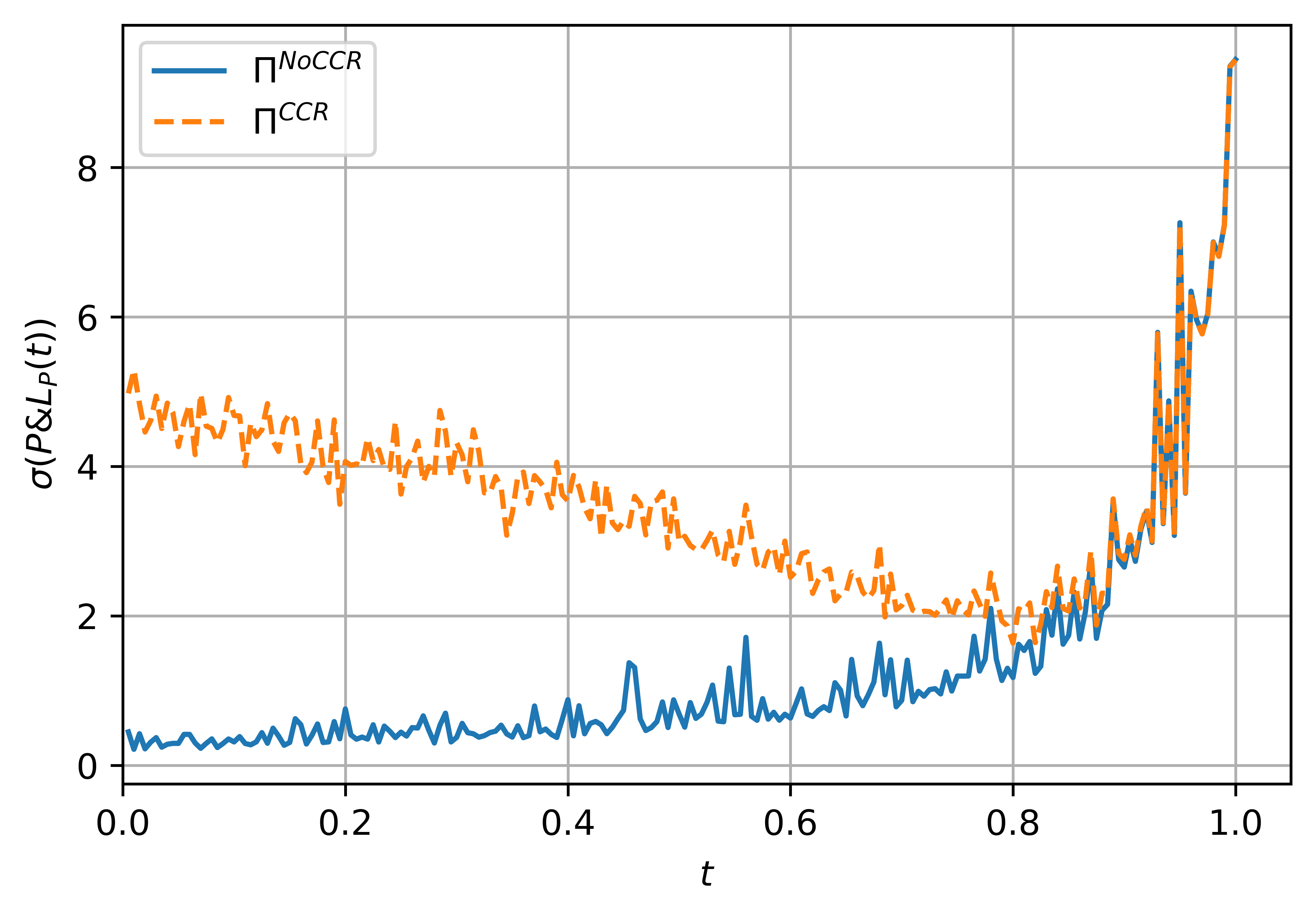}
    \caption{Volatility of $\PnLPortfolio(t)$.}
    \label{fig:mertonSingleOptHedgeNewDealPnLPortfolioVolatility}
  \end{subfigure}
  \caption{Comparison of $\strategyWithoutCCR$ and $\strategyWithCCR$ using a Merton market and valuation model. An additional option is added to the hedging portfolio. $\CVA$ is hedged using both the underlying stock and the additional option.}
  \label{fig:mertonSingleOptHedgeNewDeal}
\end{figure}

As before, the $\CVA$ hedge has no effect on $\strategy(t) + \wealth(t)$ or $\PnLUnexplained$.
The effect of the $\CVA$ hedge on $\PnLPortfolio$, see Figure~\ref{fig:mertonSingleOptHedgeNewDeal}, is similar to the effect in the pure Black-Scholes case.
Especially the volatility, which is the dominating factor, is comparable to the pure Black-Scholes case (see Figure~\ref{fig:bsHedgeNewDealPnLPortfolioVolatility}).
For the average we observe in Figure~\ref{fig:mertonSingleOptHedgeNewDealPnLPortfolioMean} that the $\strategyWithCCR$ is slightly above $\strategyWithoutCCR$, which is also observed for the pure Black-Scholes case (see Figure~\ref{fig:bsHedgeNewDealPnLPortfolioMean}).

Adding the option to the hedging strategy has a clear effect of smaller $\PnLPortfolio$ volatility.
Increasing the number of options in the hedging strategy will likely increase the hedging performance, however, the effect of the first option is already significant~\cite{HeKennedyColemanForsythLiVetzal200601,KennedyForsythVetzal200905}.
The increase in $\PnLUnexplained$ volatility can be explained by the hedging option having a fixed strike.
For some of the simulated paths, this hedge option may become significantly OTM, meaning the sensitivity to the jumps is close to zero.
This results in a less effective hedge of the jump risk.

\rem{In addition to the experiments done so far, we perform the experiments under another set of parameters which represent a stressed market.
For the Black-Scholes case we use an increased volatility of $\vol = 0.35$.
We observe the same patterns but with a different scaling of the results, all conclusions remain the same.
For the Merton jump-diffusion case the stressed market is represented by the following choice of parameters: $\jumpVol = 0.2$, $\jumpMean = -0.4$, and $\jumpIntensity = 0.2$
This can be interpreted as a jump of average size $-31.6\%$ that is expected every 5 years, meaning we expect larger and more frequent jumps than before.
When the stock jump risk is not hedged with an extra option, this risk dominates the CCR effect.
This is no longer the case when adding an option to the set of hedging instruments.
In general we observe comparable patterns, and an increase in level with more variability due to the increased randomness in the stock dynamics.
On the other hand, the $\PnLUnexplained$ volatility is shifted upwards and the peak close to maturity increases significantly.
So for this setting with larger and more frequent jumps, a single hedging option seems to be less sufficient to hedge the randomness in the stock.
However, the effect of the $\CVA$ hedge remains unchanged.
}

\section{Conclusion} \label{sec:conclusion}

In conclusion, dynamic hedging of $\CVA$ market risk is now better understood, both in a Black-Scholes and a Merton jump-diffusion setting.
Starting from a theoretical hedging framework, we have examined the mechanics of a trading strategy which included $\CVA$ pricing and hedging.
We visualized cash-flows and exchanges of traded instruments of the portfolio.
For a case study of a portfolio containing European options and the underlying stock, we used a Monte Carlo simulation to study hedging.
Hedging performance was assessed by analyzing the trading strategy balance, including the corresponding wealth account.
Furthermore, we studied the $\PnL$ behaviour of the strategy, as well as the performance of the $\PnL$ explain.
In particular, analytic results helped us to explain and analyze $\PnL$ behaviour that was observed from the simulation.

First, in a Black-Scholes setting we showed that failing to charge $\CVA$ to a credit risky counterparty will result in an expected loss, motivating the vision of $\CVA$ as fair compensation for this credit risk.
We have shown that $\CVA$ market risk in a Black-Scholes context can satisfactorily be hedged using the underlying stock, resulting in an improved stability of the trading strategy in the form of a lower $\PnLPortfolio$ volatility.
For the $\PnLPortfolio$, we saw a significant increase in volatility as the option maturity approached, which we understood as the result of gamma instability close to maturity.
We conclude that $\CVA$ market risk hedging is necessary for a more stable trading strategy.

The residual risk after the $\CVA$ market risk hedge and $\PnL$ explain can be hedged using CDSs as opposed to the current credit risk warehousing.
This will be interesting to study when taking into account Wrong Way Risk effects, for example by choosing a credit process dynamics which is correlated to the market risks.
We leave this for future research.

Next, we examined our case study portfolio when the underlying stock is driven by a Merton jump-diffusion process.
Adding an option as a hedging instrument mitigates a significant part of the jump risk.
Hedging the $\CVA$ market risk in this situation is still a must.
Extending the hedging instruments with more options may result in an even more stable strategy.

The framework can easily be extended for more advanced models for the stock dynamics like a stochastic (local) volatility model.
This is relevant for exotic trading instruments, e.g., a path-dependent option.
In this case, increase the number of hedging instruments, as now the model is calibrated to a full range of option strikes and maturities.
Computing future exposures will require extra effort, but can be done working in a Least-Squares Monte-Carlo setting for example.

All in all, the theoretical hedging framework in which dynamic $\CVA$ hedging has been studied allows one to consider a broad selection of trading strategies, but above all different and more $\xva$s.
Understanding the mechanics of $\CVA$ hedging is a crucial first step for future research on $\xva$ hedging.

\section*{Acknowledgements} This work has been financially supported by Rabobank.
The authors would like to thank the two anonymous referees for their helpful comments and feedback, which contributed to an improvement of the manuscript.

\bibliographystyle{abbrv}
\bibliography{bib/MacroStrings,bib/Articles,bib/Books,bib/Regulation} 

\appendix
\section{The Merton jump-diffusion model} \label{sec:merton}

\subsection{Analytical European option prices and sensitivities} \label{sec:mertonOptionPrice}
Here we re-iterate a result from~\cite{Merton197601,OosterleeGrzelak201911}, namely the analytical expression for a European call option price on stock $\stock$, maturity $T$, and strike $\strike$, under the Merton jump-diffusion model:
\begin{align}
    \tradeVal(t)
        &= \expPower{-\shortRate(T-t)} \sum_{n \geq 0} \frac{\left[ \jumpIntensity(T-t)\right]^n \expPower{-\jumpIntensity(T-t)}}{n!}\overline{V}(n), \label{eq:mertonPrice} \\
    \overline{V}(n)
        &= \expPower{\hat{\mu}_X(n) + \half \hat{\sigma}^2_X(n)(T-t)}\normCDF(d_1) - \strike \normCDF(d_2), \nonumber \\ 
    \hat{\mu}_X(n)
        &= \log (\stock(t)) + \left[ \shortRate - \jumpIntensity \left( \expPower{\jumpMean+ \half \jumpVol^2} - 1\right) - \half \vol^2 \right](T-t) + n\jumpMean, \nonumber \\ 
    \hat{\sigma}_X(n)
        &= \sqrt{\vol^2 + \frac{n \jumpVol^2}{T-t}},  \ \ 
    d_2
        = d_1 - \hat{\sigma}_X(n)\sqrt{T-t}, \nonumber \\ 
    d_1
        &= \frac{\log \left( \frac{\stock(t)}{\strike} \right) + \left[ \shortRate - \jumpIntensity \left( \expPower{\jumpMean+ \half \jumpVol^2} - 1\right) - \half \vol^2 + \hat{\sigma}^2_X(n)\right](T-t) + n\jumpMean}{\hat{\sigma}_X(n)\sqrt{T-t}}, \nonumber 
\end{align}
where $\shortRate$ is the risk-free interest rate, $\vol$ is the stock volatility.
Furthermore, $\jumpMean$, $\jumpVol$, and $\jumpIntensity$ are respectively the jump mean, volatility and intensity.
In the case of a put option, one can simply apply the \textit{put-call parity} to the call option price from Equation~\eqref{eq:mertonPrice}.

It can be shown that call price $\tradeVal(t)$ from Equation~\eqref{eq:mertonPrice} has the following analytical expressions of derivatives w.r.t. jump parameters $\jumpMean$, $\jumpVol$ and $\jumpIntensity$:
\begin{align}
    \pderiv{\tradeVal(t)}{\jumpMean}
        &= \expPower{-\shortRate(T-t)} \sum_{n \geq 0} \frac{\left[ \jumpIntensity(T-t)\right]^n \expPower{-\jumpIntensity(T-t)}}{n!} \left(  n - \jumpIntensity  \expPower{\jumpMean + \half \jumpVol^2} (T-t) \right) \expPower{\hat{\mu}_X(n) + \half \hat{\sigma}^2_X(n)(T-t)}\normCDF(d_1), \label{eq:dVdMuJ3} \\
    \pderiv{\tradeVal(t)}{\jumpVol}
        &= \expPower{-\shortRate(T-t)} \sum_{n \geq 0} \frac{\left[ \jumpIntensity(T-t)\right]^n \expPower{-\jumpIntensity(T-t)}}{n!} \nonumber \\
        &\quad \left( \jumpVol \left(n -  \jumpIntensity  \expPower{\jumpMean + \half \jumpVol^2} (T-t)\right) \expPower{\hat{\mu}_X(n) + \half \hat{\sigma}^2_X(n)(T-t)}\normCDF(d_1) + K \normPDF(d_2) \frac{1}{\hat{\sigma}_X(n)} \frac{n\jumpVol}{\sqrt{T-t}}  \right),  \label{eq:dVdSigmaJ3} \\
    \pderiv{\tradeVal(t)}{\jumpIntensity}
        &= \expPower{-\shortRate(T-t)} \sum_{n \geq 0} \frac{\left[ \jumpIntensity(T-t)\right]^n \expPower{-\jumpIntensity(T-t)}}{n!} \nonumber \\
        &\quad \left( \overline{V}(n) \left[ \frac{n}{\jumpIntensity} - (T-t)\right] - \left( \expPower{\jumpMean + \half \jumpVol^2} -1\right)(T-t) \expPower{\hat{\mu}_X(n) + \half \hat{\sigma}^2_X(n)(T-t)}\normCDF(d_1) \right).\label{eq:dVdXiJ3}
\end{align}
These results are valid for both call and put options, which can be seen from the put-call parity.

In a similar fashion we can derive the following expression for the call option \textit{delta} and \textit{gamma} under the Merton jump-diffusion model:
\begin{align}
    \pderiv{\tradeVal(t)}{\stock}
        &= \expPower{-\shortRate(T-t)} \sum_{n \geq 0} \frac{\left[ \jumpIntensity(T-t)\right]^n \expPower{-\jumpIntensity(T-t)}}{n!}  \cdot  \expPower{\hat{\mu}_X(n) - \log(\stock(t))+ \half \hat{\sigma}^2_X(n)(T-t)}\normCDF(d_1), \label{eq:dVdS}\\
    \ppderiv{\tradeVal(t)}{\stock}
        &= \expPower{-\shortRate(T-t)} \sum_{n \geq 0} \frac{\left[ \jumpIntensity(T-t)\right]^n \expPower{-\jumpIntensity(T-t)}}{n!}  \cdot  \expPower{\hat{\mu}_X(n) - \log(\stock(t))+ \half \hat{\sigma}^2_X(n)(T-t)}\frac{\normPDF(d_1)}{\stock(t) \hat{\sigma}_X(n)\sqrt{T - t}}, \label{eq:dV2dS2}
\end{align}

The option pricing formula~\eqref{eq:mertonPrice} contains an infinite sum of scaled Black-Scholes option prices, being a direct result of the jump size following a continuous distribution.
This clearly shows that if one attempts to hedge the jump risk introduced by the model, one would theoretically need infinitely many options to do so, which is infeasible from a practical perspective.

\subsection{Jump parameter impact on implied volatilities} \label{sec:mertonJumpImpact}
We assess the impact of the jump parameters on the \textit{Black-Scholes implied volatilities} $\impliedVol$.
We do so by varying parameters $\jumpMean$, $\jumpVol$, and $\jumpIntensity$ one by one while keeping all the others constant.

\begin{figure}[!h]
  \centering
  \begin{subfigure}[b]{\impliedVolFigureSize}
    \includegraphics[width=\linewidth]{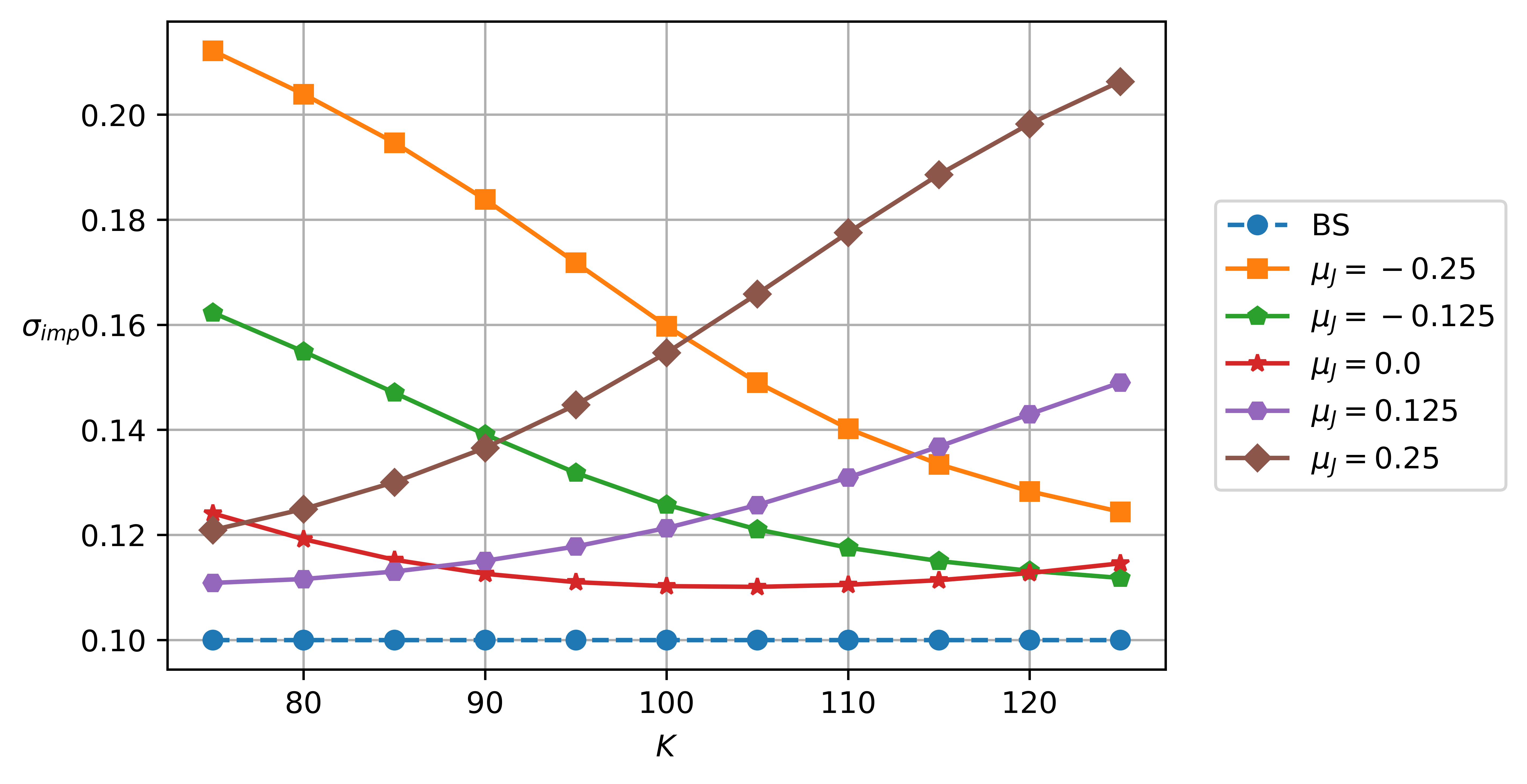}
    \caption{Impact of jump size mean $\jumpMean$.}
    \label{fig:muImpact}
  \end{subfigure}
  \begin{subfigure}[b]{\impliedVolFigureSize}
    \includegraphics[width=\linewidth]{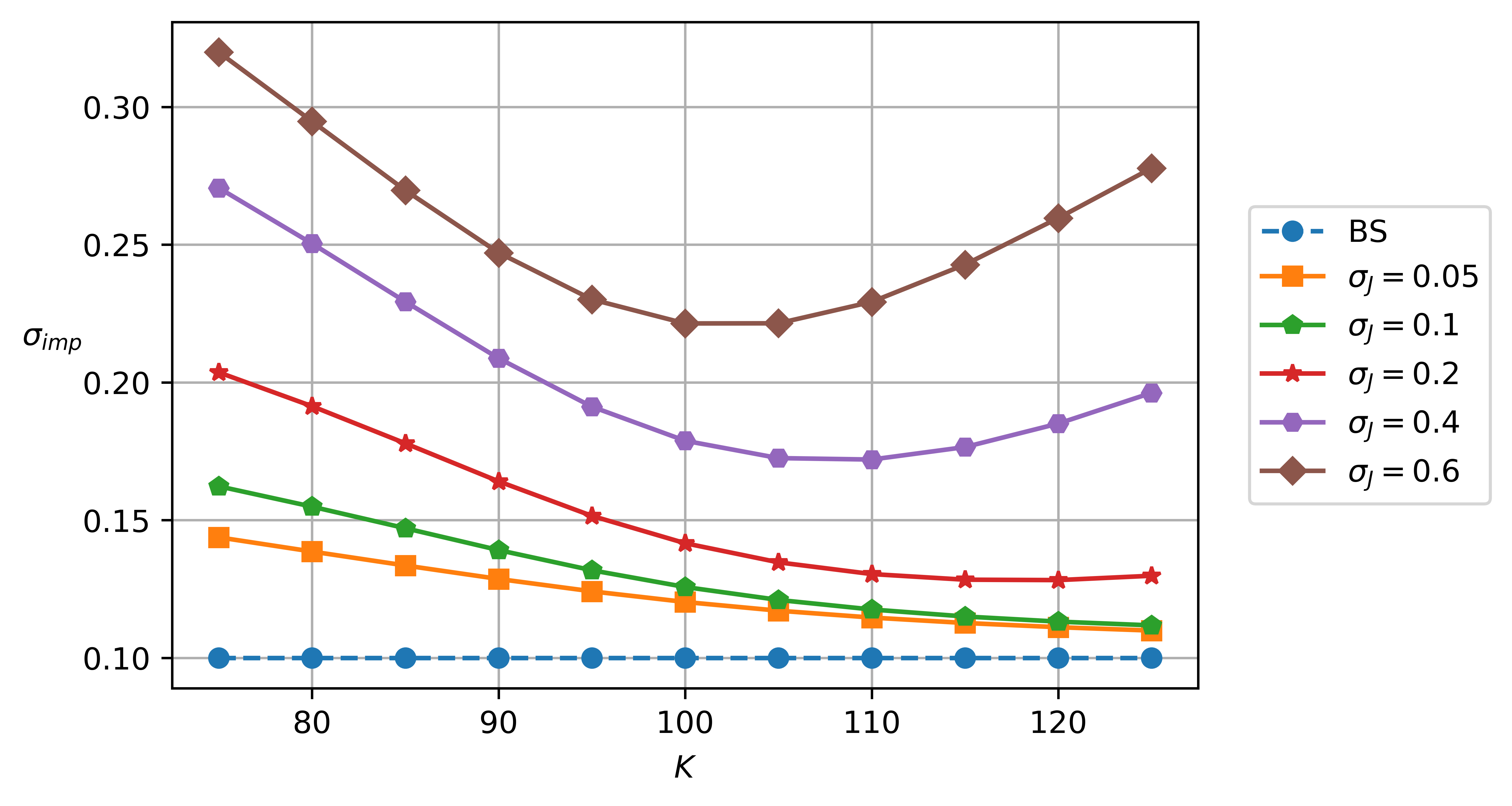}
    \caption{Impact of jump size volatility $\jumpVol$.}
    \label{fig:sigmaImpact}
  \end{subfigure}
  \begin{subfigure}[b]{\impliedVolFigureSize}
    \includegraphics[width=\linewidth]{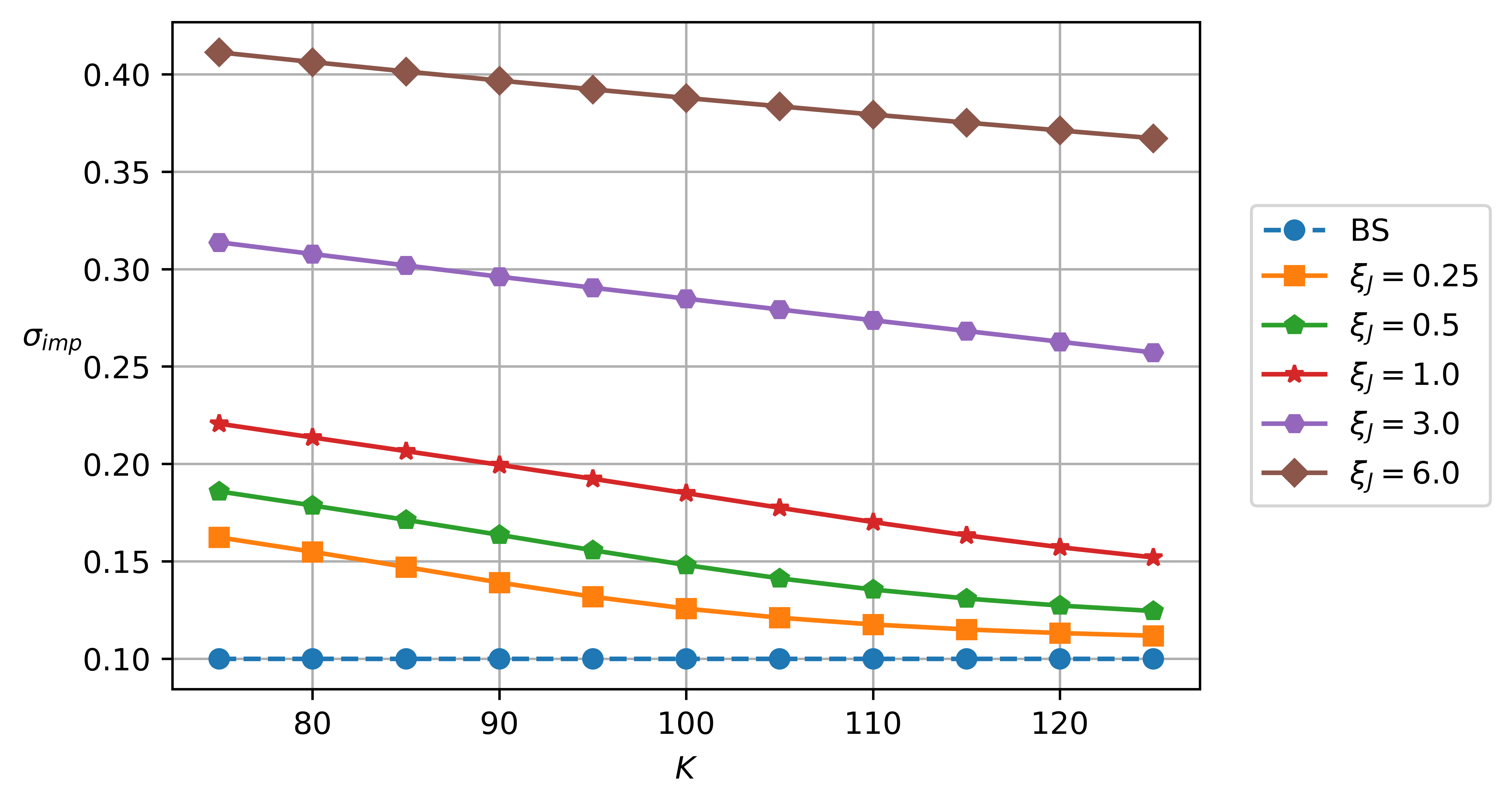}
    \caption{Impact of jump intensity $\jumpIntensity$.}
    \label{fig:xiImpact}
  \end{subfigure}
  \caption{Merton jump parameter impact on $\impliedVol$. Parameters: $\stock(t_0) = 100$, $\shortRate=0.05$, $\vol = 0.1$, $t_0 = 0$, $T = 1$. Jump parameters base values: $\jumpMean = -0.125$, $\jumpVol = 0.1$, $\jumpIntensity = 0.25$. $\jumpMean \in \left\{ -0.25, -0.125, 0.0, 0.125, 0.25 \right\}$, $\jumpVol \in \left\{ 0.05, 0.1, 0.2, 0.4, 0.6 \right\}$, $\jumpIntensity \in \left\{ 0.25, 0.5, 1.0, 3.0, 6.0 \right\}$.}
  \label{fig:parameterImpact}
\end{figure}

In Figure~\ref{fig:muImpact} we observe that $\jumpMean$ has a \textit{level effect}. In addition the \textit{slope}, i.e., \textit{skew}, and direction of the slope are affected.
For the simulations we choose a negative value of $\jumpMean$, which implies that the stock will likely go down.
The reason of this choice is that typically in the market we see that OTM options are cheaper, which matches the higher likelihood of a downward jump in stock.
From Figure~\ref{fig:sigmaImpact} we conclude that $\jumpVol$ has an impact on both the level and the \textit{curvature} of the implied volatility.
A higher $\jumpVol$ means more uncertainty, hence higher option prices, so also a higher $\impliedVol$.
The curvature effect can be observed by the OTM strike region exhibiting a more pronounced \textit{hockey-stick} effect which becomes larger as $\jumpVol$ increases.
Finally, Figure~\ref{fig:xiImpact} tells us that $\jumpIntensity$ impacts the level only.
For higher $\jumpIntensity$, jumps will occur more frequently, introducing more uncertainty, resulting in higher $\impliedVol$.

\subsection{Convergence of number of terms in expansion} \label{sec:mertonPriceConvergence}
We have seen that the option price~\eqref{eq:mertonPrice} as well as sensitivities~(\ref{eq:dVdMuJ3}-\ref{eq:dV2dS2}) can be written as an infinite sum of scaled Black-Scholes option prices.
The question arises how to approximate these infinite sums in a practical situation by means of a numerical implementation.
Each of the quantities we need to compute contain the following weights in the infinite summation:
\begin{align*}
    \frac{\left[ \jumpIntensity(T-t)\right]^n \expPower{-\jumpIntensity(T-t)}}{n!}.
\end{align*}
These weights appear to be exponentially decaying, indicating that in a numerical implementation the sum can simply be cut off at some point at which the desired accuracy is achieved.

As an example we cut off the option price at a \textit{basis-point (bps)} level, i.e., $10^{-4}$.
\begin{figure}[!h]
  \centering
  \begin{subfigure}[b]{\mertonConvergenceFigureSize}
    \includegraphics[width=\linewidth]{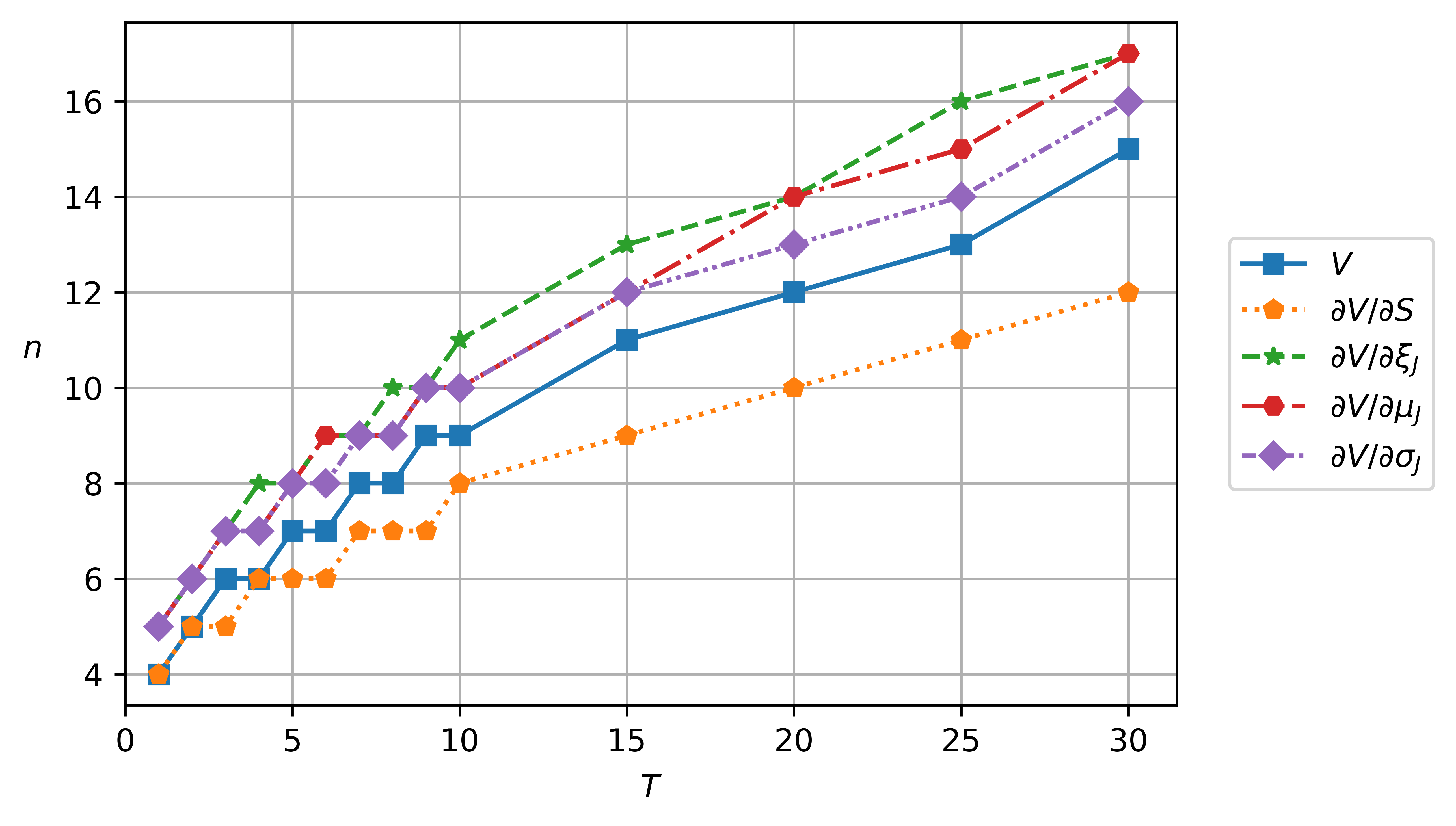}
    \caption{Impact of maturity $T$.}
    \label{fig:mertonImpactMaturity1e-4}
  \end{subfigure}
  \begin{subfigure}[b]{\mertonConvergenceFigureSize}
    \includegraphics[width=\linewidth]{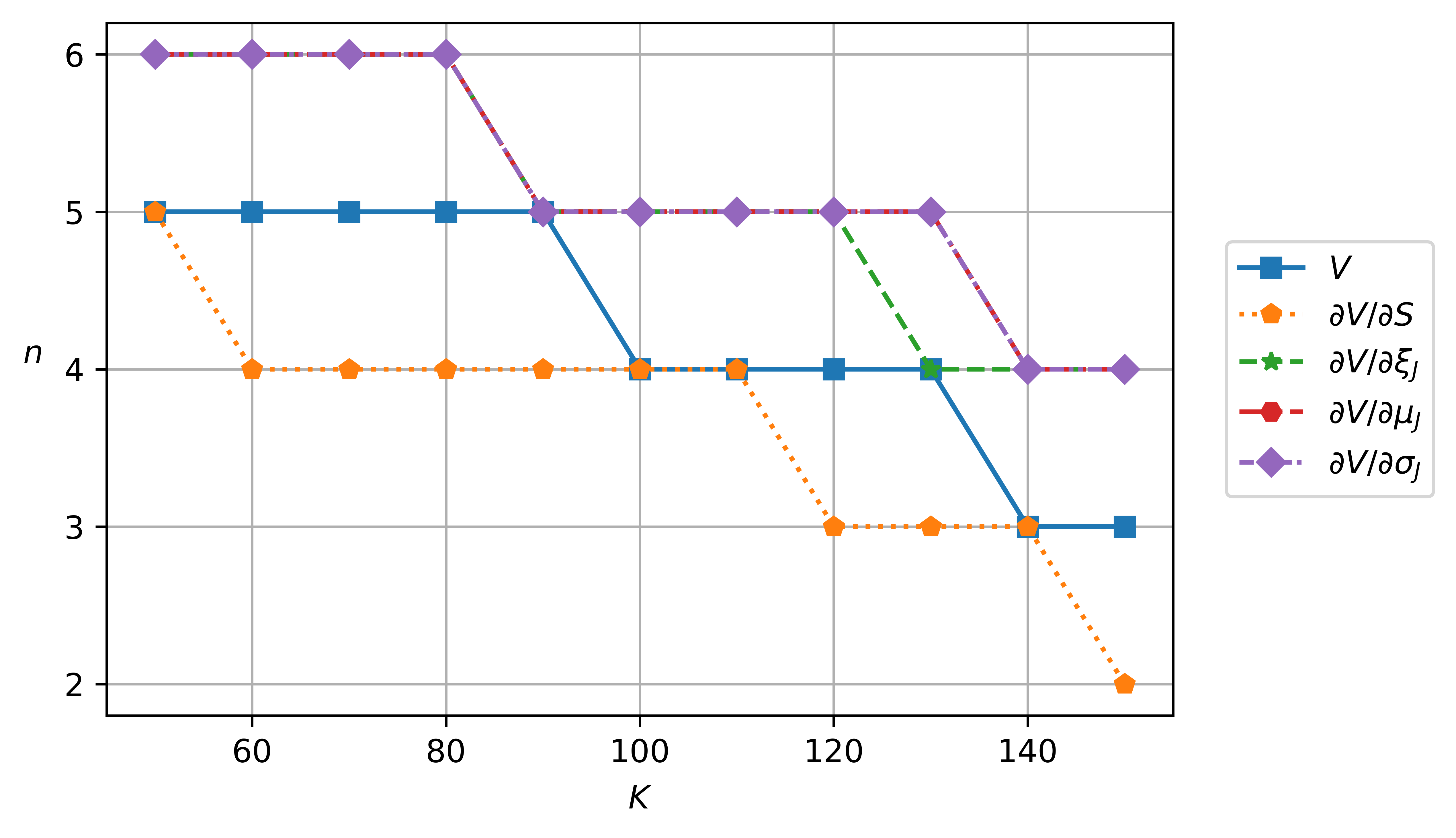}
    \caption{Impact of strike $\strike$.}
    \label{fig:mertonImpactStrike1e-4}
  \end{subfigure}
  \begin{subfigure}[b]{\mertonConvergenceFigureSize}
    \includegraphics[width=\linewidth]{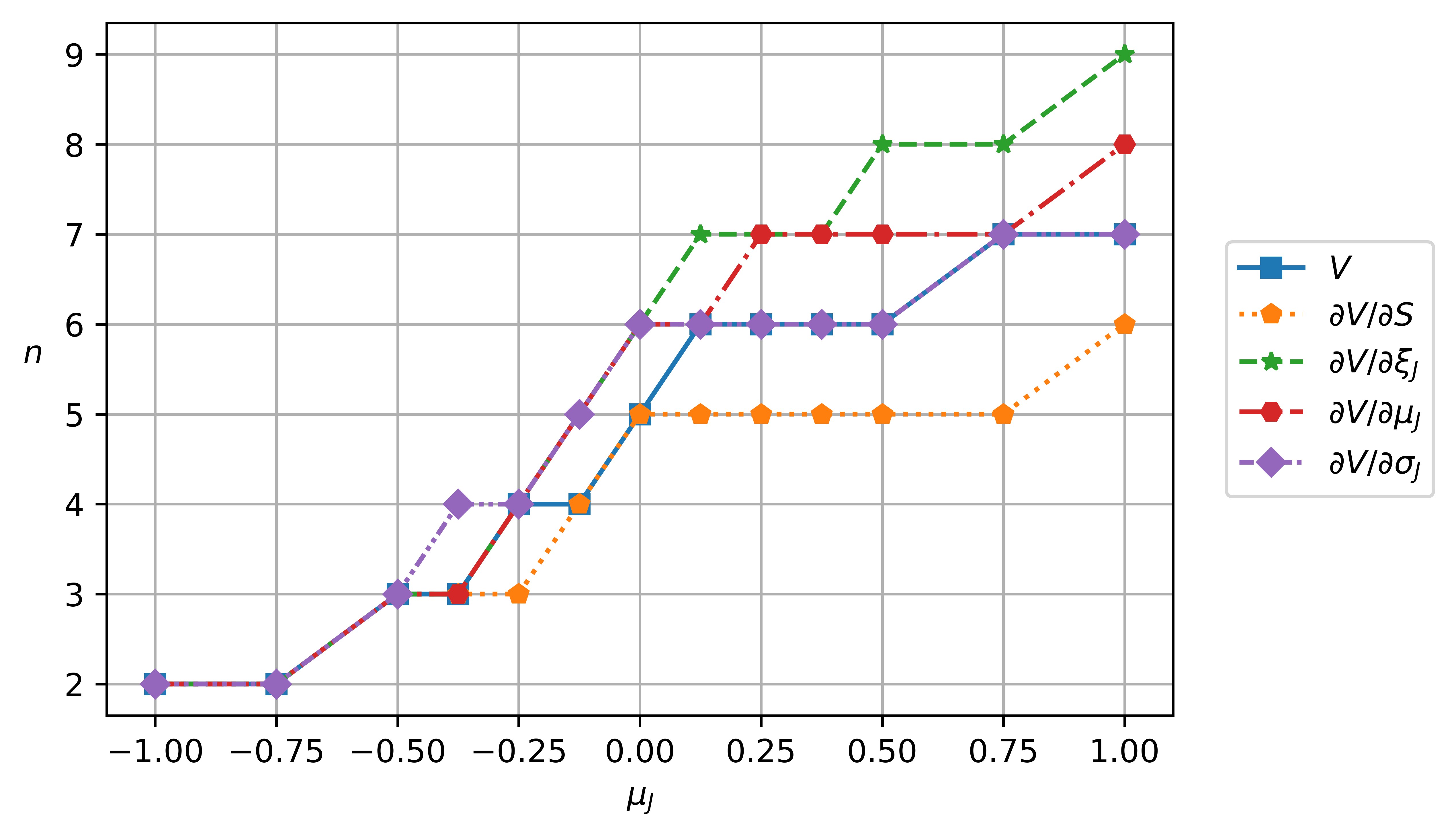}
    \caption{Impact of jump size mean $\jumpMean$.}
    \label{fig:mertonImpactMu1e-4}
  \end{subfigure}
    \begin{subfigure}[b]{\mertonConvergenceFigureSize}
    \includegraphics[width=\linewidth]{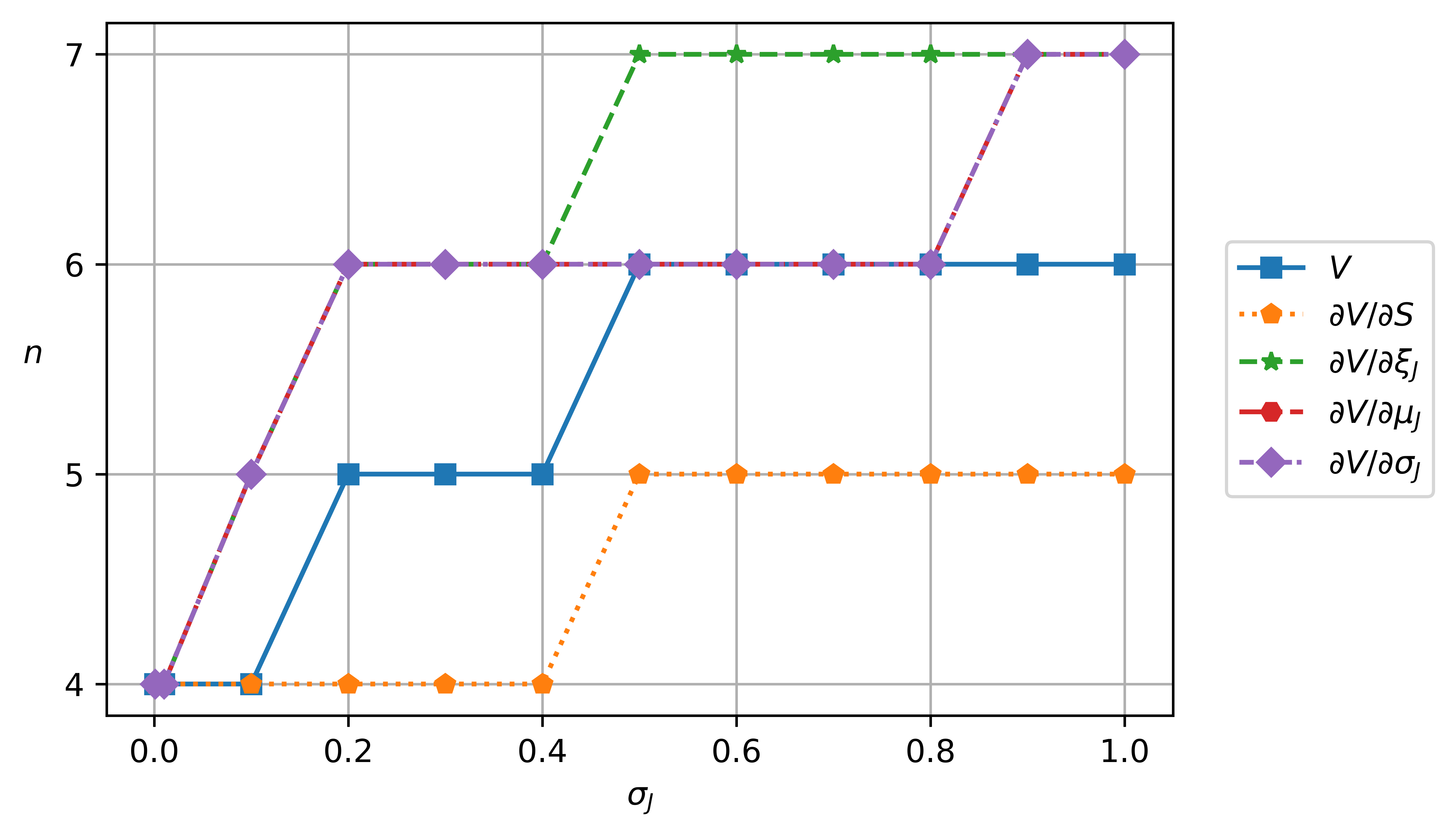}
    \caption{Impact of jump size volatility $\jumpVol$.}
    \label{fig:mertonImpactSigma1e-4}
  \end{subfigure}
  \begin{subfigure}[b]{\mertonConvergenceFigureSize}
    \includegraphics[width=\linewidth]{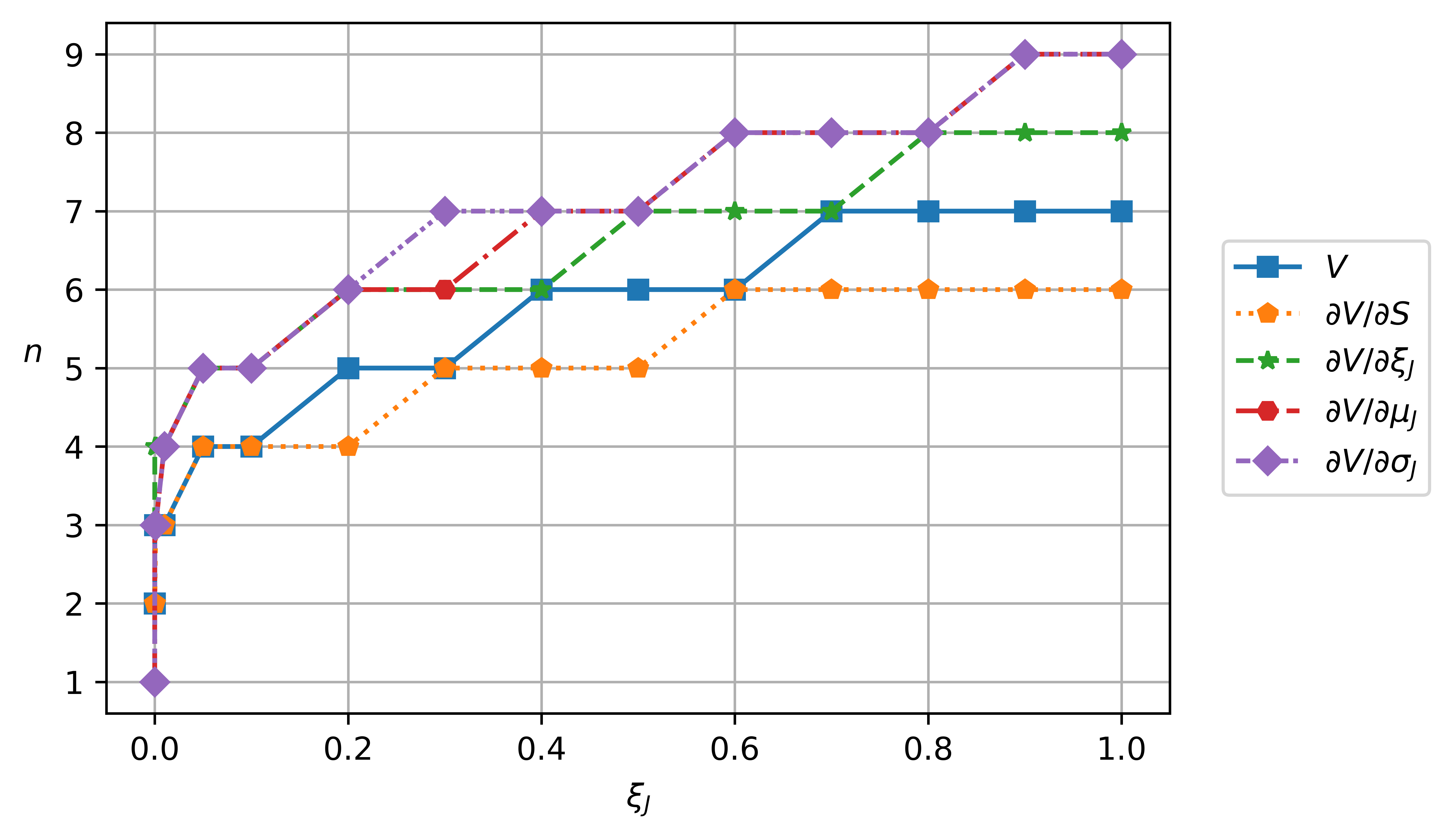}
    \caption{Impact of jump intensity $\jumpIntensity$.}
    \label{fig:mertonImpactXi1e-4}
  \end{subfigure}
  \caption{Impact on number of terms in the Merton option price expansion. Tolerance $10^{-4}$ has been used to determine the cutoff point.
  Parameters: $t_0 = 0$, $T = 1$, $\stock(t_0) = 100$, $\shortRate = 0.1$, $\vol = 0.2$, $\strike=100$, $\jumpMean = -0.125$, $\jumpVol = 0.1$, $\jumpIntensity = 0.1$.}
  \label{fig:mertonImpact1e-4}
\end{figure}

\begin{figure}[!h]
  \centering
  \begin{subfigure}[b]{\mertonConvergenceFigureSize}
    \includegraphics[width=\linewidth]{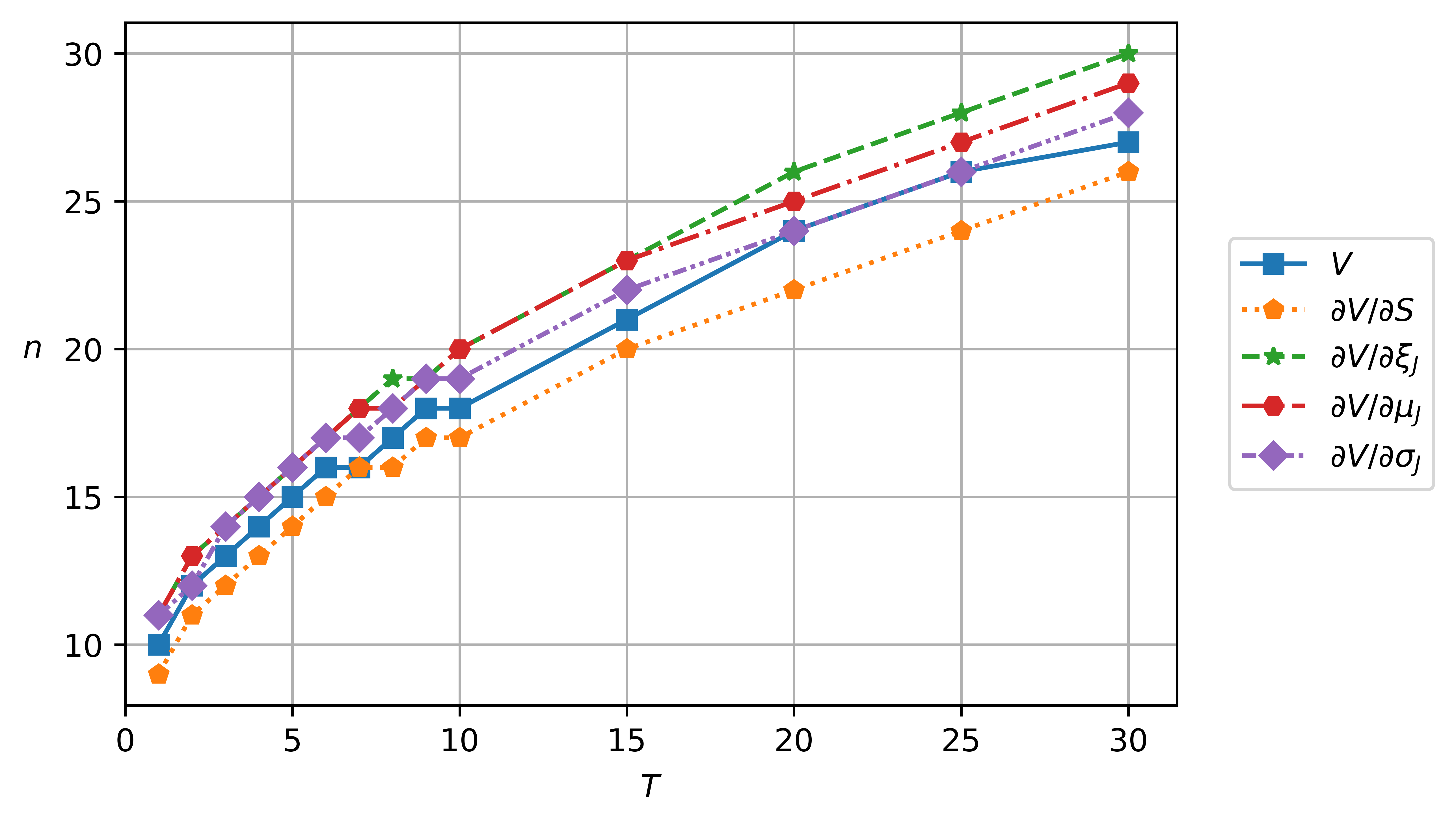}
    \caption{Impact of maturity $T$.}
    \label{fig:mertonImpactMaturity1e-15}
  \end{subfigure}
  \begin{subfigure}[b]{\mertonConvergenceFigureSize}
    \includegraphics[width=\linewidth]{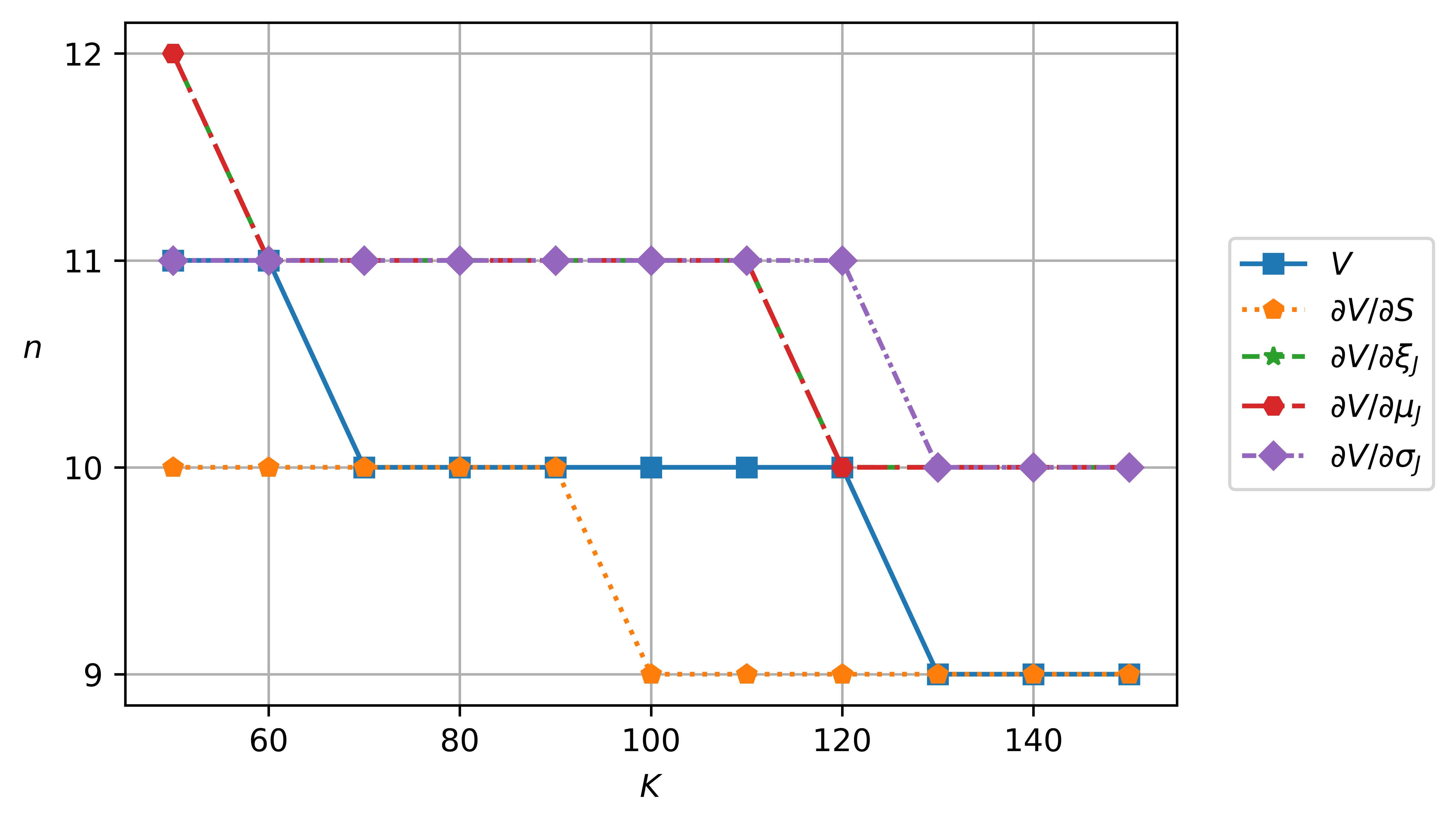}
    \caption{Impact of strike $\strike$.}
    \label{fig:mertonImpactStrike1e-15}
  \end{subfigure}
  \begin{subfigure}[b]{\mertonConvergenceFigureSize}
    \includegraphics[width=\linewidth]{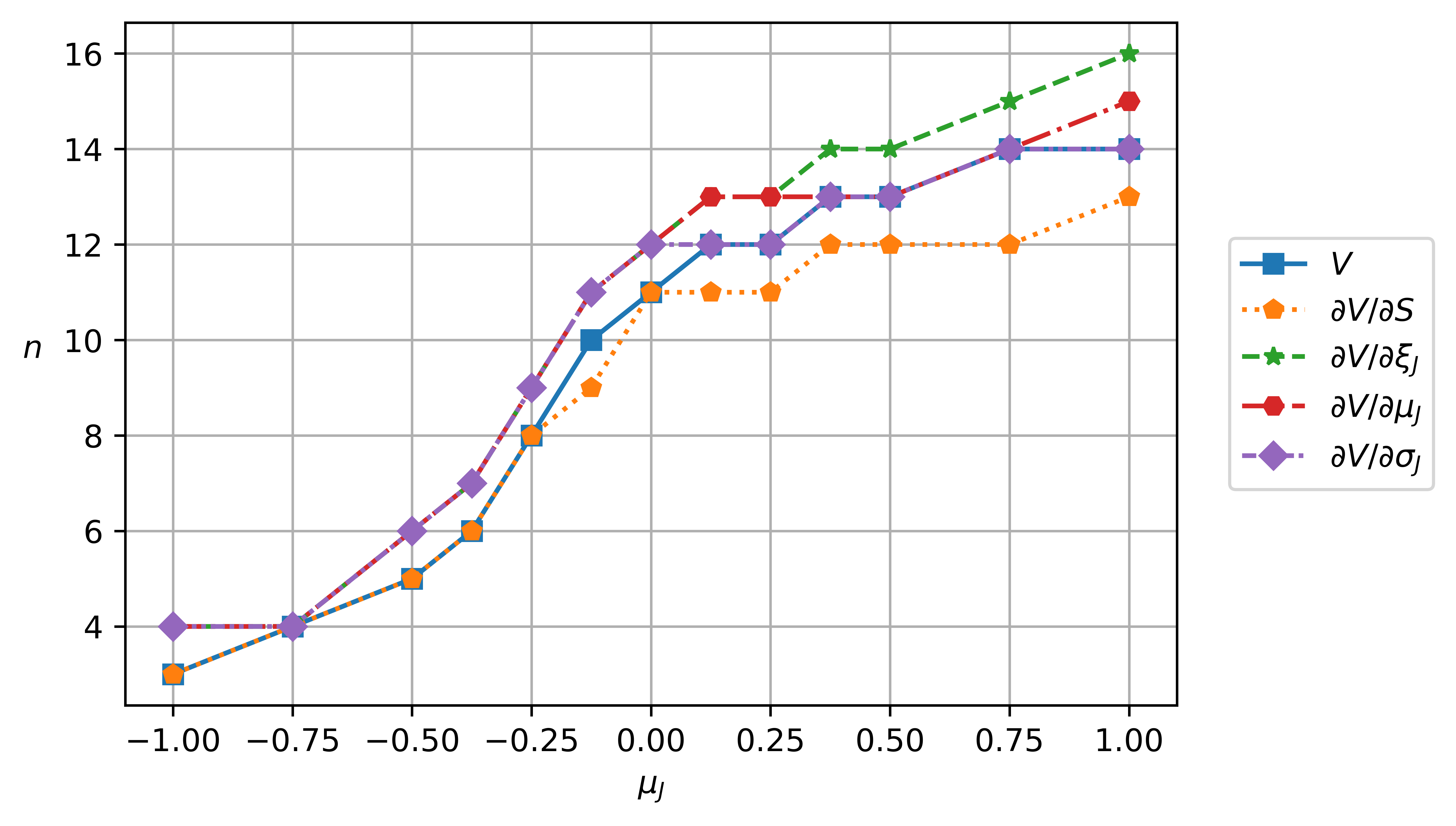}
    \caption{Impact of jump size mean $\jumpMean$.}
    \label{fig:mertonImpactMu1e-15}
  \end{subfigure}
  \begin{subfigure}[b]{\mertonConvergenceFigureSize}
    \includegraphics[width=\linewidth]{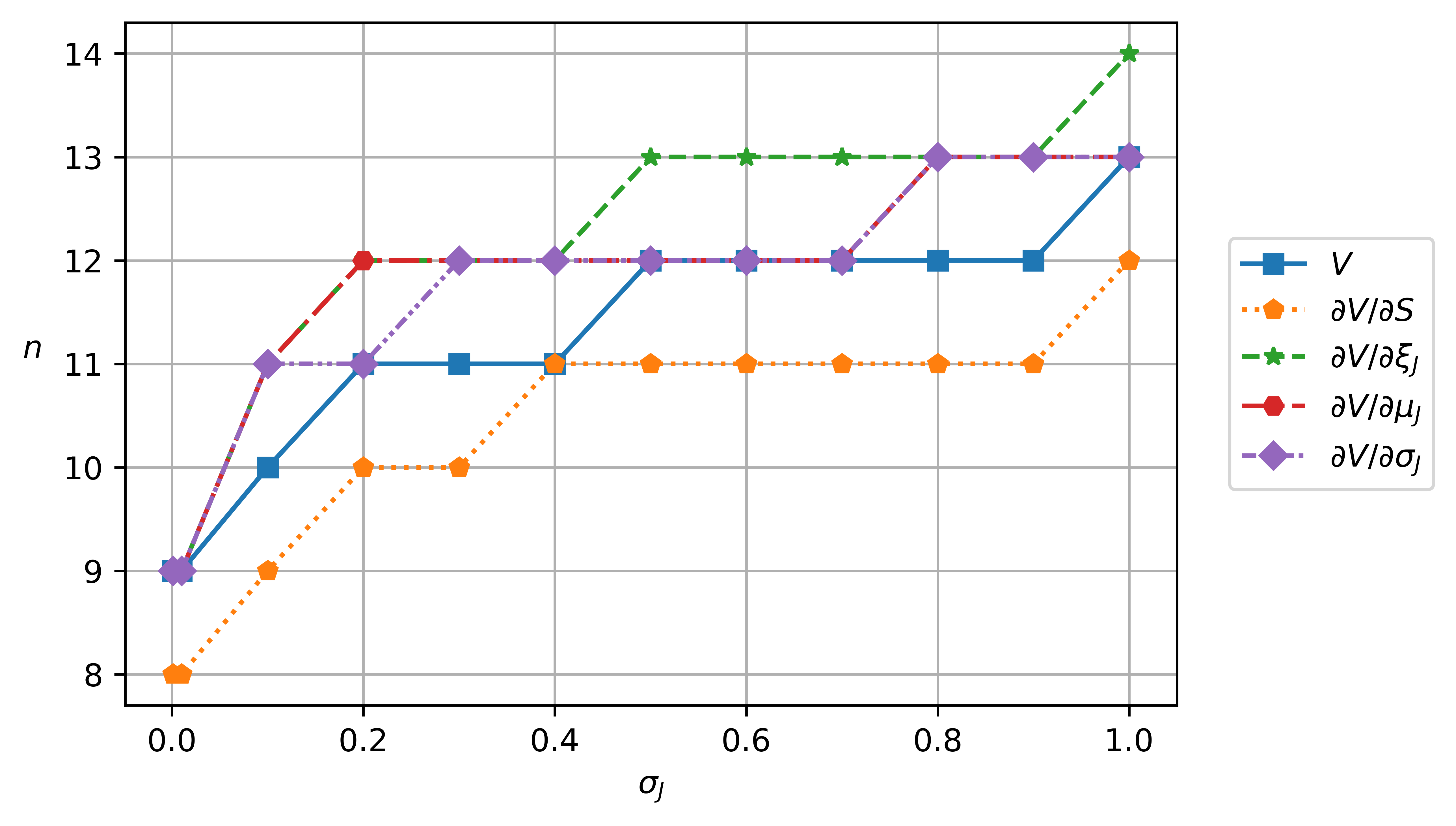}
    \caption{Impact of jump size volatility $\jumpVol$.}
    \label{fig:mertonImpactSigma1e-15}
  \end{subfigure}
  \begin{subfigure}[b]{\mertonConvergenceFigureSize}
    \includegraphics[width=\linewidth]{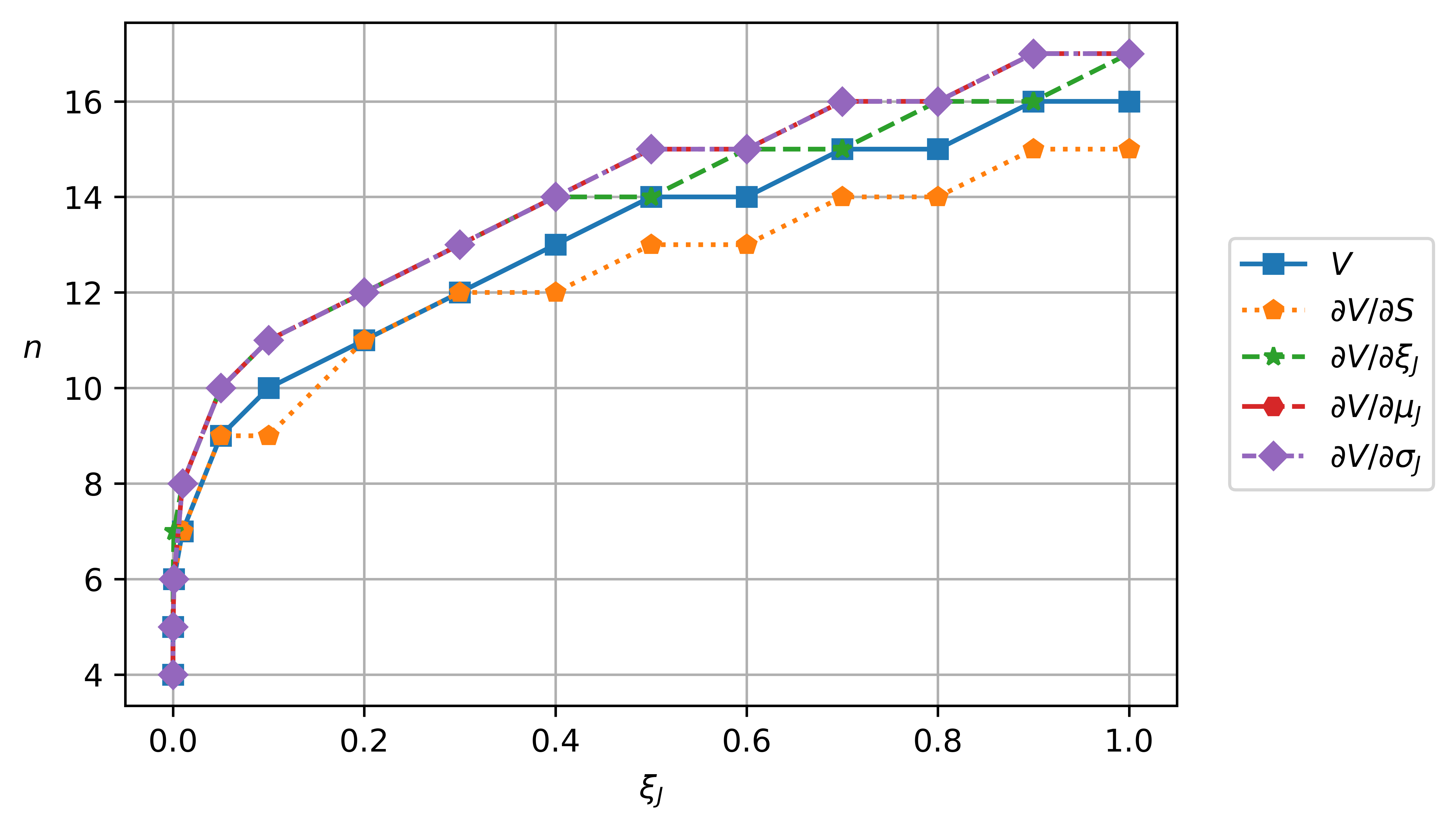}
    \caption{Impact of jump intensity $\jumpIntensity$.}
    \label{fig:mertonImpactXi1e-15}
  \end{subfigure}
  \caption{Impact on number of terms in the Merton option price expansion. Tolerance $10^{-15}$ has been used to determine the cutoff point.
  Parameters: $t_0 = 0$, $T = 1$, $\stock(t_0) = 100$, $\shortRate = 0.1$, $\vol = 0.2$, $\strike=100$, $\jumpMean = -0.125$, $\jumpVol = 0.1$, $\jumpIntensity = 0.1$.}
  \label{fig:mertonImpact1e-15}
\end{figure}

From the experiment in Figure~\ref{fig:mertonImpact1e-4} we can conclude that for a chosen impact ($T$, $\strike$, $\jumpMean$, $\jumpVol$, $\jumpIntensity$) all examined quantities (price, delta, jump parameter sensitivities) exhibit the same pattern regarding the number of terms required to achieve the desired bps accuracy.
Indeed, using a higher tolerance for the cutoff point yields a higher number of terms required in the expansion, see Figure~\ref{fig:mertonImpact1e-15} where we cut off at $10^{-15}$.

These results are in line with those presented by Boen on European rainbow option values under the two-asset Merton jump-diffusion model~\cite{Boen202001}.
Boen derives a semi-closed analytical formula for European rainbow options, under the two-asset Merton jump-diffusion model.
Depending on the size of $\jumpIntensity(T-t)$ the number of terms used in the sum needs to be sufficiently large to guarantee accurate option prices.
Convergence of this formula in the case of a European put-on-the-min option is illustrated.

\section{Variance analysis derivations} \label{sec:hedgeErrorDerivations}
As is stated in Section~\ref{sec:hedingError}, we derive an analytic expression for the mean and variance of $\PnLPortfolio$ from Equation~\eqref{eq:pnlPortfolioFD5}.
But before that, we need the following two results.
First,
\begin{align}
    \stock(t_k)
        &\equalDistr \stock(t_{k-1}) \expPower{X}, \label{eq:stock1}\\
    X
        &= \left(\shortRate - \half \vol^2\right) [t_k - t_{k-1}] + \vol \sqrt{t_k - t_{k-1}} Z \nonumber \\
        &\sim \N\left( \left(\shortRate - \half \vol^2\right) [t_k - t_{k-1}], \vol^2 [t_k - t_{k-1}] \right)
        \rdef \N(\mu_X, \sigma^2_X), \nonumber
\end{align}
where $Z \sim \N(0,1)$.
Second, in a similar fashion as Equation~\eqref{eq:stock1} we write:
\begin{align}
    \stock(t_{k-1})
        &\equalDistr \stock(t_0) \expBrace{ \left(\shortRate - \half \vol^2\right) [t_{k-1} - t_0] + \vol \sqrt{t_{k-1} - t_0} \tilde{Z}}, \label{eq:stock2}
\end{align}
where $\tilde{Z} \sim \N(0,1)$, independent of $Z$.
Using Equation~\eqref{eq:stock2} and $\tau \ldef T - t_{k-1}$ we write $d_2$ as:
\begin{align}
    d_2
        &= \frac{\ln \frac{\stock(t_0)}{K}  + \left(\shortRate - \half \vol^2\right) [T - t_0] }{\vol \sqrt{\tau}} + \sqrt{\frac{t_{k-1} - t_0}{\tau}}  \tilde{Z} \nonumber \\
        &\sim \N\left(\frac{\ln \frac{\stock(t_0)}{K}  + \left(\shortRate - \half \vol^2\right) [T - t_0] }{\vol \sqrt{\tau}}, \frac{t_{k-1} - t_0}{\tau} \right)
        \rdef \N(\mu_{d_2}, \sigma^2_{d_2}). \nonumber
\end{align}
\subsection{Mean} \label{sec:hedgeErrorDerivationsMean}

We are interested in how $\PnLPortfolio$ from Equation~\eqref{eq:pnlPortfolioFD5} is distributed conditional on the information known at today ($t_0$).
First, look at the mean:
\begin{align}
    \E_{t_0}\left[\PnLPortfolio(t_k)\right]
        &\approx \E_{t_0} \left[ \frac{\strike \expPower{-\shortRate\tau} }{2\vol\sqrt{\tau}}  \left[\expPower{X} - 1\right]^2 \normPDF(d_2) \right]
        =  \frac{\strike \expPower{-\shortRate\tau} }{2\vol\sqrt{\tau}} \E_{t_0} \left[ \left[\expPower{X} - 1\right]^2 \right] \E_{t_0} \left[\normPDF(d_2) \right],
        \label{eq:pnlPortfolioFDMean2}
\end{align}
where we use the independency of $\left[\expPower{X} - 1\right]^2$ and $\normPDF(d_2)$ due to the independent increments.
The first expectation from Equation~\eqref{eq:pnlPortfolioFDMean2} can be written as:
\begin{align}
    \E_{t_0} \left[ \left[\expPower{X} - 1\right]^2 \right]
        &= \Var_{t_0} \left( \expPower{X} - 1 \right) + \left( \E_{t_0}  \left[\expPower{X} - 1\right] \right)^2 \nonumber \\
        &= \left( \expPower{\sigma^2_X} - 1 \right) \expPower{2\mu_X + \sigma^2_X} + \left( \expPower{\mu_X + \half \sigma^2_X} - 1 \right)^2, \label{eq:pnlPortfolioFDMeanSub1}
\end{align}
where we use the lognormality of $\expPower{X}$.
The second expectation can be written as:
\begin{align}
    \E_{t_0} \left[ \normPDF(d_2) \right]
        &= \frac{\E_{t_0} \left[ \expPower{-\half d_2^2} \right]}{\sqrt{2\pi}}
        = \frac{ \expPower{-\frac{\mu^2_{d_2}}{2\left(1 + \sigma^2_{d_2} \right)}}}{\sqrt{2\pi}\sqrt{1+\sigma^2_{d_2}}},\label{eq:pnlPortfolioFDMeanSub2}
\end{align}
where we use the fact that $d_2^2$ follows $\sigma^2_{d_2}$ times a non-central chi-squared distribution with one degree of freedom ($k=1$), and with non-centrality parameter $\lambda = \frac{\mu^2_{d_2}}{\sigma^2_{d_2}}$, i.e., $d_2^2 \sim \sigma^2_{d_2}\chi^2\left(1, \frac{\mu^2_{d_2}}{\sigma^2_{d_2}}\right)$.
Using results~\eqref{eq:pnlPortfolioFDMeanSub1} and~\eqref{eq:pnlPortfolioFDMeanSub2} allows us to rewrite Equation~\eqref{eq:pnlPortfolioFDMean2} as:
\begin{align}
    \E_{t_0}\left[\PnLPortfolio(t_k)\right]
        &\approx \frac{\strike \expPower{-\shortRate\tau} }{2\vol\sqrt{\tau}}
        \frac{\expPower{-\frac{\mu^2_{d_2}}{2\left(1+\sigma^2_{d_2} \right)}}}{\sqrt{2\pi}\sqrt{1+\sigma^2_{d_2}}}
        \left( \left( \expPower{\sigma^2_X} - 1 \right) \expPower{2\mu_X + \sigma^2_X} + \left( \expPower{\mu_X + \half \sigma^2_X} - 1 \right)^2 \right).
        \label{eq:pnlPortfolioFDMean3}
\end{align}

\subsection{Variance} \label{sec:hedgeErrorDerivationsVar}
For the variance of $\PnLPortfolio$ from Equation~\eqref{eq:pnlPortfolioFD5} we have:
\begin{align}
    \Var_{t_0}\left(\PnLPortfolio(t_k)\right)
        &= \E_{t_0}\left[\left(\PnLPortfolio(t_k)\right)^2\right] - \left(\E_{t_0}\left[\PnLPortfolio(t_k)\right]\right)^2,
        \label{eq:pnlPortfolioFDVar1}
\end{align}
where the second term can be computed using Equation~\eqref{eq:pnlPortfolioFDMean3}.
By the independent increment argument, the first term in Equation~\eqref{eq:pnlPortfolioFDVar1} can be written as:
\begin{align}
    \E_{t_0}\left[\left(\PnLPortfolio(t_k)\right)^2\right]
        &\approx  \E_{t_0}\left[\left(\frac{\strike \expPower{-\shortRate\tau} }{2\vol\sqrt{\tau}}  \left[\expPower{X} - 1\right]^2 \normPDF(d_2)\right)^2\right] \nonumber \\
        &= \frac{\strike^2 \expPower{-2\shortRate\tau}}{4\vol^2\tau} \E_{t_0}\left[\left[\expPower{X} - 1\right]^4 \right] \E_{t_0}\left[\left(\normPDF(d_2)\right)^2\right]. \label{eq:pnlPortfolioFDVarSub1}
\end{align}
For the first expectation in Equation~\eqref{eq:pnlPortfolioFDVarSub1} we write:
\begin{align}
    \E_{t_0}\left[\left[\expPower{X} - 1\right]^4 \right]
        &= \E_{t_0}\left[ \expPower{4X} - 4 \expPower{3X}  + 6 \expPower{2X} - 4 \expPower{X} + 1 \right] \nonumber \\
        &= \expPower{4 \mu_X + \frac{16}{2} \sigma^2_X} - 4 \expPower{3 \mu_X + \frac{9}{2} \sigma^2_X} + 6 \expPower{2 \mu_X + \frac{4}{2} \sigma^2_X} - 4 \expPower{ \mu_X + \frac{1}{2} \sigma^2_X} + 1, \label{eq:pnlPortfolioFDVarSub2}
\end{align}
using the properties of the moment generating function of $X$.
The second expectation in Equation~\eqref{eq:pnlPortfolioFDVarSub1} can be written as:
\begin{align}
    \E_{t_0}\left[\left(\normPDF(d_2)\right)^2 \right]
        &= \frac{\E_{t_0} \left[ \expPower{- d_2^2} \right]}{2\pi}
        =  \frac{\expPower{-\frac{\mu^2_{d_2}}{1 + 2\sigma^2_{d_2}}}}{2\pi\sqrt{1+2\sigma^2_{d_2}}} ,
        \label{eq:pnlPortfolioFDVarSub3}
\end{align}
where we once more use that $d_2^2$ follows $\sigma^2_{d_2}$ times a non-central chi-squared distribution with parameters as stated before.
Results~\eqref{eq:pnlPortfolioFDVarSub2} and~\eqref{eq:pnlPortfolioFDVarSub3} allow us to rewrite Equation~\eqref{eq:pnlPortfolioFDVarSub1} as:
\begin{align}
    \E_{t_0}\left[\left(\PnLPortfolio(t_k)\right)^2\right]
        &\approx \frac{\strike^2 \expPower{-2\shortRate\tau}}{4\vol^2\tau}  \frac{\expPower{-\frac{\mu^2_{d_2}}{1 + 2\sigma^2_{d_2}}}}{2\pi\sqrt{1+2\sigma^2_{d_2}}}  \nonumber \\
        &\quad  \left(\expPower{4 \mu_X + \frac{16}{2} \sigma^2_X} - 4 \expPower{3 \mu_X + \frac{9}{2} \sigma^2_X} + 6 \expPower{2 \mu_X + \frac{4}{2} \sigma^2_X} - 4 \expPower{ \mu_X + \frac{1}{2} \sigma^2_X} + 1 \right) . \label{eq:pnlPortfolioFDVarSub4}
\end{align}
Using the results from Equations~\eqref{eq:pnlPortfolioFDMean3} and~\eqref{eq:pnlPortfolioFDVarSub4} in Equation~\eqref{eq:pnlPortfolioFDVar1} yields:
\begin{align}
    \Var_{t_0}\left(\PnLPortfolio(t_k)\right)
        &\approx \frac{\strike^2 \expPower{-2\shortRate\tau}}{8\pi\vol^2\tau} & \left[
        f(\mu_X, \sigma_X) \cdot \frac{\expPower{-\frac{\mu^2_{d_2}}{1 + 2\sigma^2_{d_2}}}}{\sqrt{1+2\sigma^2_{d_2}}} -  g(\mu_X, \sigma_X) \cdot \frac{\expPower{-\frac{\mu^2_{d_2}}{1+\sigma^2_{d_2}}}}{1+\sigma^2_{d_2}} \right], \label{eq:pnlPortfolioFDVar2Appendix}
\end{align}
where
\begin{align}
    f(\mu_X, \sigma_X)
        &\ldef \expPower{4 \mu_X + \frac{16}{2} \sigma^2_X} - 4 \expPower{3 \mu_X + \frac{9}{2} \sigma^2_X} + 6 \expPower{2 \mu_X + \frac{4}{2} \sigma^2_X} - 4 \expPower{ \mu_X + \frac{1}{2} \sigma^2_X} + 1, \label{eq:pnlPortfolioFDVar2f} \\
    g(\mu_X, \sigma_X)
        &\ldef \left( \left( \expPower{\sigma^2_X} - 1 \right) \expPower{2\mu_X + \sigma^2_X} + \left( \expPower{\mu_X + \half \sigma^2_X} - 1 \right)^2 \right)^2. \label{eq:pnlPortfolioFDVar2g}
\end{align}

\section{Desk structure and flow of cash} \label{sec:deskStructureCashFlows}
The simulation is illustrated by a series of schematic drawings that indicate the flow of cash and instruments.
First, in~\ref{sec:flowsBankRiskFree}, we consider the case without CCR to get a basic understanding.
In~\ref{sec:flowsBankRisky}, CCR is introduced, and the trading desk and xVA desk are represented as a single entity, referred to as the trading desk.
Finally, in~\ref{sec:flowsBankRiskyTwoDesks}, we remove this assumption by examining the internal exchange of cash-flows and products between the desks.

In all figures to follow, dotted lines indicate flows of cash, hence this relates to wealth accounts $\wealth$, $\wealthTrading$, and $\wealthXVA$.
Furthermore, solid lines indicate the exchange of a product/asset, hence this relates to portfolios $\strategy$, $\strategyTrading$, and $\strategyXVA$.
Values are seen from the perspective of the desks, i.e., an arrow away from (towards) a desk indicates the desk needs to pay (receives).
We assume that the wealth account at time $t_{k-1}$ is positive, in the sense that this can be seen as a deposit with the Treasury department for which interest is received.~\footnote{In the case that $\wealth(t_{k-1})$ is negative we have to pay interest to the Treasury desk on the borrowed amount and the flows between the Treasury department and wealth account would be in the opposite direction.}

\subsection{Case without CCR} \label{sec:flowsBankRiskFree}
At this stage the $\xva$ desk is not involved as we assume no $\xva$s are required.
Here
\begin{align}
  \hedgePos_1(t) &= - \Delta(t) = - \pderiv{\tradeVal(t)}{\stock}. \label{eq:hedgeSingleDeskWithoutCCR}
\end{align}

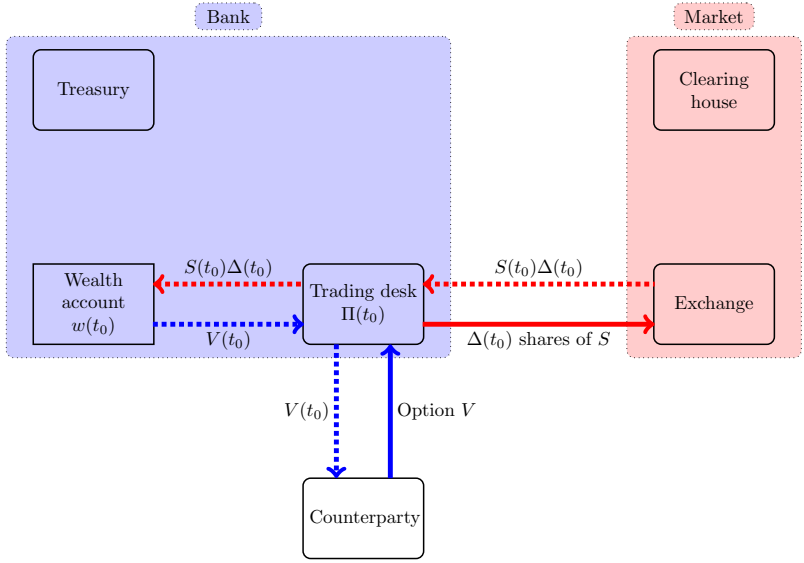
\begin{figure}[!h]
    \centering
    \begin{subfigure}[b]{\cashFlowFigureSize}
    \scalebox{\scaleSubfigure}{
    \begin{tikzpicture}
        \node[rectangle, draw, dotted, text width=8cm, text centered, rounded corners, minimum height=6cm, fill=\bankColor] (Bank) at (1,0) {};
        \node[rectangle, draw, dotted, text width=1cm, text centered, rounded corners, minimum height=0.5cm, fill=\bankColor] (Bank) at (1,3.375) {Bank};
        \node[rectangle, draw, thick, text width=2.0cm, text centered, minimum height=1.5cm] (WealthAccount) at (-1.5,-2) {Wealth account $\wealth(t_0)$};
        \node[rectangle, draw, thick, text width=2.0cm, text centered, rounded corners, minimum height=1.5cm] (Treasury) at (-1.5,2) {Treasury};
        \node[rectangle, draw, thick, text width=2.0cm, text centered, rounded corners, minimum height=1.5cm] (TradingDesk) at (3.5,-2) {Trading desk \\ $\strategy(t_0)$};

        \node[rectangle, draw, dotted, text width=3cm, text centered, rounded corners, minimum height=6cm, fill=\marketColor] (Market) at (10,0) {};
        \node[rectangle, draw, dotted, text width=1.25cm, text centered, rounded corners, minimum height=0.5cm, fill=\marketColor] (Market) at (10,3.375) {Market};
        \node[rectangle, draw, thick, text width=2.0cm, text centered, rounded corners, minimum height=1.5cm] (Exchange) at (10,-2) {Exchange};
        \node[rectangle, draw, thick, text width=2.0cm, text centered, rounded corners, minimum height=1.5cm] (ClearingHouse) at (10,2) {Clearing house};

        \node[rectangle, draw, thick, text width=2.0cm, text centered, rounded corners, minimum height=1.5cm] (Counterparty) at (3.5,-6) {Counterparty};



        \draw[->, line width=2.6pt, dotted, color=\tradeColor] (-0.4,-2.375) -- (2.4,-2.375);
        \node[align=center, below] (WealthToTradingDesk) at (1.0,-2.375) {$\tradeVal(t_0)$};
        \draw[<-, line width=2.6pt, dotted, color=\hedgeColor] (-0.4,-1.625) -- (2.4,-1.625);
        \node[align=center, above] (TradingDeskToWealth) at (1.0,-1.625) {$\stock(t_0) \Delta(t_0)$};


        \draw[<-, line width=2.6pt, dotted, color=\hedgeColor] (4.6,-1.625) --  (8.9,-1.625);
        \node[align=center, above] (ExchangeToTradingDesk) at (6.75,-1.625) {$\stock(t_0) \Delta(t_0)$};
        \draw[->, line width=2.6pt, color=\hedgeColor] (4.6,-2.375) -- (8.9,-2.375);
        \node[align=center, below] (TradingDeskToExchange) at (6.75,-2.375) {$\Delta(t_0)$ shares of $\stock$};

        \draw[<-, line width=2.6pt, color=\tradeColor] (4,-2.75) -- (4,-5.25);
        \node[align=center, right] (CounterpartyToTradingDesk) at (4,-4) {Option $\tradeVal$};
        \draw[->, line width=2.6pt, dotted, color=\tradeColor] (3,-2.75) -- (3,-5.25);
        \node[align=center, left] (TradingDeskToCounterparty) at (3,-4) {$\tradeVal(t_0)$};
    \end{tikzpicture}
    }
    \end{subfigure}
    \caption{Case without CCR, situation at $t_0$. The \tradeColor \ lines correspond to the trade; the \hedgeColor \ lines correspond to the hedge.}
    \label{fig:t0BankRiskFree}
\end{figure}

At $t_0$, see Figure~\ref{fig:t0BankRiskFree}, the option is bought from the counterparty (\tradeColor \ lines) and the Black-Scholes delta hedge~\eqref{eq:hedgeSingleDeskWithoutCCR} is constructed (\hedgeColor \ lines).

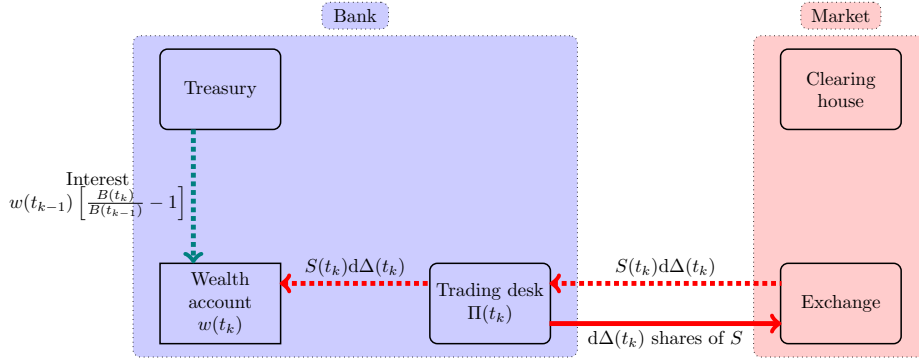
\begin{figure}[!h]
    \centering
    \begin{subfigure}[b]{\cashFlowFigureSize}
    \scalebox{\scaleSubfigure}{
    \begin{tikzpicture}
        \node[rectangle, draw, dotted, text width=8cm, text centered, rounded corners, minimum height=6cm, fill=\bankColor] (Bank) at (1,0) {};
        \node[rectangle, draw, dotted, text width=1cm, text centered, rounded corners, minimum height=0.5cm, fill=\bankColor] (Bank) at (1,3.375) {Bank};
        \node[rectangle, draw, thick, text width=2.0cm, text centered, minimum height=1.5cm] (WealthAccount) at (-1.5,-2) {Wealth account $\wealth(t_k)$};
        \node[rectangle, draw, thick, text width=2.0cm, text centered, rounded corners, minimum height=1.5cm] (Treasury) at (-1.5,2) {Treasury};
        \node[rectangle, draw, thick, text width=2.0cm, text centered, rounded corners, minimum height=1.5cm] (TradingDesk) at (3.5,-2) {Trading desk \\ $\strategy(t_k)$};

        \node[rectangle, draw, dotted, text width=3cm, text centered, rounded corners, minimum height=6cm, fill=\marketColor] (Market) at (10,0) {};
        \node[rectangle, draw, dotted, text width=1.25cm, text centered, rounded corners, minimum height=0.5cm, fill=\marketColor] (Market) at (10,3.375) {Market};
        \node[rectangle, draw, thick, text width=2.0cm, text centered, rounded corners, minimum height=1.5cm] (Exchange) at (10,-2) {Exchange};
        \node[rectangle, draw, thick, text width=2.0cm, text centered, rounded corners, minimum height=1.5cm] (ClearingHouse) at (10,2) {Clearing house};




        \draw[<-, line width=2.6pt, dotted, color=\hedgeColor] (-0.4,-1.625) -- (2.4,-1.625);
        \node[align=center, above] (TradingDeskToWealth) at (1.0,-1.625) {$\stock(t_k) \d\Delta(t_k) $};

        \draw[->, line width=2.6pt, dotted, color=\interestColor] (-2,1.25) -- (-2,-1.25);
        \node[align=center, left] (TreasuryToWealth) at (-2,0) {Interest \\ $\wealth(t_{k-1}) \left[\frac{\bank(t_k)}{\bank(t_{k-1})} - 1\right]$};

        \draw[<-, line width=2.6pt, dotted, color=\hedgeColor] (4.6,-1.625) --  (8.9,-1.625);
        \node[align=center, above] (ExchangeToTradingDesk) at (6.75,-1.625) {$\stock(t_k) \d\Delta(t_k) $};
        \draw[->, line width=2.6pt, color=\hedgeColor] (4.6,-2.375) -- (8.9,-2.375);
        \node[align=center, below] (TradingDeskToExchange) at (6.75,-2.375) {$\d\Delta(t_k)$ shares of $\stock$};

    \end{tikzpicture}
    }
    \end{subfigure}
    \caption{Case without CCR, situation at $t_0 < t_k < t_K$. The \hedgeColor \ lines correspond to the hedge; the \interestColor \ lines correspond to the interest.}
    \label{fig:tkBankRiskFree}
\end{figure}

Then for $t_0 < t_k < t_K$, see Figure~\ref{fig:tkBankRiskFree}, interest on the wealth account is received from the Treasury department (\interestColor \ lines) and the hedge is rebalanced (\hedgeColor \ lines).

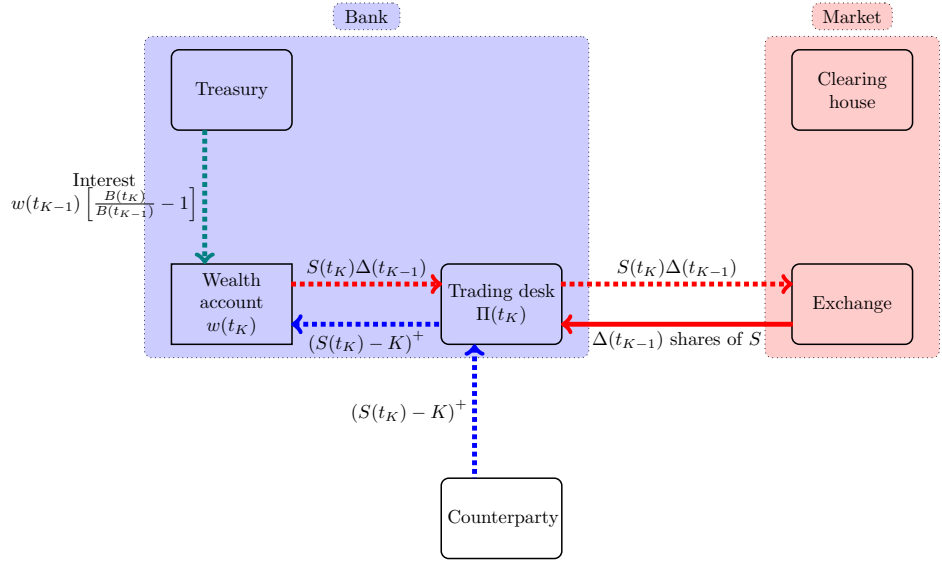
\begin{figure}[!h]
    \centering
    \begin{subfigure}[b]{\cashFlowFigureSize}
    \scalebox{\scaleSubfigure}{
    \begin{tikzpicture}
        \node[rectangle, draw, dotted, text width=8cm, text centered, rounded corners, minimum height=6cm, fill=\bankColor] (Bank) at (1,0) {};
        \node[rectangle, draw, dotted, text width=1cm, text centered, rounded corners, minimum height=0.5cm, fill=\bankColor] (Bank) at (1,3.375) {Bank};
        \node[rectangle, draw, thick, text width=2.0cm, text centered, minimum height=1.5cm] (WealthAccount) at (-1.5,-2) {Wealth account $\wealth(t_K)$};
        \node[rectangle, draw, thick, text width=2.0cm, text centered, rounded corners, minimum height=1.5cm] (Treasury) at (-1.5,2) {Treasury};
        \node[rectangle, draw, thick, text width=2.0cm, text centered, rounded corners, minimum height=1.5cm] (TradingDesk) at (3.5,-2) {Trading desk \\ $\strategy(t_K)$};

        \node[rectangle, draw, dotted, text width=3cm, text centered, rounded corners, minimum height=6cm, fill=\marketColor] (Market) at (10,0) {};
        \node[rectangle, draw, dotted, text width=1.25cm, text centered, rounded corners, minimum height=0.5cm, fill=\marketColor] (Market) at (10,3.375) {Market};
        \node[rectangle, draw, thick, text width=2.0cm, text centered, rounded corners, minimum height=1.5cm] (Exchange) at (10,-2) {Exchange};
        \node[rectangle, draw, thick, text width=2.0cm, text centered, rounded corners, minimum height=1.5cm] (ClearingHouse) at (10,2) {Clearing house};

        \node[rectangle, draw, thick, text width=2.0cm, text centered, rounded corners, minimum height=1.5cm] (Counterparty) at (3.5,-6) {Counterparty};



        \draw[<-, line width=2.6pt, dotted, color=\tradeColor] (-0.4,-2.375) -- (2.4,-2.375);
        \node[align=center, below] (TradingDeskToWealth) at (1.0,-2.375) {$\left(\stock(t_K) - \strike\right)^+$};
        \draw[->, line width=2.6pt, dotted, color=\hedgeColor] (-0.4,-1.625) -- (2.4,-1.625);
        \node[align=center, above] (WealthToTradingDesk) at (1.0,-1.625) {$\stock(t_K) \Delta(t_{K-1})$};

        \draw[->, line width=2.6pt, dotted, color=\interestColor] (-2,1.25) -- (-2,-1.25);
        \node[align=center, left] (TreasuryToWealth) at (-2,0) {Interest \\ $\wealth(t_{K-1}) \left[\frac{\bank(t_K)}{\bank(t_{K-1})} - 1\right]$};

        \draw[->, line width=2.6pt, dotted, color=\hedgeColor] (4.6,-1.625) --  (8.9,-1.625);
        \node[align=center, above] (ExchangeToTradingDesk) at (6.75,-1.625) {$\stock(t_K) \Delta(t_{K-1})$};
        \draw[<-, line width=2.6pt, color=\hedgeColor] (4.6,-2.375) -- (8.9,-2.375);
        \node[align=center, below] (TradingDeskToExchange) at (6.75,-2.375) {$\Delta(t_{K-1})$ shares of $\stock$};

       \draw[<-, line width=2.6pt, dotted, color=\tradeColor] (3,-2.75) -- (3,-5.25);
       \node[align=center, left] (CounterpartyToTradingDesk) at (3,-4) {$\left(\stock(t_K) - \strike\right)^+$};
    \end{tikzpicture}
    }
    \end{subfigure}
    \caption{Case without CCR, situation at $t_K$. The \tradeColor \ lines correspond to the trade; the \hedgeColor \ lines correspond to the hedge; the \interestColor \ lines correspond to the interest.}
    \label{fig:tKBankRiskFree}
\end{figure}

Finally, the situation maturity $t_K$ is displayed in Figure~\ref{fig:tKBankRiskFree}.
We first again receive interest (\interestColor \ lines), then close the hedge (\hedgeColor \ lines) and settle the payoff of the option in cash (\tradeColor \ lines).

\subsection{Case with CCR, single desk within the bank} \label{sec:flowsBankRisky}
Assume that the trading desk and $\xva$ desk are represented as a single entity, still referred to as the trading desk.
Here
\begin{align}
  \hedgePos_1(t) &= -\hat{\Delta}(t) = - \pderiv{\tradeVal(t)}{\stock}\left[ 1 - (1-\recovRate)\PD(t,t_K)\right], \label{eq:hedgeSingleDeskWithCCR}
\end{align}
meaning that also the $\CVA$ market risk is hedged.

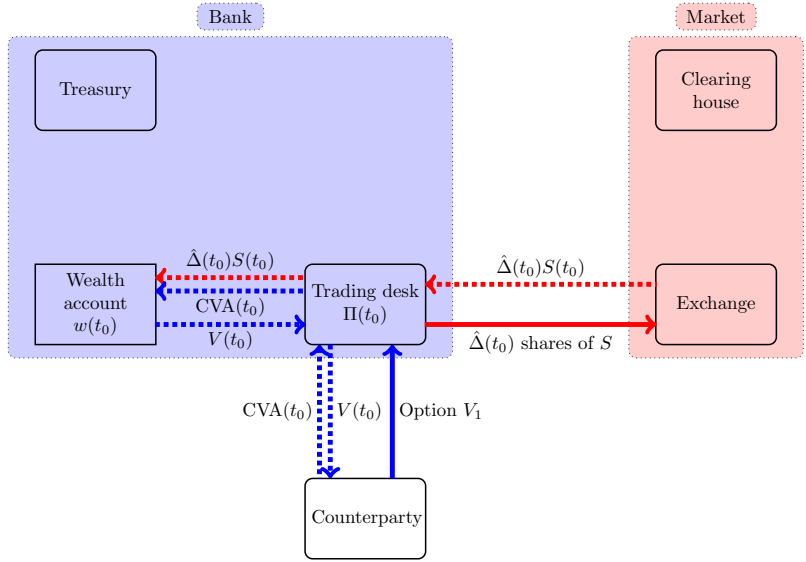
\begin{figure}[!h]
    \centering
    \begin{subfigure}[b]{\cashFlowFigureSize}
    \scalebox{\scaleSubfigure}{
    \begin{tikzpicture}
        \node[rectangle, draw, dotted, text width=8cm, text centered, rounded corners, minimum height=6cm, fill=\bankColor] (Bank) at (1,0) {};
        \node[rectangle, draw, dotted, text width=1cm, text centered, rounded corners, minimum height=0.5cm, fill=\bankColor] (Bank) at (1,3.375) {Bank};
        \node[rectangle, draw, thick, text width=2.0cm, text centered, minimum height=1.5cm] (WealthAccount) at (-1.5,-2) {Wealth account $\wealth(t_0)$};
        \node[rectangle, draw, thick, text width=2.0cm, text centered, rounded corners, minimum height=1.5cm] (Treasury) at (-1.5,2) {Treasury};
        \node[rectangle, draw, thick, text width=2.0cm, text centered, rounded corners, minimum height=1.5cm] (TradingDesk) at (3.5,-2) {Trading desk \\ $\strategy(t_0)$};

        \node[rectangle, draw, dotted, text width=3cm, text centered, rounded corners, minimum height=6cm, fill=\marketColor] (Market) at (10,0) {};
        \node[rectangle, draw, dotted, text width=1.25cm, text centered, rounded corners, minimum height=0.5cm, fill=\marketColor] (Market) at (10,3.375) {Market};
        \node[rectangle, draw, thick, text width=2.0cm, text centered, rounded corners, minimum height=1.5cm] (Exchange) at (10,-2) {Exchange};
        \node[rectangle, draw, thick, text width=2.0cm, text centered, rounded corners, minimum height=1.5cm] (ClearingHouse) at (10,2) {Clearing house};

        \node[rectangle, draw, thick, text width=2.0cm, text centered, rounded corners, minimum height=1.5cm] (Counterparty) at (3.5,-6) {Counterparty};



        \draw[->, line width=2.6pt, dotted, color=\tradeColor] (-0.4,-2.375) -- (2.4,-2.375);
        \node[align=center, below] (WealthToTradingDesk) at (1.0,-2.375) {$\tradeVal(t_0)$};
        \draw[<-, line width=2.6pt, dotted, color=\hedgeColor] (-0.4,-1.5) -- (2.4,-1.5);
        \node[align=center, above] (TradingDeskToWealth) at (1.0,-1.5) {$\hat{\Delta}(t_0)\stock(t_0)$};
        \draw[<-, line width=2.6pt, dotted, color=\tradeColor] (-0.4,-1.75) -- (2.4,-1.75);
        \node[align=center, below] (TradingDeskToWealth) at (1.0,-1.75) {$\CVA(t_0)$};


        \draw[<-, line width=2.6pt, dotted, color=\hedgeColor] (4.6,-1.625) --  (8.9,-1.625);
        \node[align=center, above] (ExchangeToTradingDesk) at (6.75,-1.625) {$\hat{\Delta}(t_0)\stock(t_0)$};
        \draw[->, line width=2.6pt, color=\hedgeColor] (4.6,-2.375) -- (8.9,-2.375);
        \node[align=center, below] (TradingDeskToExchange) at (6.75,-2.375) {$\hat{\Delta}(t_0)$ shares of $\stock$};

        \draw[<-, line width=2.6pt, color=\tradeColor] (4,-2.75) -- (4,-5.25);
        \node[align=center, right] (CounterpartyToTradingDesk) at (4,-4) {Option $\tradeVal_1$};
        \draw[->, line width=2.6pt, dotted, color=\tradeColor] (2.85,-2.75) -- (2.85,-5.25);
        \node[align=center, right] (TradingDeskToCounterparty) at (2.85,-4) {$\tradeVal(t_0)$};
        \draw[<-, line width=2.6pt, dotted, color=\tradeColor] (2.65,-2.75) -- (2.65,-5.25);
        \node[align=center, left] (CounterpartyToTradingDeskCash) at (2.65,-4) {$\CVA(t_0)$};
    \end{tikzpicture}
    }
    \end{subfigure}
    \caption{Case with CCR, single desk within the bank, situation at $t_0$.
    The \tradeColor \ lines correspond to the trade; the \hedgeColor \ lines correspond to the market risk hedge.}
    \label{fig:t0BankRisky}
\end{figure}

Initially at $t_0$, see Figure~\ref{fig:t0BankRisky}, the risky option $\tradeVal_1$ is bought from the counterparty (\tradeColor \ lines).
As in the previous case, we construct a market risk hedge (\hedgeColor \ lines), but now with a position that takes into account the hedging of $\CVA$ market risk, see Equation~\eqref{eq:hedgeSingleDeskWithCCR}.
In this example the desk decides to only actively manage the $\CVA$ market risk and warehouse the residual credit risk that can manifest itself upon the jump to default of the counterparty.

\begin{figure}[!h]
    \centering
    \begin{subfigure}[b]{\cashFlowFigureSize}
    \scalebox{\scaleSubfigure}{
    \begin{tikzpicture}
        \node[rectangle, draw, dotted, text width=8cm, text centered, rounded corners, minimum height=6cm, fill=\bankColor] (Bank) at (1,0) {};
        \node[rectangle, draw, dotted, text width=1cm, text centered, rounded corners, minimum height=0.5cm, fill=\bankColor] (Bank) at (1,3.375) {Bank};
        \node[rectangle, draw, thick, text width=2.0cm, text centered, minimum height=1.5cm] (WealthAccount) at (-1.5,-2) {Wealth account $\wealth(t_k)$};
        \node[rectangle, draw, thick, text width=2.0cm, text centered, rounded corners, minimum height=1.5cm] (Treasury) at (-1.5,2) {Treasury};
        \node[rectangle, draw, thick, text width=2.0cm, text centered, rounded corners, minimum height=1.5cm] (TradingDesk) at (3.5,-2) {Trading desk \\ $\strategy(t_k)$};

        \node[rectangle, draw, dotted, text width=3cm, text centered, rounded corners, minimum height=6cm, fill=\marketColor] (Market) at (10,0) {};
        \node[rectangle, draw, dotted, text width=1.25cm, text centered, rounded corners, minimum height=0.5cm, fill=\marketColor] (Market) at (10,3.375) {Market};
        \node[rectangle, draw, thick, text width=2.0cm, text centered, rounded corners, minimum height=1.5cm] (Exchange) at (10,-2) {Exchange};
        \node[rectangle, draw, thick, text width=2.0cm, text centered, rounded corners, minimum height=1.5cm] (ClearingHouse) at (10,2) {Clearing house};

        \node[rectangle, draw, thick, text width=2.0cm, text centered, rounded corners, minimum height=1.5cm] (Counterparty) at (3.5,-6) {Counterparty};



         \draw[->, line width=2.6pt, dotted, color=\newTradeColor] (-0.4,-2.375) -- (2.4,-2.375);
         \node[align=center, below] (WealthToTradingDesk) at (1.0,-2.375) {$\tradeVal(t_d)$};
        \draw[<-, line width=2.6pt, dotted, color=\hedgeColor] (-0.4,-1.5) -- (2.4,-1.5);
        \node[align=center, above] (TradingDeskToWealth) at (1.0,-1.5) {$\Delta(t_d) \stock(t_d)$\\$- \hat{\Delta}(t_{d-1})\stock(t_d)$};
        \draw[<-, line width=2.6pt, dotted, color=\defaultColor] (-0.4,-1.75) -- (2.4,-1.75);
        \node[align=center, below] (TradingDeskToWealth) at (1.0,-1.75) {$\recovRate \cdot \tradeVal(t_d)$};

        \draw[->, line width=2.6pt, dotted, color=\interestColor] (-2,1.25) -- (-2,-1.25);
        \node[align=center, left] (TreasuryToWealth) at (-2,0) {Interest \\ $\wealth(t_{d-1}) \left[\frac{\bank(t_d)}{\bank(t_{d-1})} - 1\right]$};

        \draw[<-, line width=2.6pt, dotted, color=\hedgeColor] (4.6,-1.625) --  (8.9,-1.625);
        \node[align=center, above] (ExchangeToTradingDesk) at (6.75,-1.625) {$\Delta(t_d) \stock(t_d)$\\$- \hat{\Delta}(t_{d-1})\stock(t_d)$};
        \draw[->, line width=2.6pt, color=\hedgeColor] (4.6,-2.375) -- (8.9,-2.375);
        \node[align=center, below] (TradingDeskToExchange) at (6.75,-2.375) {$\Delta(t_d) - \hat{\Delta}(t_{d-1})$\\ shares of $\stock$};

        \draw[->, line width=2.6pt, color=\defaultColor] (4,-2.75) -- (4,-5.25);
        \node[align=center, right] (TradingDeskToCounterparty) at (4,-4) {Option $\tradeVal_1$};
        \draw[<-, line width=2.6pt, dotted, color=\defaultColor] (3,-2.75) -- (3,-5.25);
        \node[align=center, left] (CounterpartyToTradingDesk) at (3,-4) {$\recovRate \cdot \tradeVal(t_d)$};

        \draw[<-, line width=2.6pt, color=\newTradeColor] (4,-1.25) -- (4,1.625) -- (8.9, 1.625);
        \node[align=center, below] (ClearingHouseToTradingDesk) at (6.75, 1.625) {Option $\tradeVal$};
        \draw[->, line width=2.6pt, dotted, color=\newTradeColor] (3,-1.25) -- (3,2.375) -- (8.9, 2.375);
        \node[align=center, above] (TradingDeskToClearingHouse) at (6.75, 2.375) {$\tradeVal(t_d)$};
    \end{tikzpicture}
    }
    \end{subfigure}
    \caption{Case with CCR, single desk within the bank, situation at $t_0 < \default = t_d < t_K = T$.
    The \hedgeColor \ lines correspond to the market risk hedge; the \interestColor \ lines correspond to the interest; the \defaultColor \ lines correspond to the default; the \newTradeColor \ lines correspond to the new trade entered upon default.}
    \label{fig:tkBankRiskyDefault}
\end{figure}
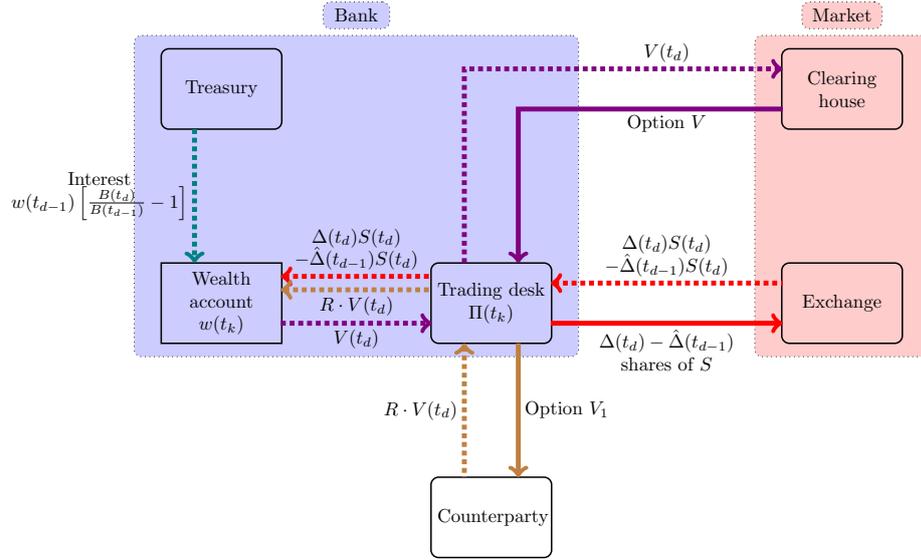

For $t_0 < t_k < t_K$ if there is no default, we rebalance our hedge as before and receive interest on the wealth account.
If a default does occur, see Figure~\ref{fig:tkBankRiskyDefault}, the risk-free closeout takes place (\defaultColor \ lines) and the same risk-neutral option is bought from the clearing house (\newTradeColor \ lines).~\footnote{Here we assume the default takes place before maturity. The case of default at maturity is not discussed for sake of brevity. The remaining case naturally extends from the demonstrated material.}
As always, we receive interest (\interestColor \ lines).
We close the $\CVA$ market risk hedge (\hedgeColor \ lines), see Equation~\eqref{eq:hedgeSingleDeskWithCCR}, and construct a new hedge for the option bought from the clearing house.
This new hedge is equivalent to the Black-Scholes hedge for the risk-free case in Equation~\eqref{eq:hedgeSingleDeskWithoutCCR}.

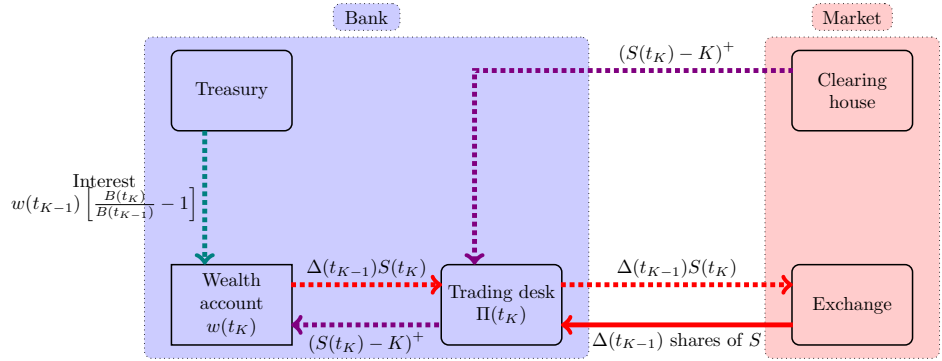
\begin{figure}[!h]
    \centering
    \begin{subfigure}[b]{\cashFlowFigureSize}
    \scalebox{\scaleSubfigure}{
    \begin{tikzpicture}
        \node[rectangle, draw, dotted, text width=8cm, text centered, rounded corners, minimum height=6cm, fill=\bankColor] (Bank) at (1,0) {};
        \node[rectangle, draw, dotted, text width=1cm, text centered, rounded corners, minimum height=0.5cm, fill=\bankColor] (Bank) at (1,3.375) {Bank};
        \node[rectangle, draw, thick, text width=2.0cm, text centered, minimum height=1.5cm] (WealthAccount) at (-1.5,-2) {Wealth account $\wealth(t_K)$};
        \node[rectangle, draw, thick, text width=2.0cm, text centered, rounded corners, minimum height=1.5cm] (Treasury) at (-1.5,2) {Treasury};
        \node[rectangle, draw, thick, text width=2.0cm, text centered, rounded corners, minimum height=1.5cm] (TradingDesk) at (3.5,-2) {Trading desk \\ $\strategy(t_K)$};

        \node[rectangle, draw, dotted, text width=3cm, text centered, rounded corners, minimum height=6cm, fill=\marketColor] (Market) at (10,0) {};
        \node[rectangle, draw, dotted, text width=1.25cm, text centered, rounded corners, minimum height=0.5cm, fill=\marketColor] (Market) at (10,3.375) {Market};
        \node[rectangle, draw, thick, text width=2.0cm, text centered, rounded corners, minimum height=1.5cm] (Exchange) at (10,-2) {Exchange};
        \node[rectangle, draw, thick, text width=2.0cm, text centered, rounded corners, minimum height=1.5cm] (ClearingHouse) at (10,2) {Clearing house};




        \draw[<-, line width=2.6pt, dotted, color=\newTradeColor] (-0.4,-2.375) -- (2.4,-2.375);
        \node[align=center, below] (TradingDeskToWealth) at (1.0,-2.375) {$\left(\stock(t_K) - \strike\right)^+$};
        \draw[->, line width=2.6pt, dotted, color=\hedgeColor] (-0.4,-1.625) -- (2.4,-1.625);
        \node[align=center, above] (WealthToTradingDesk) at (1.0,-1.625) {$\Delta(t_{K-1})\stock(t_K)$};

        \draw[->, line width=2.6pt, dotted, color=\interestColor] (-2,1.25) -- (-2,-1.25);
        \node[align=center, left] (TreasuryToWealth) at (-2,0) {Interest \\ $\wealth(t_{K-1}) \left[\frac{\bank(t_K)}{\bank(t_{K-1})} - 1\right]$};

        \draw[->, line width=2.6pt, dotted, color=\hedgeColor] (4.6,-1.625) --  (8.9,-1.625);
        \node[align=center, above] (ExchangeToTradingDesk) at (6.75,-1.625) {$\Delta(t_{K-1})\stock(t_K)$};
        \draw[<-, line width=2.6pt, color=\hedgeColor] (4.6,-2.375) -- (8.9,-2.375);
        \node[align=center, below] (TradingDeskToExchange) at (6.75,-2.375) {$\Delta(t_{K-1})$ shares of $\stock$};


        \draw[<-, line width=2.6pt, dotted, color=\newTradeColor] (3,-1.25) -- (3,2.375) -- (8.9, 2.375);
        \node[align=center, above] (ClearingHouseToTradingDesk) at (6.75, 2.375) {$\left(\stock(t_K) - \strike \right)^+$};

    \end{tikzpicture}
    }
    \end{subfigure}
    \caption{Case with CCR, single desk within the bank, situation at $t_K$.
    The \hedgeColor \ lines correspond to the market risk hedge; the \interestColor \ lines correspond to the interest; the \newTradeColor \ lines correspond to the new trade entered upon default.}
    \label{fig:tKBankRiskyDefault}
\end{figure}

The situation at maturity, see Figure~\ref{fig:tKBankRiskyDefault}, is then the same as for the case without CCR, apart from the payoff of the option now being settled with the clearing house rather than the counterparty.

\subsection{Case with CCR, separate trading desk and xVA desk within the bank} \label{sec:flowsBankRiskyTwoDesks}
We now remove the assumption of a single desk, and examine the internal exchange of cash-flows and products between the trading desk and $\xva$ desk.
Here
\begin{align*}
  \hedgePos_1(t) &= \underbrace{-\pderiv{\tradeVal(t)}{\stock}}_{-\Delta(t)} + \underbrace{\pderiv{\tradeVal(t)}{\stock}(1-\recovRate)\PD(t,t_K)}_{-\overline{\Delta}(t)},
\end{align*}
meaning that also the $\CVA$ market risk is hedged.

\begin{figure}[!h]
    \centering
    \begin{subfigure}[b]{\cashFlowFigureSize}
    \scalebox{\scaleSubfigure}{
    \begin{tikzpicture}
        \node[rectangle, draw, dotted, text width=13cm, text centered, rounded corners, minimum height=7cm, fill=\bankColor] (Bank) at (-1.5,0) {};
        \node[rectangle, draw, dotted, text width=1cm, text centered, rounded corners, minimum height=0.5cm, fill=\bankColor] (Bank) at (-1.5,3.875) {Bank};
        \node[rectangle, draw, thick, text width=2.0cm, text centered, minimum height=1.5cm] (WealthAccountTrading) at (-1.5,-2) {Wealth account $\wealthTrading(t_0)$};
        \node[rectangle, draw, thick, text width=2.0cm, text centered, minimum height=1.5cm] (WealthAccountXva) at (-1.5,2) {Wealth account $\wealthXVA(t_0)$};
        \node[rectangle, draw, thick, text width=2.0cm, text centered, rounded corners, minimum height=1.5cm] (Treasury) at (-6.5,0) {Treasury};
        \node[rectangle, draw, thick, text width=2.0cm, text centered, rounded corners, minimum height=1.5cm] (TradingDesk) at (3.5,-2) {Trading desk \\ $\strategyTrading(t_0)$};
        \node[rectangle, draw, thick, text width=2.0cm, text centered, rounded corners, minimum height=1.5cm] (XvaDesk) at (3.5,2) {$\xva$ desk\\ $\strategyXVA(t_0)$};

        \node[rectangle, draw, dotted, text width=3cm, text centered, rounded corners, minimum height=6.5cm, fill=\marketColor] (Market) at (10,2.25) {};
        \node[rectangle, draw, dotted, text width=1.25cm, text centered, rounded corners, minimum height=0.5cm, fill=\marketColor] (Market) at (10,5.875) {Market};
        \node[rectangle, draw, thick, text width=2.0cm, text centered, rounded corners, minimum height=1.5cm] (Exchange) at (10,0) {Exchange};
        \node[rectangle, draw, thick, text width=2.0cm, text centered, rounded corners, minimum height=1.5cm] (ClearingHouse) at (10,4.5) {Clearing house};

        \node[rectangle, draw, thick, text width=2.0cm, text centered, rounded corners, minimum height=1.5cm] (Counterparty) at (3.5,-6) {Counterparty};


       \draw[<-, line width=2.6pt, color=\tradeColor] (4,-1.25) -- (4,1.25);
       \node[align=center, right] (XvaDeskToTradingDesk) at (4,0) {$\CVA$};
       \draw[->, line width=2.6pt, dotted, color=\tradeColor] (3,-1.25) -- (3,1.25);
       \node[align=center, left] (TradingDeskToXvaDesk) at (3,0) {$\CVA(t_0)$};

        \draw[->, line width=2.6pt, dotted, color=\tradeColor] (-0.4,-2.375) -- (2.4,-2.375);
        \node[align=center, below] (WealthTradingToTradingDesk) at (1.0,-2.375) {$\tradeVal(t_0)$};
        \draw[<-, line width=2.6pt, dotted, color=\hedgeColor] (-0.4,-1.625) -- (2.4,-1.625);
        \node[align=center, above] (TradingDeskToWealthTrading) at (1.0,-1.625) {$\Delta(t_0)\stock(t_0)$};


        \draw[->, line width=2.6pt, dotted, color=\hedgeCvaColor] (-0.4,1.625) -- (2.4,1.625);
        \node[align=center, below] (WealthXVAToXVADesk) at (1.0,1.625) {$\overline{\Delta}(t_0)\stock(t_0)$};
        \draw[<-, line width=2.6pt, dotted, color=\tradeColor] (-0.4,2.375) -- (2.4,2.375);
        \node[align=center, above] (XVADeskToWealthXVA) at (1.0,2.375) {$\CVA(t_0)$};


        \draw[<-, line width=2.6pt, dotted, color=\hedgeColor] (4.6,-1.625) --  (9.5,-1.625) -- (9.5,-0.75);
        \node[align=center, above] (ExchangeToTradingDesk) at (6.75,-1.625) {$\Delta(t_0)\stock(t_0)$};
        \draw[->, line width=2.6pt, color=\hedgeColor] (4.6,-2.375) -- (10.5,-2.375) -- (10.5,-0.75);
        \node[align=center, below] (TradingDeskToExchange) at (6.75,-2.375) {$\Delta(t_0)$ shares of $\stock$};

        \draw[<-, line width=2.6pt, color=\hedgeCvaColor] (4.6,2.375) --  (10.5,2.375) -- (10.5,0.75);
        \node[align=center, above] (ExchangeToXVADesk) at (6.75,2.375) {$\overline{\Delta}(t_0)$ shares of $\stock$};
        \draw[->, line width=2.6pt, dotted, color=\hedgeCvaColor] (4.6,1.625) -- (9.5,1.625) -- (9.5,0.75);
        \node[align=center, below] (XVADeskToExchange) at (6.75,1.625) {$\overline{\Delta}(t_0)\stock(t_0)$};


        \draw[<-, line width=2.6pt, color=\tradeColor] (4,-2.75) -- (4,-5.25);
        \node[align=center, right] (CounterpartyToTradingDesk) at (4,-4) {Option $\tradeVal_1$};
        \draw[->, line width=2.6pt, dotted, color=\tradeColor] (2.85,-2.75) -- (2.85,-5.25);
        \node[align=center, right] (TradingDeskToCounterparty) at (2.85,-4) {$\tradeVal(t_0)$};
        \draw[<-, line width=2.6pt, dotted, color=\tradeColor] (2.65,-2.75) -- (2.65,-5.25);
        \node[align=center, left] (CounterpartyToTradingDeskCash) at (2.65,-4) {$\CVA(t_0)$};
    \end{tikzpicture}
    }
    \end{subfigure}
    \caption{Case with CCR, separate trading desk and $\xva$ desk within the bank, situation at $t_0$.
    The \tradeColor \ lines correspond to the trade; the \hedgeColor \ lines correspond to the market risk hedge; the \hedgeCvaColor \ lines correspond to the $\CVA$ market risk hedge.}
    \label{fig:t0BankRiskyXVADesk}
\end{figure}
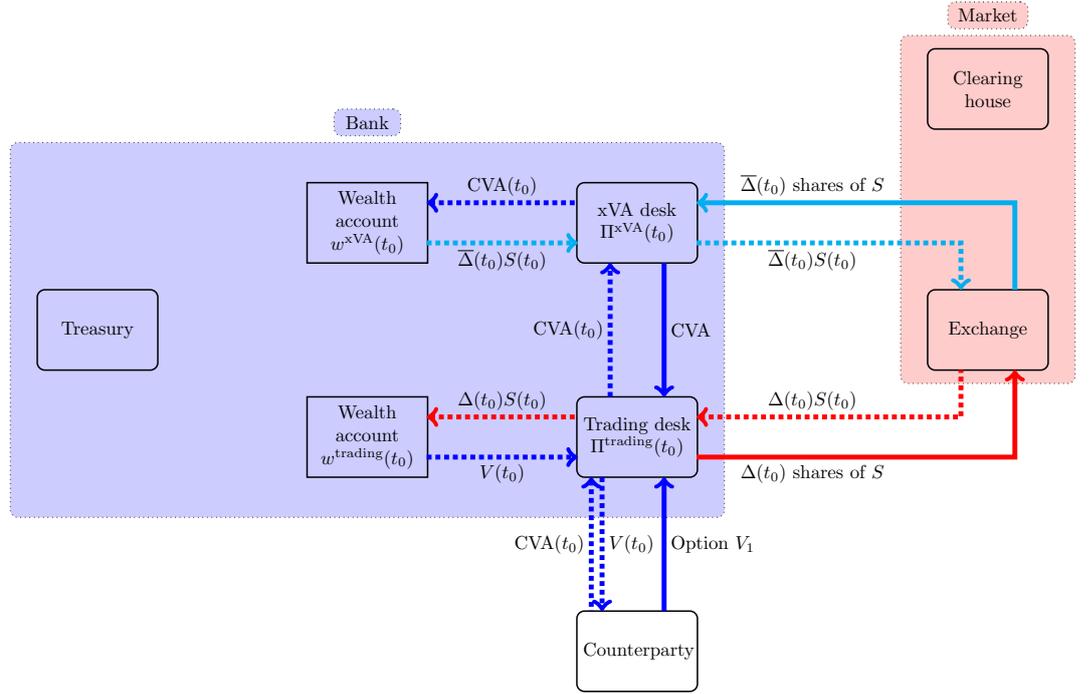

Initially, the situation at $t_0$ is displayed in Figure~\ref{fig:t0BankRiskyXVADesk}, where the positions in trading and hedging instruments are assumed.
Note the transfer of the $\CVA$ between the trading and $\xva$ desk, indicating that the trading desk manages the risk-free component of the transaction, whereas the $\xva$ desk takes responsibility for the $\CVA$.
In this example the $\xva$ desk decides to only actively manage the $\CVA$ market risk and warehouse the residual credit risk that can manifest itself upon the jump to default of the counterparty.
When a default occurs this will result in a loss for the $\xva$ desk, leaving the trading desk unaffected.

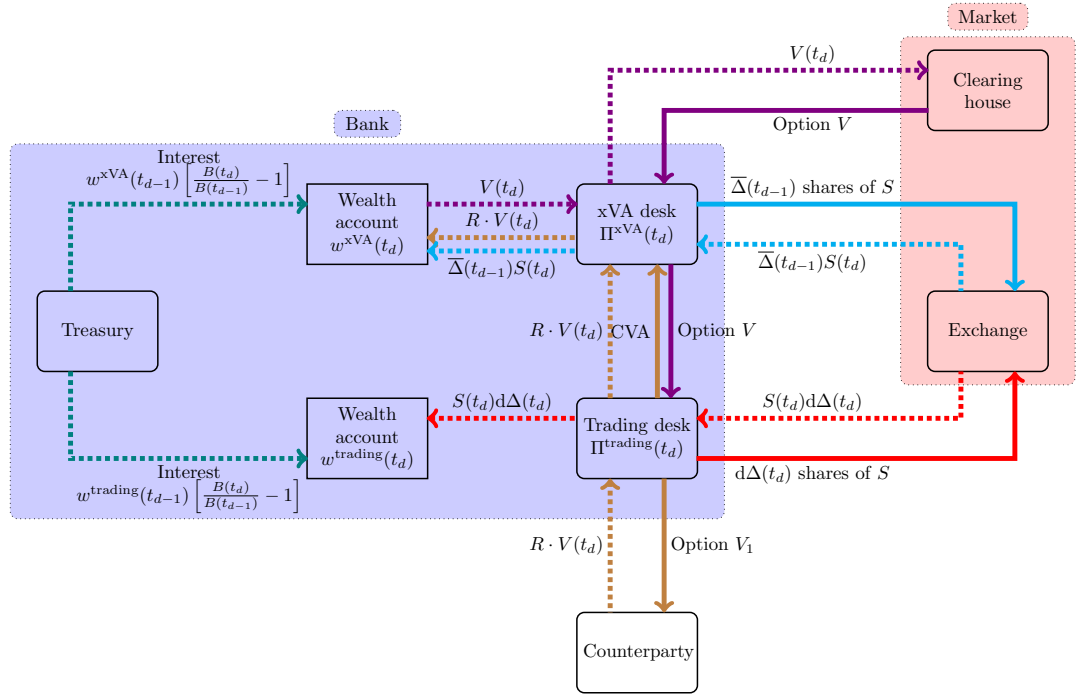
\begin{figure}[!h]
    \centering
    \begin{subfigure}[b]{\cashFlowFigureSize}
    \scalebox{\scaleSubfigure}{
    \begin{tikzpicture}
        \node[rectangle, draw, dotted, text width=13cm, text centered, rounded corners, minimum height=7cm, fill=\bankColor] (Bank) at (-1.5,0) {};
        \node[rectangle, draw, dotted, text width=1cm, text centered, rounded corners, minimum height=0.5cm, fill=\bankColor] (Bank) at (-1.5,3.875) {Bank};
        \node[rectangle, draw, thick, text width=2.0cm, text centered, minimum height=1.5cm] (WealthAccountTrading) at (-1.5,-2) {Wealth account $\wealthTrading(t_d)$};
        \node[rectangle, draw, thick, text width=2.0cm, text centered, minimum height=1.5cm] (WealthAccountXva) at (-1.5,2) {Wealth account $\wealthXVA(t_d)$};
        \node[rectangle, draw, thick, text width=2.0cm, text centered, rounded corners, minimum height=1.5cm] (Treasury) at (-6.5,0) {Treasury};
        \node[rectangle, draw, thick, text width=2.0cm, text centered, rounded corners, minimum height=1.5cm] (TradingDesk) at (3.5,-2) {Trading desk \\ $\strategyTrading(t_d)$};
        \node[rectangle, draw, thick, text width=2.0cm, text centered, rounded corners, minimum height=1.5cm] (XvaDesk) at (3.5,2) {$\xva$ desk\\ $\strategyXVA(t_d)$};

        \node[rectangle, draw, dotted, text width=3cm, text centered, rounded corners, minimum height=6.5cm, fill=\marketColor] (Market) at (10,2.25) {};
        \node[rectangle, draw, dotted, text width=1.25cm, text centered, rounded corners, minimum height=0.5cm, fill=\marketColor] (Market) at (10,5.875) {Market};
        \node[rectangle, draw, thick, text width=2.0cm, text centered, rounded corners, minimum height=1.5cm] (Exchange) at (10,0) {Exchange};
        \node[rectangle, draw, thick, text width=2.0cm, text centered, rounded corners, minimum height=1.5cm] (ClearingHouse) at (10,4.5) {Clearing house};

        \node[rectangle, draw, thick, text width=2.0cm, text centered, rounded corners, minimum height=1.5cm] (Counterparty) at (3.5,-6) {Counterparty};


       \draw[<-, line width=2.6pt, color=\newTradeColor] (4.125,-1.25) -- (4.125,1.25);
       \node[align=center, right] (XvaDeskToTradingDesk) at (4.125,0) {Option $\tradeVal$};
       \draw[->, line width=2.6pt, color=\defaultColor] (3.875,-1.25) -- (3.875,1.25);
       \node[align=center, left] (XvaDeskToTradingDesk) at (3.875,0) {$\CVA$};
       \draw[->, line width=2.6pt, dotted, color=\defaultColor] (3,-1.25) -- (3,1.25);
       \node[align=center, left] (TradingDeskToXvaDesk) at (3,0) {$\recovRate \cdot \tradeVal(t_d)$};

        \draw[<-, line width=2.6pt, dotted, color=\hedgeColor] (-0.4,-1.625) -- (2.4,-1.625);
        \node[align=center, above] (TradingDeskToWealthTrading) at (1.0,-1.625) {$\stock(t_d)\d\Delta(t_d) $};

        \draw[->, line width=2.6pt, dotted, color=\interestColor] (-7,-0.75) -- (-7,-2.375) -- (-2.6,-2.375);
        \node[align=center, below] (TreasuryToWealthTrading) at (-4.8,-2.375) {Interest \\ $\wealthTrading(t_{d-1}) \left[\frac{\bank(t_d)}{\bank(t_{d-1})} - 1\right]$};

        \draw[<-, line width=2.6pt, dotted, color=\defaultColor] (-0.4,1.75) -- (2.4,1.75);
        \node[align=center, above] (XVADeskToWealthXVA) at (1.0,1.75) {$\recovRate \cdot \tradeVal(t_d)$};
        \draw[<-, line width=2.6pt, dotted, color=\hedgeCvaColor] (-0.4,1.5) -- (2.4,1.5);
        \node[align=center, below] (XVADeskToWealthXVA) at (1.0,1.5) {$\overline{\Delta}(t_{d-1})\stock(t_d)$};
        \draw[->, line width=2.6pt, dotted, color=\newTradeColor] (-0.4,2.375) -- (2.4,2.375);
        \node[align=center, above] (WealthXVAToXVADesk) at (1.0,2.375) {$\tradeVal(t_d)$};

        \draw[->, line width=2.6pt, dotted, color=\interestColor] (-7,0.75) -- (-7,2.375) -- (-2.6,2.375);
        \node[align=center, above] (TreasuryToWealthXVA) at (-4.8,2.375) {Interest \\ $\wealthXVA(t_{d-1}) \left[\frac{\bank(t_d)}{\bank(t_{d-1})} - 1\right]$};

        \draw[<-, line width=2.6pt, dotted, color=\hedgeColor] (4.6,-1.625) --  (9.5,-1.625) -- (9.5,-0.75);
        \node[align=center, above] (ExchangeToTradingDesk) at (6.75,-1.625) {$\stock(t_d)\d\Delta(t_d)$};
        \draw[->, line width=2.6pt, color=\hedgeColor] (4.6,-2.375) -- (10.5,-2.375) -- (10.5,-0.75);
        \node[align=center, below] (TradingDeskToExchange) at (6.75,-2.375) {$\d\Delta(t_d)$\ shares of $\stock$};

        \draw[->, line width=2.6pt, color=\hedgeCvaColor] (4.6,2.375) --  (10.5,2.375) -- (10.5,0.75);
        \node[align=center, above] (XVADeskToExchange) at (6.75,2.375) {$\overline{\Delta}(t_{d-1})$ shares of $\stock$};
        \draw[<-, line width=2.6pt, dotted, color=\hedgeCvaColor] (4.6,1.625) -- (9.5,1.625) -- (9.5,0.75);
        \node[align=center, below] (ExchangeToXVADesk) at (6.75,1.625) {$\overline{\Delta}(t_{d-1})\stock(t_d)$};

        \draw[->, line width=2.6pt, dotted, color=\newTradeColor] (3,2.75) -- (3,4.875) -- (8.9,4.875);
        \node[align=center, above] (XVADeskToClearingHouse) at (6.75,4.875) {$\tradeVal(t_d)$};
        \draw[<-, line width=2.6pt, color=\newTradeColor] (4,2.75) -- (4,4.125) -- (8.9,4.125);
        \node[align=center, below] (ClearingHouseToXVADesk) at (6.75,4.125) {Option $\tradeVal$};

        \draw[->, line width=2.6pt, color=\defaultColor] (4,-2.75) -- (4,-5.25);
        \node[align=center, right] (TradingDeskToCounterparty) at (4,-4) {Option $\tradeVal_1$};
        \draw[<-, line width=2.6pt, dotted, color=\defaultColor] (3,-2.75) -- (3,-5.25);
        \node[align=center, left] (CounterpartyToTradingDesk) at (3,-4) {$\recovRate \cdot \tradeVal(t_d)$};
    \end{tikzpicture}
    }
    \end{subfigure}
    \caption{Case with CCR, separate trading desk and $\xva$ desk within the bank, situation at $t_0 < \default = t_d < t_K$.
    The \hedgeColor \ lines correspond to the market risk hedge; the \hedgeCvaColor \ lines correspond to the $\CVA$ market risk hedge; the \interestColor \ lines correspond to the interest; the \defaultColor \ lines correspond to the default; the \newTradeColor \ lines correspond to the new trade entered upon default.}
    \label{fig:tkBankRiskyXVADeskDefault}
\end{figure}

At $t_0 < t_k < t_K$, if there is no default, both desks rebalance their hedging positions and receive interest on their wealth account.
In case of a default at $t_0 < \default = t_d < t_K$, see Figure~\ref{fig:tkBankRiskyXVADeskDefault}, a risk-free closeout takes place between the bank and the counterparty.
Furthermore, the $\xva$ desk enters the same risk-free contract $\tradeVal$ with a clearing house and closes its hedging position on the $\CVA$.
The $\xva$ desk then gives the new contract $\tradeVal$ to the trading desk, so that they are immune to the default.
The `damage' of the default is thus visible at the $\xva$ desk level, which is precisely the place handling this risk.

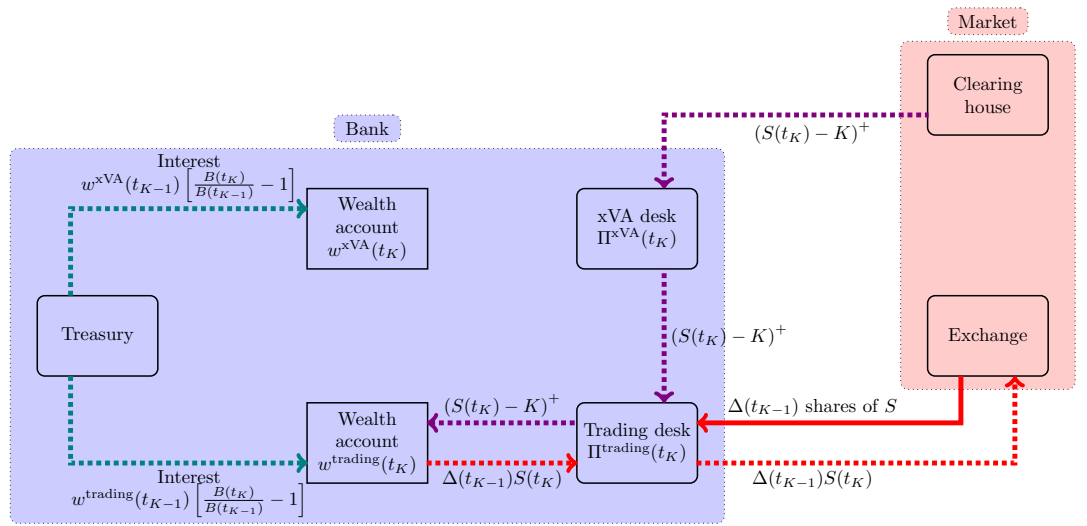
\begin{figure}[!h]
    \centering
    \begin{subfigure}[b]{\cashFlowFigureSize}
    \scalebox{\scaleSubfigure}{
    \begin{tikzpicture}

        \node[rectangle, draw, dotted, text width=13cm, text centered, rounded corners, minimum height=7cm, fill=\bankColor] (Bank) at (-1.5,0) {};
        \node[rectangle, draw, dotted, text width=1cm, text centered, rounded corners, minimum height=0.5cm, fill=\bankColor] (Bank) at (-1.5,3.875) {Bank};
        \node[rectangle, draw, thick, text width=2.0cm, text centered, minimum height=1.5cm] (WealthAccountTrading) at (-1.5,-2) {Wealth account $\wealthTrading(t_K)$};
        \node[rectangle, draw, thick, text width=2.0cm, text centered, minimum height=1.5cm] (WealthAccountXva) at (-1.5,2) {Wealth account $\wealthXVA(t_K)$};
        \node[rectangle, draw, thick, text width=2.0cm, text centered, rounded corners, minimum height=1.5cm] (Treasury) at (-6.5,0) {Treasury};
        \node[rectangle, draw, thick, text width=2.0cm, text centered, rounded corners, minimum height=1.5cm] (TradingDesk) at (3.5,-2) {Trading desk \\ $\strategyTrading(t_K)$};
        \node[rectangle, draw, thick, text width=2.0cm, text centered, rounded corners, minimum height=1.5cm] (XvaDesk) at (3.5,2) {$\xva$ desk\\ $\strategyXVA(t_K)$};

        \node[rectangle, draw, dotted, text width=3cm, text centered, rounded corners, minimum height=6.5cm, fill=\marketColor] (Market) at (10,2.25) {};
        \node[rectangle, draw, dotted, text width=1.25cm, text centered, rounded corners, minimum height=0.5cm, fill=\marketColor] (Market) at (10,5.875) {Market};
        \node[rectangle, draw, thick, text width=2.0cm, text centered, rounded corners, minimum height=1.5cm] (Exchange) at (10,0) {Exchange};
        \node[rectangle, draw, thick, text width=2.0cm, text centered, rounded corners, minimum height=1.5cm] (ClearingHouse) at (10,4.5) {Clearing house};



       \draw[<-, line width=2.6pt, dotted, color=\newTradeColor] (4,-1.25) -- (4,1.25);
       \node[align=center, right] (XvaDeskToTradingDesk) at (4,0) {$\left(\stock(t_K) - \strike\right)^+$};

        \draw[->, line width=2.6pt, dotted, color=\hedgeColor] (-0.4,-2.375) -- (2.4,-2.375);
        \node[align=center, below] (WealthTradingToTradingDesk) at (1.0,-2.375) {$\Delta(t_{K-1})\stock(t_K)$};
        \draw[<-, line width=2.6pt, dotted, color=\newTradeColor] (-0.4,-1.625) -- (2.4,-1.625);
        \node[align=center, above] (TradingDeskToWealthTrading) at (1.0,-1.625) {$\left(\stock(t_K) - \strike\right)^+$};

        \draw[->, line width=2.6pt, dotted, color=\interestColor] (-7,-0.75) -- (-7,-2.375) -- (-2.6,-2.375);
        \node[align=center, below] (TreasuryToWealthTrading) at (-4.8,-2.375) {Interest \\ $\wealthTrading(t_{K-1}) \left[\frac{\bank(t_K)}{\bank(t_{K-1})} - 1\right]$};


        \draw[->, line width=2.6pt, dotted, color=\interestColor] (-7,0.75) -- (-7,2.375) -- (-2.6,2.375);
        \node[align=center, above] (TreasuryToWealthXVA) at (-4.8,2.375) {Interest \\ $\wealthXVA(t_{K-1}) \left[\frac{\bank(t_K)}{\bank(t_{K-1})} - 1\right]$};

        \draw[<-, line width=2.6pt, color=\hedgeColor] (4.6,-1.625) --  (9.5,-1.625) -- (9.5,-0.75);
        \node[align=center, above] (ExchangeToTradingDesk) at (6.75,-1.625) {$\Delta(t_{K-1})$ shares of $\stock$};
        \draw[->, line width=2.6pt, dotted, color=\hedgeColor] (4.6,-2.375) -- (10.5,-2.375) -- (10.5,-0.75);
        \node[align=center, below] (TradingDeskToExchange) at (6.75,-2.375) {$\Delta(t_{K-1})\stock(t_K)$};


        \draw[<-, line width=2.6pt, dotted, color=\newTradeColor] (4,2.75) -- (4,4.125) -- (8.9,4.125);
        \node[align=center, below] (ClearingHouseToXVADesk) at (6.75,4.125) {$\left(\stock(t_K) - \strike\right)^+$};


    \end{tikzpicture}
    }
    \end{subfigure}
    \caption{Case with CCR, separate trading desk and $\xva$ desk within the bank, situation at $t_K$.
    The \hedgeColor \ lines correspond to the market risk hedge; the \interestColor \ lines correspond to the interest; the \newTradeColor \ lines correspond to the new trade entered upon default.}
    \label{fig:tKBankRiskyXVADeskDefault}
\end{figure}

Finally, Figure~\ref{fig:tKBankRiskyXVADeskDefault} represents the situation at $t_K$, at which the option payoff is settled in cash, and the trading desk closes its hedging positions.
The output metrics as introduced in Section~\ref{sec:outputMetrics} can be carefully analyzed to draw conclusions about the hedging strategy used.

\end{document}